\documentclass[sigconf]{acmart}
\usepackage[utf8]{inputenc}
\usepackage{microtype}
\usepackage{balance}
\usepackage{xcolor}
\usepackage{tabularx}
\usepackage{colortbl}
\usepackage{booktabs}
\usepackage{multirow}
\usepackage{paralist}
\usepackage{array}
\usepackage{bm}
\usepackage{graphicx, subfig}
\usepackage{tcolorbox}
\usepackage{tikz}
\usetikzlibrary{shadows,calc}
\graphicspath{{figures/}{plots/}}
\usepackage{cleveref}
\usepackage{enumitem}

\def\shadowshift{3pt,-3pt}
\def\shadowradius{6pt}

\colorlet{innercolor}{black!50}
\colorlet{outercolor}{gray!05}

\newcommand\drawshadow[1]{
    \begin{pgfonlayer}{shadow}
        \shade[outercolor,inner color=innercolor,outer color=outercolor] ($(#1.south west)+(\shadowshift)+(\shadowradius/2,\shadowradius/2)$) circle (\shadowradius);
        \shade[outercolor,inner color=innercolor,outer color=outercolor] ($(#1.north west)+(\shadowshift)+(\shadowradius/2,-\shadowradius/2)$) circle (\shadowradius);
        \shade[outercolor,inner color=innercolor,outer color=outercolor] ($(#1.south east)+(\shadowshift)+(-\shadowradius/2,\shadowradius/2)$) circle (\shadowradius);
        \shade[outercolor,inner color=innercolor,outer color=outercolor] ($(#1.north east)+(\shadowshift)+(-\shadowradius/2,-\shadowradius/2)$) circle (\shadowradius);
        \shade[top color=innercolor,bottom color=outercolor] ($(#1.south west)+(\shadowshift)+(\shadowradius/2,-\shadowradius/2)$) rectangle ($(#1.south east)+(\shadowshift)+(-\shadowradius/2,\shadowradius/2)$);
        \shade[left color=innercolor,right color=outercolor] ($(#1.south east)+(\shadowshift)+(-\shadowradius/2,\shadowradius/2)$) rectangle ($(#1.north east)+(\shadowshift)+(\shadowradius/2,-\shadowradius/2)$);
        \shade[bottom color=innercolor,top color=outercolor] ($(#1.north west)+(\shadowshift)+(\shadowradius/2,-\shadowradius/2)$) rectangle ($(#1.north east)+(\shadowshift)+(-\shadowradius/2,\shadowradius/2)$);
        \shade[outercolor,right color=innercolor,left color=outercolor] ($(#1.south west)+(\shadowshift)+(-\shadowradius/2,\shadowradius/2)$) rectangle ($(#1.north west)+(\shadowshift)+(\shadowradius/2,-\shadowradius/2)$);
        \filldraw ($(#1.south west)+(\shadowshift)+(\shadowradius/2,\shadowradius/2)$) rectangle ($(#1.north east)+(\shadowshift)-(\shadowradius/2,\shadowradius/2)$);
    \end{pgfonlayer}
}

\pgfdeclarelayer{shadow} 
\pgfsetlayers{shadow,main}

\newsavebox\mybox
\newlength\mylen

\newcommand\shadowimage[2][]{
\setbox0=\hbox{\includegraphics[#1]{#2}}
\setlength\mylen{\wd0}
\ifnum\mylen<\ht0
\setlength\mylen{\ht0}
\fi
\divide \mylen by 120
\def\shadowshift{\mylen,-\mylen}
\def\shadowradius{\the\dimexpr\mylen+\mylen+\mylen\relax}
\begin{tikzpicture}
\node[anchor=south west,inner sep=0] (image) at (0,0) {\includegraphics[#1]{#2}};
\drawshadow{image}
\end{tikzpicture}}

\copyrightyear{2020}
\acmYear{2020}
\setcopyright{acmlicensed}
\acmConference[SIGIR '20]{Proceedings of the 43rd International ACM SIGIR Conference on Research and Development in Information Retrieval}{July 25--30, 2020}{Virtual Event, China}
\acmBooktitle{Proceedings of the 43rd International ACM SIGIR Conference on Research and Development in Information Retrieval (SIGIR '20), July 25--30, 2020, Virtual Event, China}
\acmPrice{15.00}
\acmDOI{10.1145/3397271.3401031}
\acmISBN{978-1-4503-8016-4/20/07}

\definecolor{orange}{RGB}{255,90,20}
\definecolor{blue}{RGB}{20,90,255}

\newcolumntype{L}[1]{>{\raggedright\let\newline\\\arraybackslash\hspace{0pt}}m{#1}}
\newcolumntype{C}[1]{>{\centering\let\newline\\\arraybackslash\hspace{0pt}}m{#1}}
\newcolumntype{R}[1]{>{\raggedleft\let\newline\\\arraybackslash\hspace{0pt}}m{#1}}

\newcommand{\e}{\ensuremath{ \mbox{\scriptsize{E}} }}

\newcommand{\pl}{2.3cm}
\newcommand{\gb}{\cellcolor{gray!20}}
\newcommand{\pic}[1]{\includegraphics[width=1.35cm]{#1}}
\newcommand{\fpic}[1]{\frame{\pic{#1}}}
\newcommand{\piclarge}[1]{\includegraphics[width=3.1cm]{#1}}
\newcommand{\fpiclarge}[1]{\frame{\piclarge{#1}}}

\begin{document}
\fancyhead{}

\title[Learning Efficient Representations of Mouse Movements to Predict User Attention]{Learning Efficient Representations of Mouse Movements\\ to Predict User Attention}

\author{Ioannis Arapakis}
\affiliation{
  \institution{Telefonica Research}
  \country{Spain}}
\email{ioannis.arapakis@telefonica.com}

\author{Luis A. Leiva}
\affiliation{
  \institution{Aalto University}
  \country{Finland}}
\email{firstname.lastname@aalto.fi}

\begin{abstract}
Tracking mouse cursor movements can be used to predict 
user attention on heterogeneous page layouts like SERPs.
So far, previous work has relied heavily on handcrafted features,
which is a time-consuming approach that often requires domain expertise.
We investigate different representations of mouse cursor movements,
including time series, heatmaps, and trajectory-based images,
to build and contrast both recurrent and convolutional neural networks that can predict user attention to direct displays, such as SERP advertisements.
Our models are trained over \emph{raw} mouse cursor data and achieve competitive performance.
We conclude that neural network models should be adopted for downstream tasks involving mouse cursor movements, since they can provide 
an invaluable implicit feedback signal for re-ranking and evaluation.
\end{abstract}


\keywords{Sponsored search; Online advertising; Mouse cursor; Direct displays; User attention; Neural networks; Transfer Learning}

\maketitle

\section{Introduction}

Search engine results pages (SERPs) have become sophisticated user interfaces (UIs)
that include heterogeneous modules, or \emph{direct displays},
such as image carousels, videos, cards, and a diverse kind of advertisements.
Since users are no longer faced with a text-based linear listing of search results,
research demands more sophisticated ways of understanding how users interact and examine SERPs.
With multiple page elements competing for the user's attention,
understanding which elements do actually attract attention is key to search engines,
and has applications for ranking, search page optimization, and UI evaluation.

Researchers have shown that mouse cursor movements can be used to infer user attention~\cite{Arapakis:2016:PUE:2911451.2911505}
and information flow patterns~\cite{Navalpakkam:2013:MME:2488388.2488471} on SERPs.
While mouse tracking cannot substitute eye tracking technology,
it is nevertheless much more scalable and requires no special equipment.
Further, most queries do not result in a click
if the user can satisfy their information needs directly on the SERP~\cite{Fishkin19},
therefore search engines must rely on other behavioral signals
to understand the underlying search intent.
So, mouse tracking data can be gathered ``for free''
and can provide search engines with an implicit feedback signal for re-ranking and evaluation.
For example, a search engine can predict user attention to individual SERP components
such as the knowledge module~\cite{Arapakis:2015:KYO:2806416.2806591} and re-design it accordingly.
Similarly, predicting attention to advertisements~\cite{Arapakis20_ppaa}
can improve current auction schemes and make them more transparent to bidders.
Those are important and particularly key use cases, 
considering that previous research have assumed a uniform engagement
with a web page and do not distinguish well enough
between attended and ignored layout components~\cite{8010344, Liu:2015:DUD:2766462.2767721}.

Previous work has relied on handcrafted features to model user interaction data on SERPs.
For example, Guo and Agichtein were able to classify different query types~\cite{Guo:2008:EMM:1390334.1390462}
and infer search intent in search results~\cite{Guo:2010:RBJ:1835449.1835473}
by examining, e.g. within-distances between cursor movements, hovers, and scrolling. Similarly, \citet{Arapakis:2016:PUE:2911451.2911505} derived 638 features from mouse cursor data
to predict user attention to direct displays,
and \citet{Lagun:2014:DCM:2556195.2556265} discovered frequent subsequences, or motifs,
in mouse movements that were used to improve search results relevance.
While these are very valuable works and have contributed to our current understanding of search behavior analysis,
finding the right feature set for the task at hand is time-consuming and requires domain expertise.
To solve this, we rely on artificial neural networks (ANNs)
that are trained on different representations of mouse cursor movements.

We build and contrast both recurrent and convolutional ANNs
to predict user attention to SERP advertisements.
We thus tackle the problem of mouse movements classification
using both sequential and pixel-based representations,
the latter using different visual encoding of the temporal information embedded in mouse cursor movements, 
to be described later.
Importantly, our models are trained on \emph{raw} mouse cursor movements,
which do not depend on a particular page structure,
and achieve competitive performance while predicting user attention to different ad formats.
Taken together, our results suggest that ANN-based models should be adopted for downstream tasks involving mouse cursor movements,
as these models remove the need for handcrafted features and can capture better non-linearities within the data.

\subsection{Preliminaries} 
\label{ssec:prelim}

Arguably, ANNs are universal function approximators~\cite{Hornik91, Csaji01},
since a feedforward network (i.e. with no loops) having a single hidden layer with a sufficiently large number of neurons
can approximate \emph{any} continuous function of $n$-dimensional input variables~\cite{Lu17}.
For most Deep Learning practitioners, sequence modeling is synonymous with Recurrent Neural Networks (RNNs).
RNNs are an extension of regular ANNs that have connections feeding the hidden layers of the network back into themselves,
also know as recurrent connections or feedback loops.
Therefore it might seem clear that we should use RNNs to process mouse tracking data,
since this kind of data can be straightforwardly modeled as multivariate time series of spatial (or spatiotemporal) cursor coordinates,
where each coordinate can be assumed to depend on the previous one.
Yet recent research has suggested that convolutional neural networks (CNNs) may outperform RNNs
on tasks dealing with sequential data, such as audio synthesis and machine translation~\cite{Bai18}.
CNNs are also an extension of regular ANNs, but using feedforward connections instead of recurrent connections
and 
assembling complex hierarchical patterns via smaller and simpler patterns through convolutional operations.

A known issue of training Deep Learning models is that
gradients may either vanish or explode while they backpropagate through the network.
This problem is particularly exacerbated in RNNs,
due to their long-term dependencies within the data.
However, mouse cursor trajectories are sometimes very short, e.g. a few seconds worth of interaction,
therefore in this paper we explore both simple RNNs and more sophisticated versions thereof:
Long Short-Term Memory (LSTM) and Gate Recurrent Unit (GRU) networks.
Both networks have similar performance~\cite{Jozefowicz15}
and were designed to learn long-term information.
We also investigate the bidirectional LSTM network,
which allows RNNs to learn from past \emph{and} future timesteps,
to better understand the sequence context.

Another important limitation while training RNNs is the fine-tuning of many model parameters (network weights),
because every timestep depends on the previous one, which usually require high computational resources.
Indeed, the temporal dependencies between previous sequence elements prevents parallelizing training of RNNs~\cite{Martin18}.
Therefore, in this paper we explore alternative pixel-based representations of mouse cursor data that can be handled with CNNs.
Because CNNs usually require a large amount of training data, 
a common technique is \emph{transfer learning}: use a pre-trained network on a larger dataset
and calibrate the model architecture to the nature and characteristics of the smaller dataset.
Concretely, we used transfer learning of popular CNN architectures including
AlexNet~\cite{AlexNet}, SqueezeNet~\cite{SqueezeNet}, ResNet~\cite{ResNet}, and VGGNet~\cite{VGGNet},
all of them state-of-the-art CNNs and widely used in downstream tasks
such as image classification or video analysis.

\section{Related Work}
\label{sec:related_work}

The construct of attention has become the common currency on the Web. Objective measurements of attentional processes~\cite{Wright2008-WRIOOA} are increasingly sought after by both the media industry and scholar communities to explain or predict user behavior. Along those lines, the connection between mouse cursor movements and the underlying psychological states has been a topic of research since the early 90s~\cite{Accot1997, Accot:1999, Card:1987, MacKenzie:2001}. Some studies have investigated the utility of mouse cursor data for predicting the user’s emotional state~\cite{Zimmermann2003, Kaklauskas2009, Azcarraga:2012, Yamauchi:2013, Kapoor:2007}, but also the extent that they can help identify demographic attributes like gender~\cite{Yamauchi2014, Kratky:2016, Pentel2017} and age~\cite{Kratky:2016, Pentel2017}. The above works demonstrate that certain cognitive and motor control mechanisms are embodied and reflected, to some extent, in our mouse cursor movements and online interactions.

Recently, a large body of research~\cite{Shapira:2006:SUK:1141277.1141542, Guo:2008:EMM:1390334.1390462, Guo:2010:RBJ:1835449.1835473, Guo:2012:PWS:2396761.2398570, Huang:2012:USU:2207676.2208591, Navalpakkam:2013:MME:2488388.2488471, Lagun:2014:DCM:2556195.2556265, Liu:2015:DUD:2766462.2767721, MARTINALBO2016989, 8010344} established further the cognitive grounding for hand-eye relationship and has demonstrated the utility of mouse cursor analysis as a low-cost and scalable proxy of visual attention, especially on SERPs.
In line with this evidence, several works have investigated closely the user interactions
that stem from mouse cursor data for various use cases, such as web search~\cite{Guo:2008:EMM:1390334.1390462, Guo:2010:RBJ:1835449.1835473, Guo:2012:PWS:2396761.2398570, Lagun:2014:DCM:2556195.2556265, Liu:2015:DUD:2766462.2767721, Arapakis:2016:PUE:2911451.2911505, 8010344} or web page usability evaluation~\cite{Atterer:2006:KUM:1135777.1135811, Arroyo:2006:CPE:1125451.1125529, Leiva:2011:RWD:2037373.2037467}.
In what follows, we review previous research efforts that have focused on mouse cursor analysis to predict user interest and attention.

\subsection{User Interest in Web Search Tasks}
\label{ssec:interest}

User models of scanning behaviour in SERPs have been assumed to be linear,
as users tend to explore the list of search results from top to bottom.
However, today's SERPs include several direct displays
such as image and video search results, featured snippets, or advertisement.
To account for this SERP heterogenity, \citet{Diaz13} incorporated ancillary page modules to the classic linear scanning model,
which proved useful to help improving SERP design by anticipating searchers' engagement patterns for a given SERP arrangement.
However, this model was not designed to measure effectively user attention to specific direct displays
and does not exploit the latent information encoded in mouse cursor movements.

Another line of research considered simple, coarse-grained features derived from mouse cursor data
to be surrogate measurements of user interest, such as the amount of mouse cursor movements~\cite{Shapira:2006:SUK:1141277.1141542}
or mouse cursor's ``travel time''~\cite{Claypool:2001:III:359784.359836}.
Follow up work adopted fine-grained mouse cursor features, which have been shown to be more effective.
For example,
Guo et al.~\cite{Guo:2008:EMM:1390334.1390462, Guo:2010:RBJ:1835449.1835473}
computed within-distances between mouse cursor distances to disambiguate among informational and navigational queries,
and could identify a user's research or purchase intent based on aggregated behavioral signals
that include, among others, mouse hovering and scrolling activity.
Approaches like these have been directed at predicting general-purpose web-based tasks
like search success~\cite{Guo:2012:PWS:2396761.2398570} and satisfaction~\cite{Liu:2015:DUD:2766462.2767721},
user's frustration~\cite{Feild:2010:PSF:1835449.1835458},
relevance judgements of search results~\cite{Huang:2012:ISM:2348283.2348313, Speicher:2013:TPR:2505515.2505703},
and query abandonment~\cite{Huang:2011:NCN:1978942.1979125, Diriye:2012:LSS:2396761.2398399}.
Eventually, they lack the granularity in predicting attention with particular direct displays of a SERP,
such as advertisements, that our proposed modelling approach achieves.

\begin{figure*}[!tpb]
  \def\w{0.31\linewidth}
  \subfloat[Organic ad\label{fig:native-top-left}]{
    \hspace{-1em}\shadowimage[width=\w]{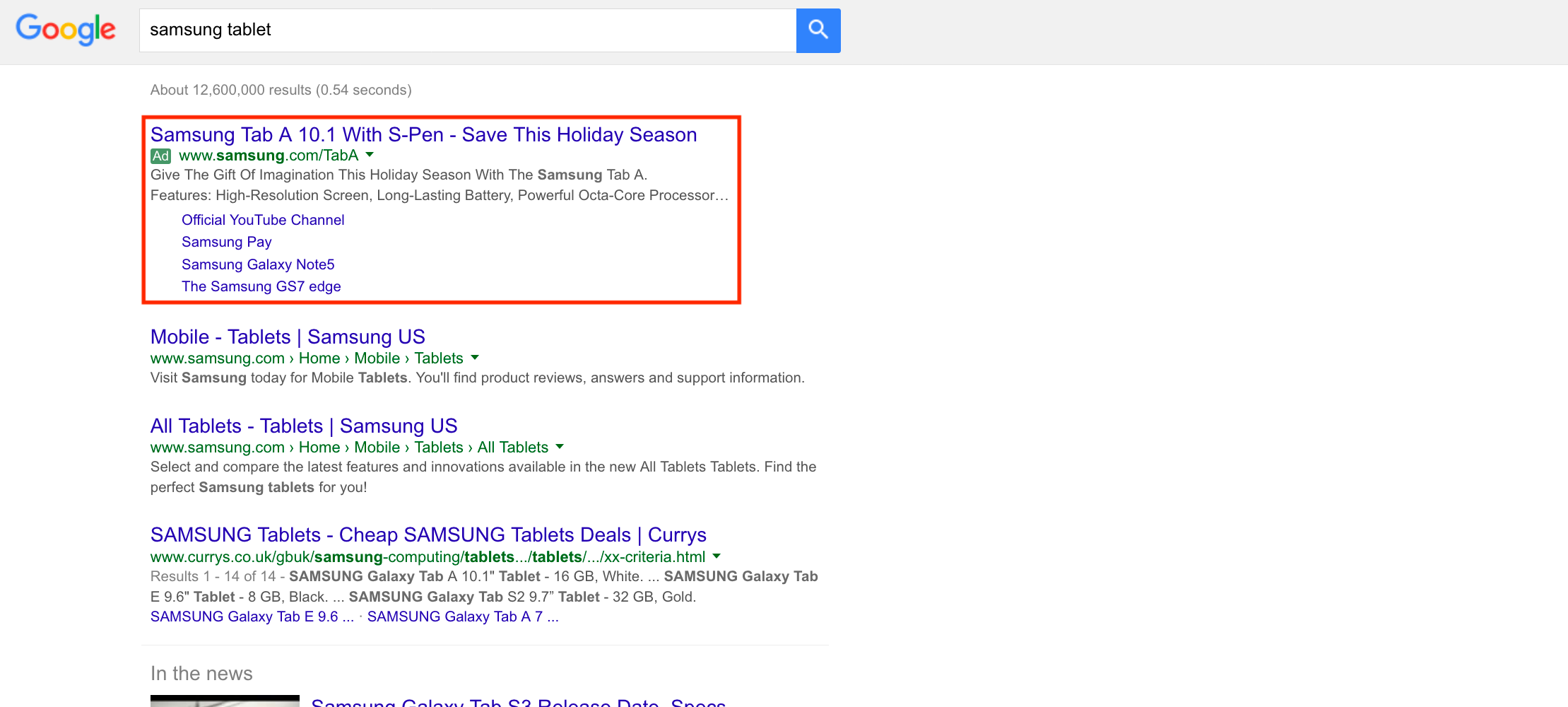}
  }
  \subfloat[Direct display ad, left-aligned\label{fig:dd-top-left}]{
    \hspace{-1em}\shadowimage[width=\w]{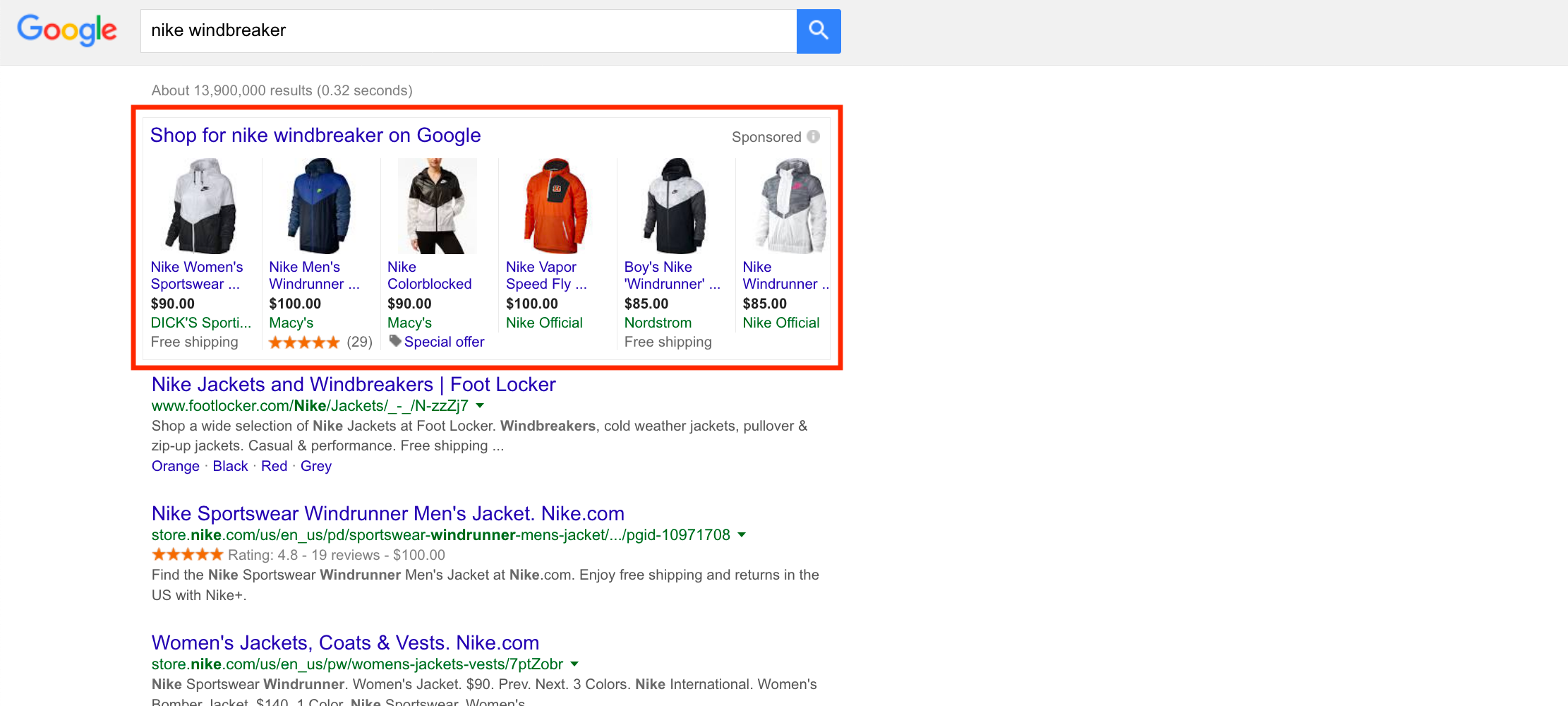}
  }
  \subfloat[Direct display ad, right-aligned\label{fig:dd-top-right}]{
    \hspace{-1em}\shadowimage[width=\w]{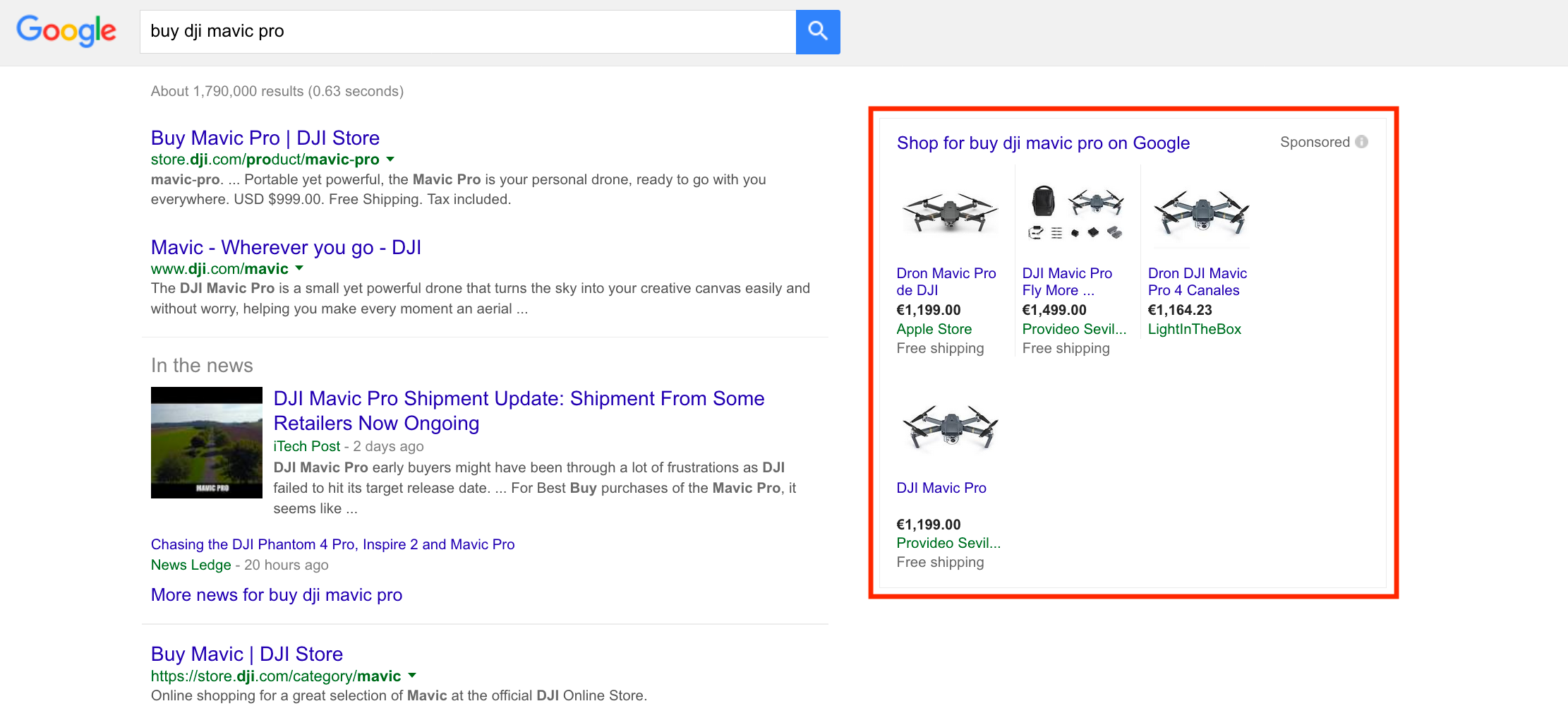}
  }
  \caption{
    Examples of the ad formats, highlighted in red, and their positions on the Google SERP:
    Organic ad~\protect\subref{fig:native-top-left}
    vs. left-aligned~\protect\subref{fig:dd-top-left}
    and right-aligned~\protect\subref{fig:dd-top-right} direct display ads.
    In our experiments, only one ad format was visible at a time.
  }
  \label{fig:display_ads}
\end{figure*}

\subsection{User Attention in Web Search Tasks}
\label{ssec:attention}

Most research studies assume that eye fixation means examination~\cite{Brightfish18}.
However, \citet{Liu:2014:SRT:2661829.2661907} reports that about half of the search results fixated by users are not actually read, since there is often a preceding skimming step in which the user quickly scans the search results.
Based on this observation, they propose a two-stage examination model:
a first ``from skimming to reading'' stage and a second ``from reading to clicking'' stage.
Interestingly, they showed that both stages can be predicted with mouse movement behaviour,
which can be collected at large scale.

Cursor movements can therefore be used to estimate user attention on SERP components,
including traditional snippets, aggregated results, maps, and advertisements, among others.
However, works that employ mouse cursor information to predict user attention with specific elements within a web page have been scarce.
Despite these challenges, some of the early work by~\citet{Arapakis:2014:UEO:3151365.3151368, Arapakis:2014:UWE:2661829.2661909} investigated the utility of mouse movement patterns to measure within-content engagement on news pages and predict reading experiences. \citet{Lagun:2014:DCM:2556195.2556265} introduced the concept of \emph{motifs} 
for estimating results relevance.
Similarly, \citet{Liu:2015:DUD:2766462.2767721} applied the motifs concept to SERPs
to predict search result utility, searcher effort, and satisfaction at a search task level.
Finally, \citet{Arapakis:2016:PUE:2911451.2911505} which investigated user engagement with direct displays on SERPs,
concretely with the Knowledge Graph~\cite{Arapakis:2015:KYO:2806416.2806591}.

Our work differs significantly from previous art in several ways.
First, we implement a predictive modelling framework to measure user attention to SERP advertisements,
which are probably the most relevant instance of direct displays for search engines, from a business perspective.
Second, previous work has used Machine Learning models which rely on ad-hoc and domain-specific features.
As previously discussed, feature engineering requires domain expertise to come up with the best discriminative features.
In contrast, we investigate several ANN architectures that use \emph{raw} mouse cursor data, represented either as time series or as visual representations, and can predict user attention with competitive performance.
Finally, we examine the performance of our predictive models w.r.t. sponsored ads served under different formats and different positions within a SERP and, thus, significantly expand on previous research and findings in the community.

\section{User Study}
\label{sec:user_study}

Online advertising comprises ads that are served under different formats (e.g. text, image, or video, or rich media),
each with its unique look and feel.
Some formats appear to be more effective than traditional online ads
in terms of user attention and purchase intent~\cite{sharethrough2013},
but also may cause ``ad blindness'' to a greater or a lesser extent~\cite{Owens:2011:TAB:2007456.2007460}.
Therefore, to understand how web search users engage with ads that appear under different formats and positions in SERPs,
we conducted a user study through the \textsc{Figure Eight}\footnote{\url{https://www.figure-eight.com}} crowdsourcing platform.
We collected feedback from participants who performed brief transactional search tasks using Google Search and aimed to predict when users notice the ads that appear on SERPs, under different conditions.
To mitigate low-quality responses, several preventive measures were put into practice,
such as introducing gold-standard questions,
selecting experienced contributors (Level 3) with high accuracy rates,
and monitoring task completion time,
thus ensuring the internal validity of our experiment.

\subsection{Experiment Design}
\label{ssec:design}

We used a between-subjects design with two independent variables:
(1)~ad format, with two levels (organic and direct display ads)
and (2)~ad position, with two levels (top-left and top-right position).
Notice that organic ads are only shown in the left part of Google SERPs; see \autoref{fig:display_ads}.
The dependent variable was ad attention.

Our experiment consisted of a brief transactional search task
where participants were presented with a predefined search query and the corresponding SERP,
and were asked to click on any element of the page that answered it best.
All search queries (\autoref{ssec:search_query_sample}) triggered both organic (\autoref{fig:native-top-left})
and direct display ads (\Cref{fig:dd-top-left,fig:dd-top-right}) on Google SERPs.
Each participant was randomly assigned a search query and could perform the task only once,
since inquiring at post-task about the presence of an ad would make them aware of it
and could introduce carry over effects.
In summary, each participant was only exposed to a unique combination of query, ad format, and ad position.

\subsection{Search Query Sample}
\label{ssec:search_query_sample}

Starting from \textsc{Google Trends},\footnote{\url{https://trends.google.com/trends/}}
we selected a subset of the Top Categories and Shopping Categories
that were suitable candidates for the transactional character of our search tasks.
From this subset of categories, we extracted the top search queries issued in the US during the last 12 months.
Next, from the resulting collection of 375 search queries,
we retained 150 for which the SERPs were showing at least one direct display ad
(50 search queries for each combination of direct display ad format and position).
Using this final selection of search queries,
we produced the static version of the corresponding Google SERPs
and injected the JavaScript code (\autoref{ssec:mouse_cursor_tracking})
that allowed us to control the ads format and capture all client-side user interactions.

\subsection{SERP Layout}
\label{ssec:serp}

All SERPs were all in English and were scraped for later instrumentation (\autoref{ssec:mouse_cursor_tracking}).
Participants accessed the instrumented SERPs through a dedicated server,
which did not alter the look and feel of the original Google SERPs.
This allowed us to capture fine-grained user interactions
while ensuring that the content of the SERPs remained consistent
and that each experimental condition was properly administered. All SERPs had both organic and direct display ads. Organic ads appeared both at the top-left and bottom-left position of the SERP,
whereas direct display ads could appear either at the top-right or top-left position
(but not both at the same time on the same SERP).
Therefore, we ensured that only one ad was visible per condition and participant,
since we are focusing on the single-slot auction case.
This was possible by instrumenting each downloaded SERP with custom JavaScript code
that removed all ads except the one that was tested in each of the experimental conditions (\autoref{fig:display_ads}).
For example, bottom-most organic ads were not shown,
since (i)~users have to scroll all way down to the bottom of the SERP to reveal them
and (ii)~these ads have the same look and feel than the organic ads shown on the top-most position.

\subsection{Mouse Cursor Tracking}
\label{ssec:mouse_cursor_tracking}

We inserted JavaScript code that captured mouse cursor movements
and associated metadata while users browsed the SERPs.
We used \textsc{EvTrack},\footnote{\url{https://github.com/luileito/evtrack}}
a general-purpose open-source JavaScript event tracking library
that allows event capturing either via event listeners (the event is captured as soon as it is fired)
or via event polling (the event is captured at fixed-time intervals).
We captured \texttt{\small{mousemove}} events via event polling,
every 150\,ms to avoid unnecessary data overhead~\cite{LEIVA2015114},
and all the other browser events (e.g., \texttt{\small{load}}, \texttt{\small{click}}, \texttt{\small{scroll}}) via event listeners.
Whenever an event was recorded, we logged the following information:
mouse cursor position ($x$ and $y$ coordinates), timestamp, event name,
and the XPath of the DOM element that relates to the event.

\subsection{Self-Reported Ground-truth Labels}
\label{ssec:self_reported_measures}
In a similar vein to previous work~\cite{Feild:2010:PSF:1835449.1835458, Liu:2015:DUD:2766462.2767721, Lagun:2014:DCM:2556195.2556265, Arapakis:2016:PUE:2911451.2911505},
we collected ground-truth labels through an online questionnaire,
which was administered at post-task
and asked the user \emph{to what extent} they paid attention to the ad
using a 5-point Likert-type scale:
``Not at all''~(1),
``Not much''~(2),
``I can't decide''~(3),
``Somewhat''~(4),
and ``Very much''~(5).
These scores would be collapsed to binary labels,
but we felt it was necessary to begin with a 5-point Likert scale for several reasons.
Unlike other scales that offer limited options (e.g. 2 or 3-point scales)
and can result in highly skewed or neutral data~\cite{Johnson82},
or scales with too many options (7-point scales) that are harder to understand,
a 5-point Likert-type scale leaves room for ``soft responses'' while remaining fairly intuitive.
Neutral scores were not considered for analysis.

\subsection{Participants}
\label{ssec:participants}

We recruited $3,206$ participants, of age $18-66$ and of mixed nationality, through the \textsc{Figure Eight} platform.
All participants were proficient in English and were experienced (Level 3) contributors,
i.e. they had a track record of successfully completed tasks and of a different variety,
thus being considered very reliable contributors.

\subsection{Procedure}
\label{ssec:procedure}

Participants were informed that they should perform the search task
from a desktop or laptop computer as long as they used a computer mouse.
They were told to deactivate any ad-blocker before proceeding with the task,
otherwise our JavaScript code would prevent them from taking part in the study.
Participants were asked to act naturally and click on anything that would best answer the search query, e.g. result links, images, etc. An example of the search task descriptions provided to the participants is the following:
``\emph{You want to buy a Rolex watch and you have submitted the search query `rolex watches' to Google Search.
Please browse the search results page and click on the element that you would normally select under this scenario.}''
The search task had to be completed in a single session and each query was performed on average by five different participants.
The SERPs were randomly assigned to the participants
and each participant could take the study only once.
Upon concluding the search task, participants were asked to complete the post-task questionnaire
which inquired about the presence of the ad (at that point the participants did not have access to the webpage). 
The payment was \$0.20 and participants could also opt out at any moment,
in which case they would not be compensated.

\subsection{Dataset}
\label{ssec:data}

After excluding those logs with incomplete mouse cursor data
(less than five mouse coordinates $\approx$ one second of user interaction data),
we concluded to $2,289$ search sessions that hold $45,082$ mouse cursor positions\footnote{The dataset is freely available here: \url{https://gitlab.com/iarapakis/the-attentive-cursor-dataset.git}}.
Of these search sessions, $763$ correspond to the organic ad condition,
$793$ correspond to the left-aligned direct display ad,
and $733$ correspond to the right-aligned direct display ad.
Ground-truth labels were converted to a binary scale using the following mapping:
``Not at all'' and ``Not much'' were assigned to the negative class,
and ``Somewhat'' and ``Very much'' were assigned to the positive class, while neutral scores were not considered in subsequent analysis.
We note that the class distribution was fairly balanced (66\% of positive cases) across the experimental conditions.
We then used 60-10-30 (\%) disjoint stratified splits to assign the observations to the training, validation, and test set, respectively.
The stratified sampling process was performed once per ad format
and preserved the original class distribution in each data partition.

\section{Data Representations}

We framed the problem of ad attention prediction as a binary classification task:
\textit{given a user's mouse cursor trajectory, did the user notice the ad?}
To this end, as introduced in \autoref{ssec:prelim},
we implemented both recurrent and convolutional neural networks
to handle different representations of mouse cursor movements on SERPs.
In what follows, we describe these data representations.

\bgroup
\def\arraystretch{0.7}
\begin{table*}[!ht]
    \caption{
        Types of visual representations used to train the CNN models.
        The top row shows representations without the ad placeholder,
        whereas the bottom row shows representations with the ad placeholder.
    }
    \vspace{-5pt}
    \label{tbl:representations}
    \centering
    {\footnotesize
    \begin{tabular}{@{}lC{3.15cm}C{3.15cm}C{3.15cm}C{3.15cm}C{3.15cm}@{}}
        \toprule
        \addlinespace
        & \textbf{Heatmap} &
          \textbf{Trajectories} &
          \textbf{Colored trajectories} &
          \textbf{Trajectories with line thickness} &
          \textbf{Colored trajectories with line thickness} \\ \addlinespace
        
        {\rotatebox[origin=c]{90}{w/o ad placeholder}} &
        \fpiclarge{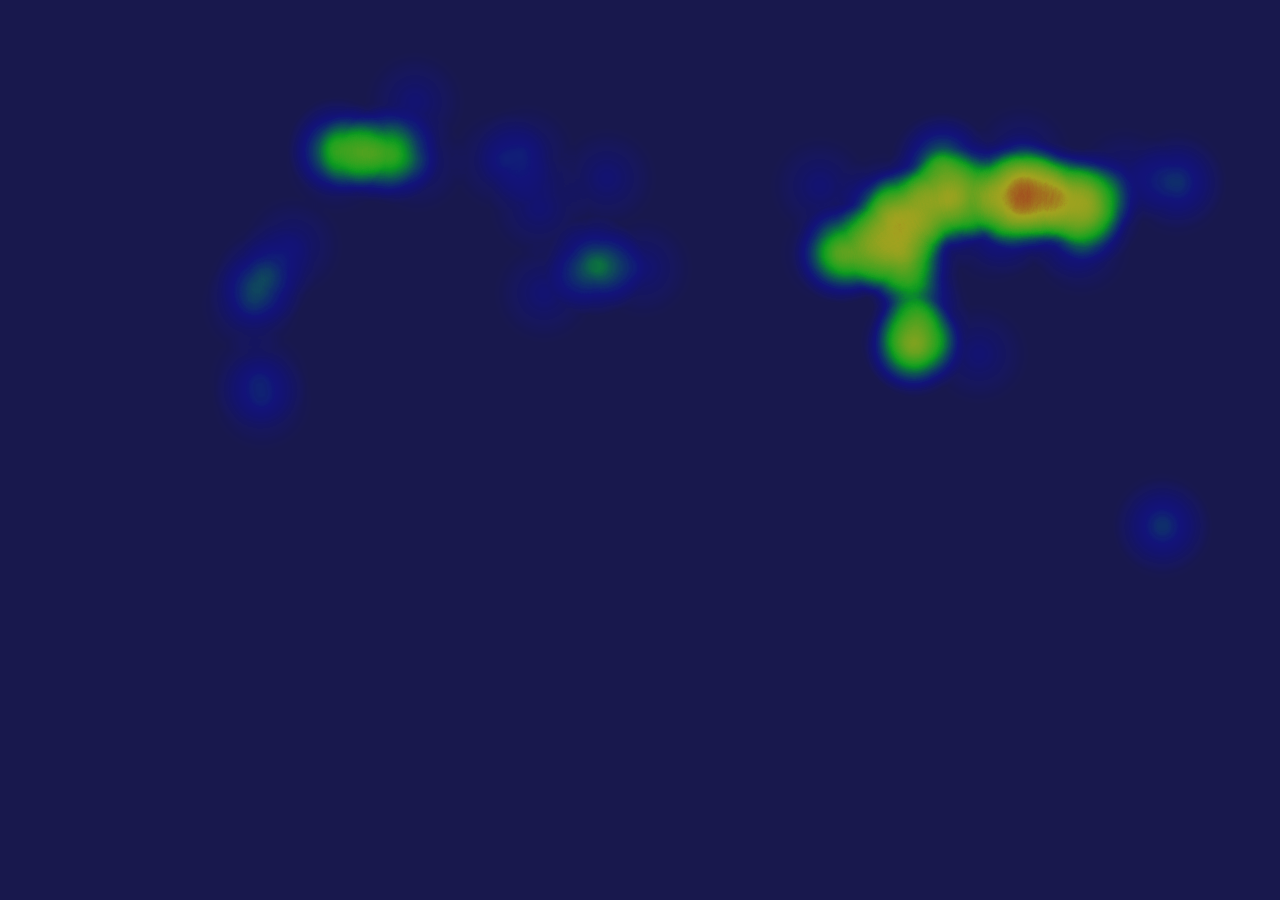}   &
        \fpiclarge{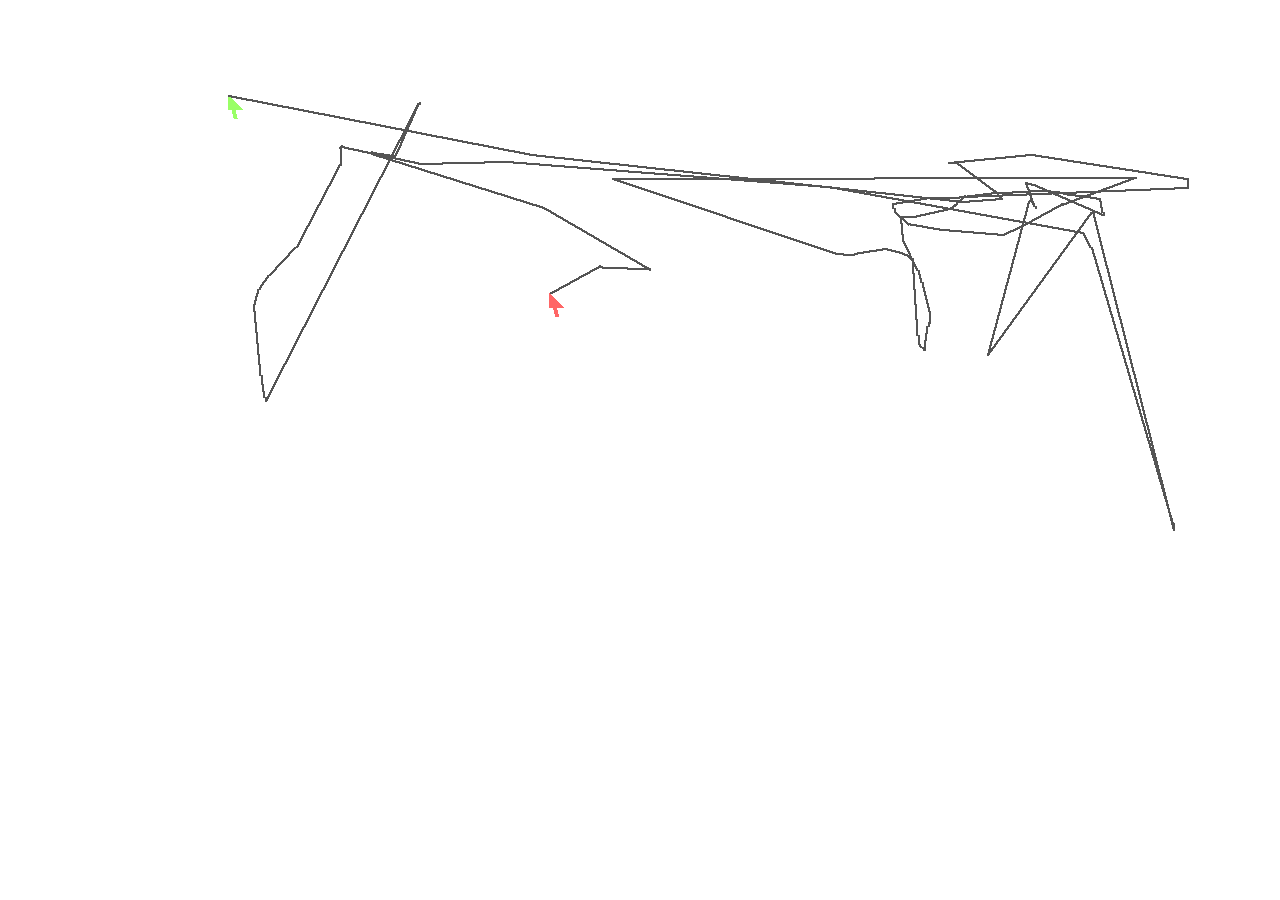}   &
        \fpiclarge{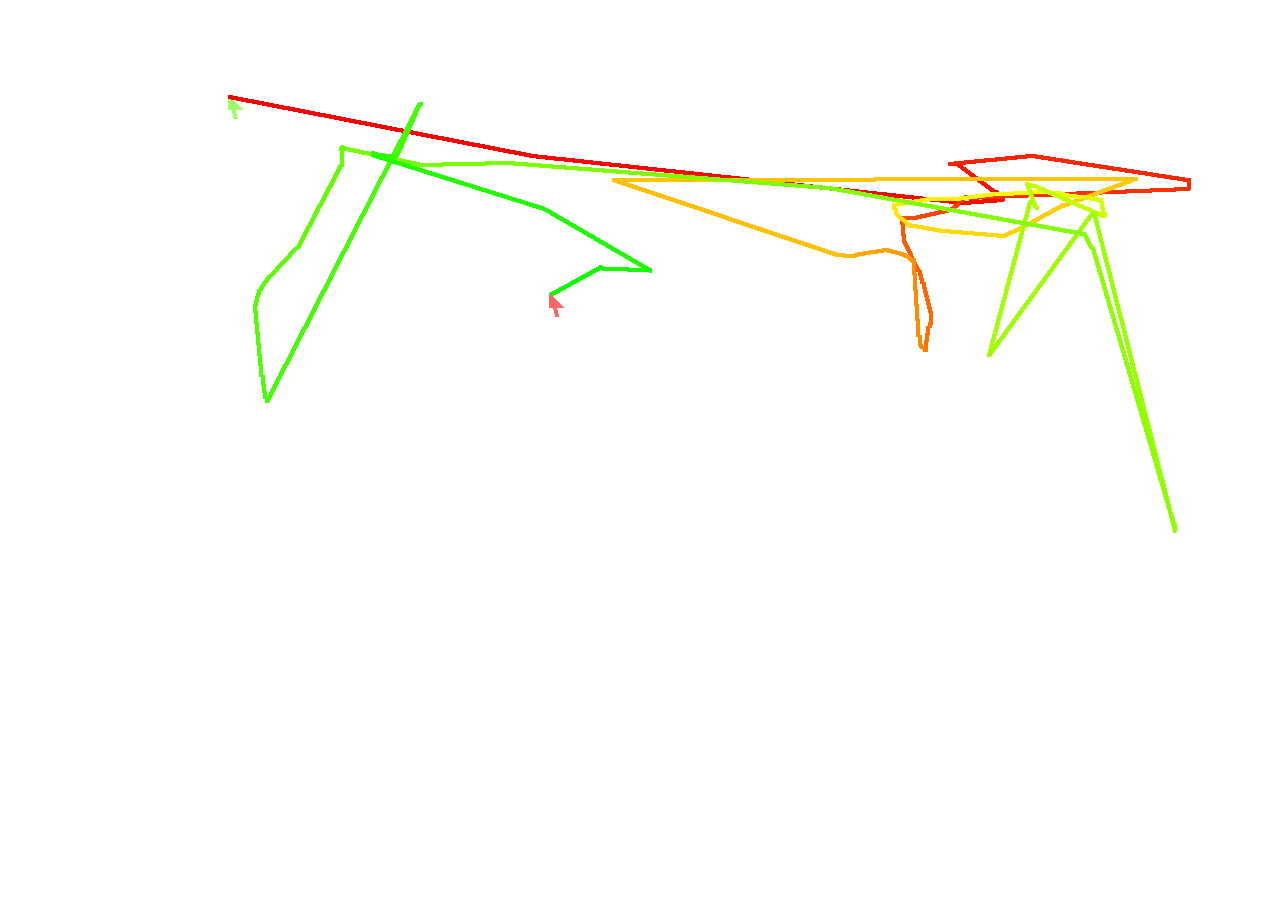}   &
        \fpiclarge{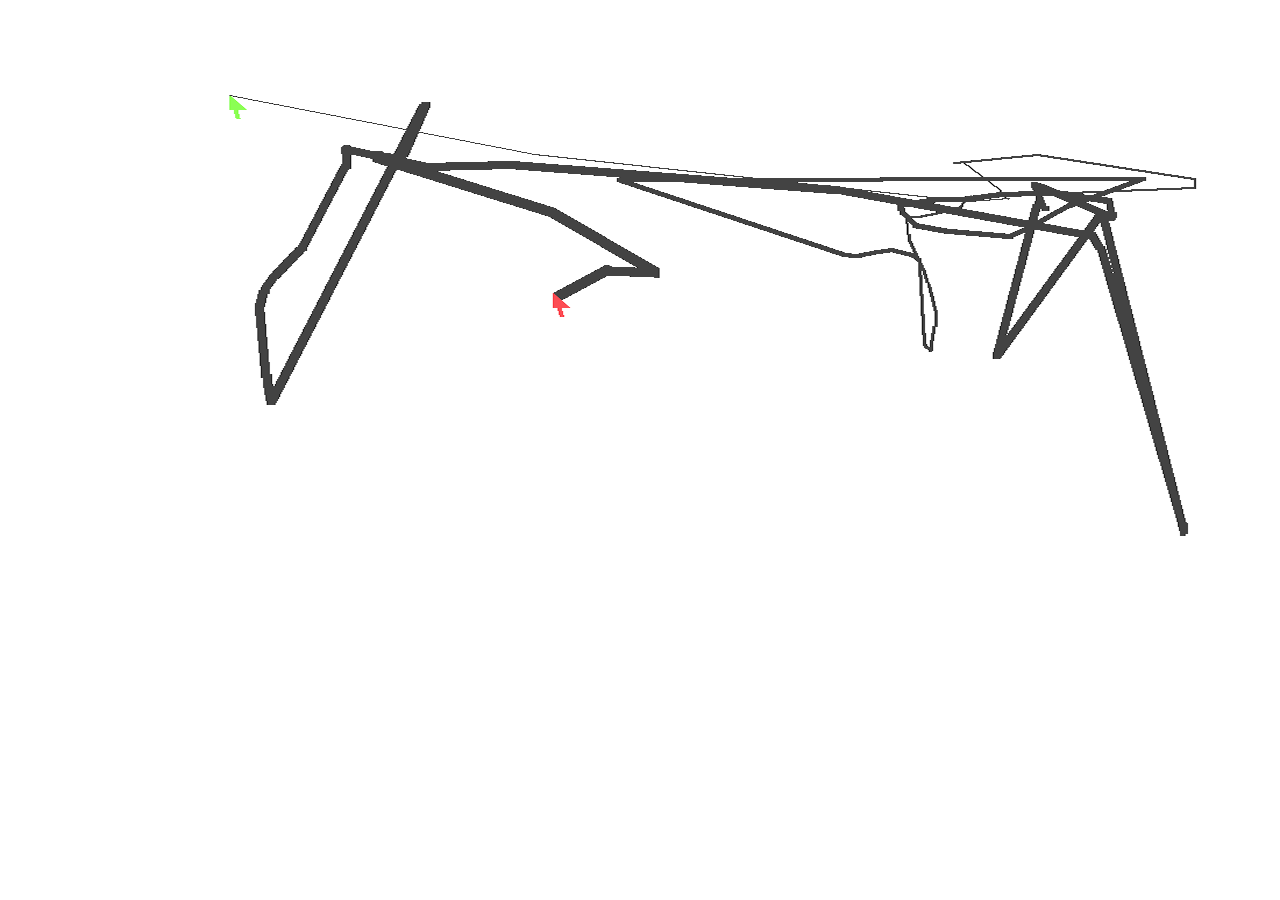}   &
        \fpiclarge{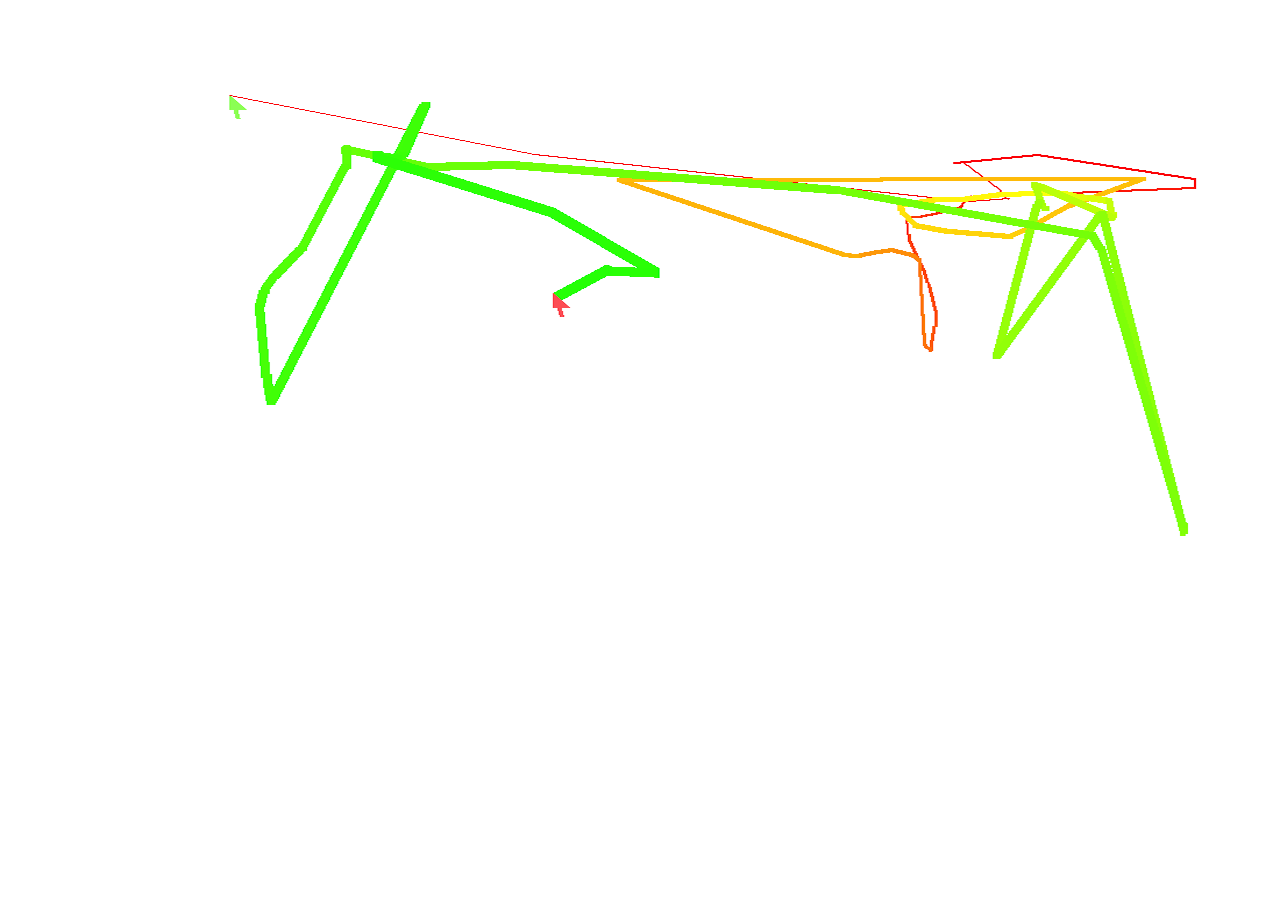} \\ \addlinespace
        & \textbf{(1)} &
          \textbf{(2)} &
          \textbf{(3)} &
          \textbf{(4)} &
          \textbf{(5)} \\[0.5em]
        
        {\rotatebox[origin=c]{90}{w/ ad placeholder}} &
        \fpiclarge{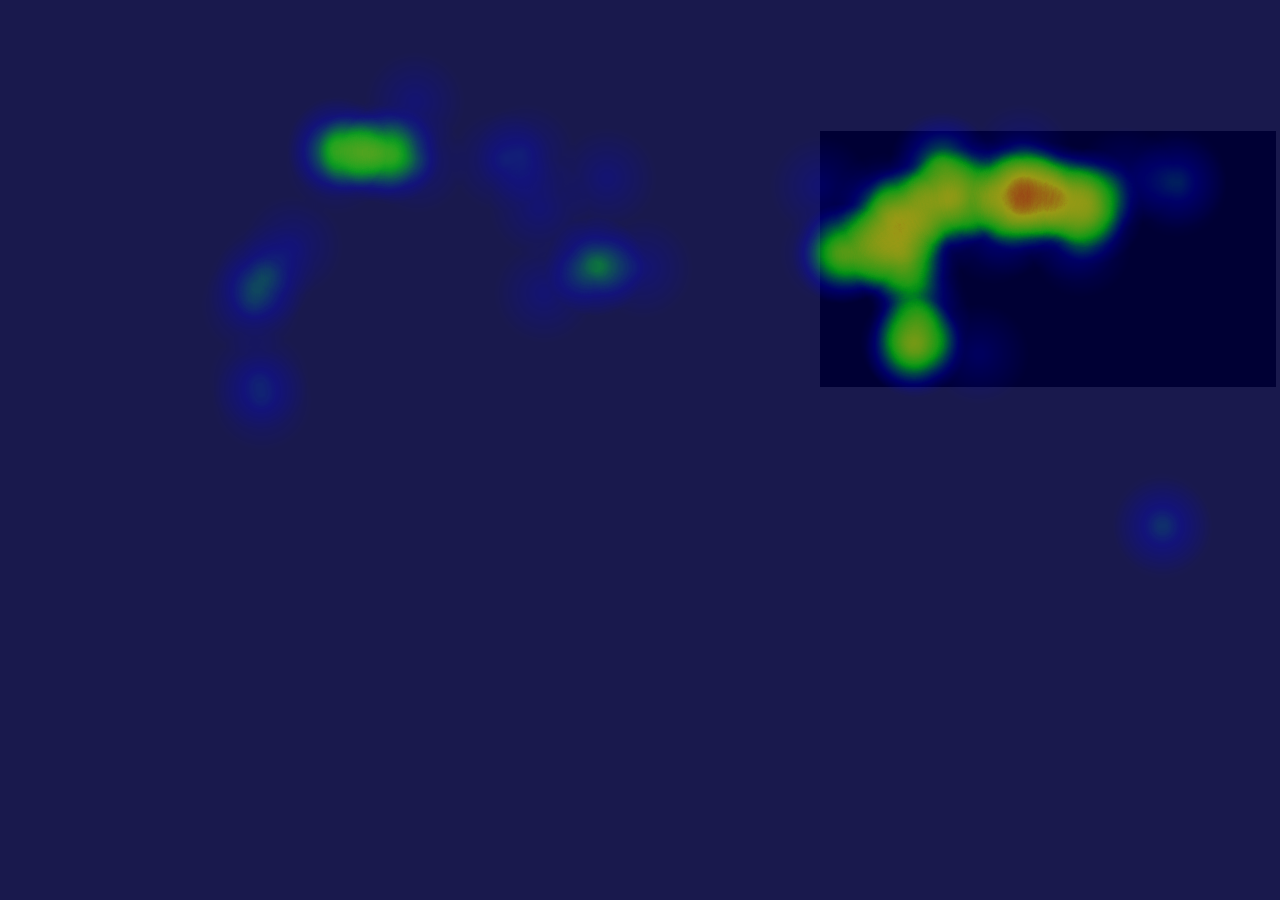}    &
        \fpiclarge{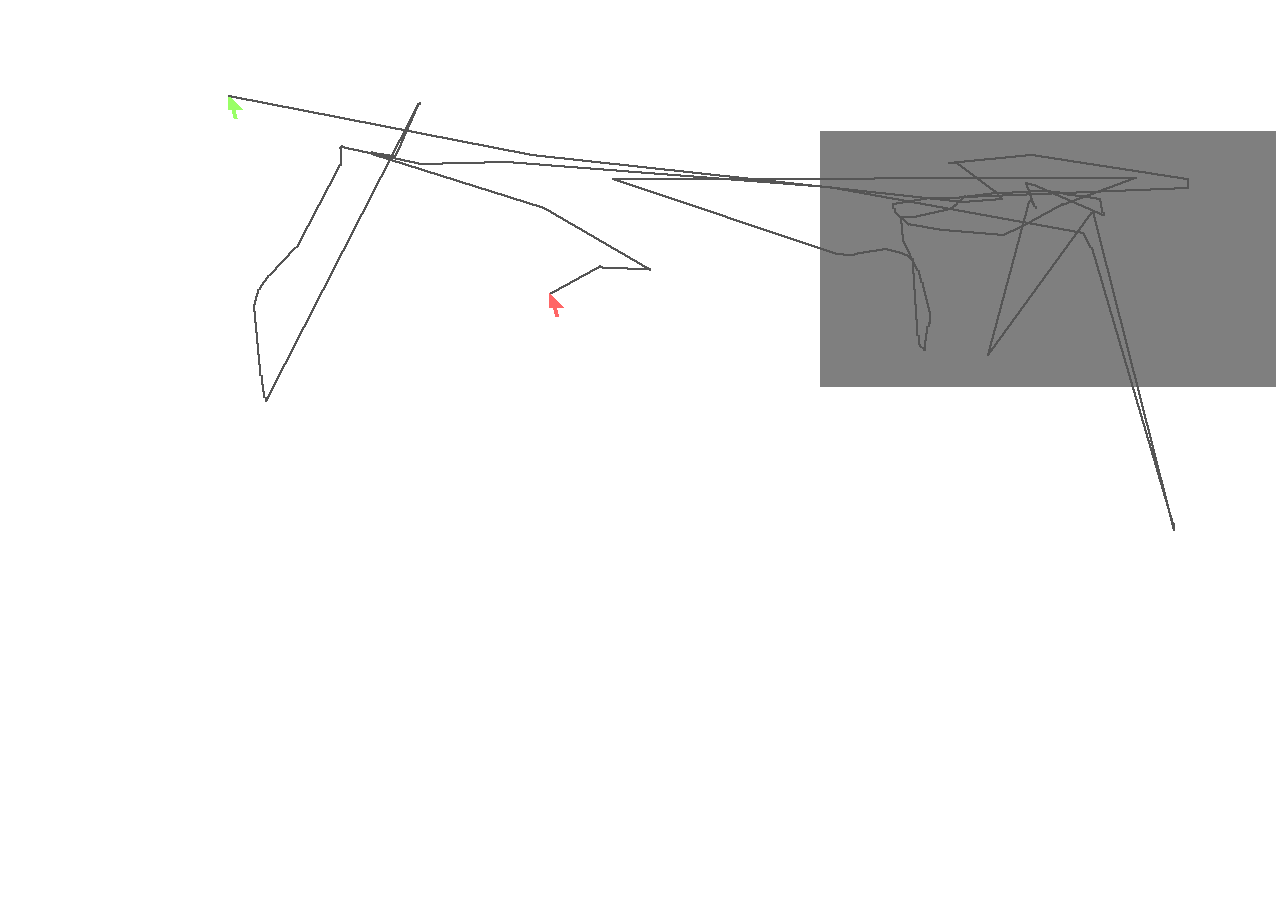}   &
        \fpiclarge{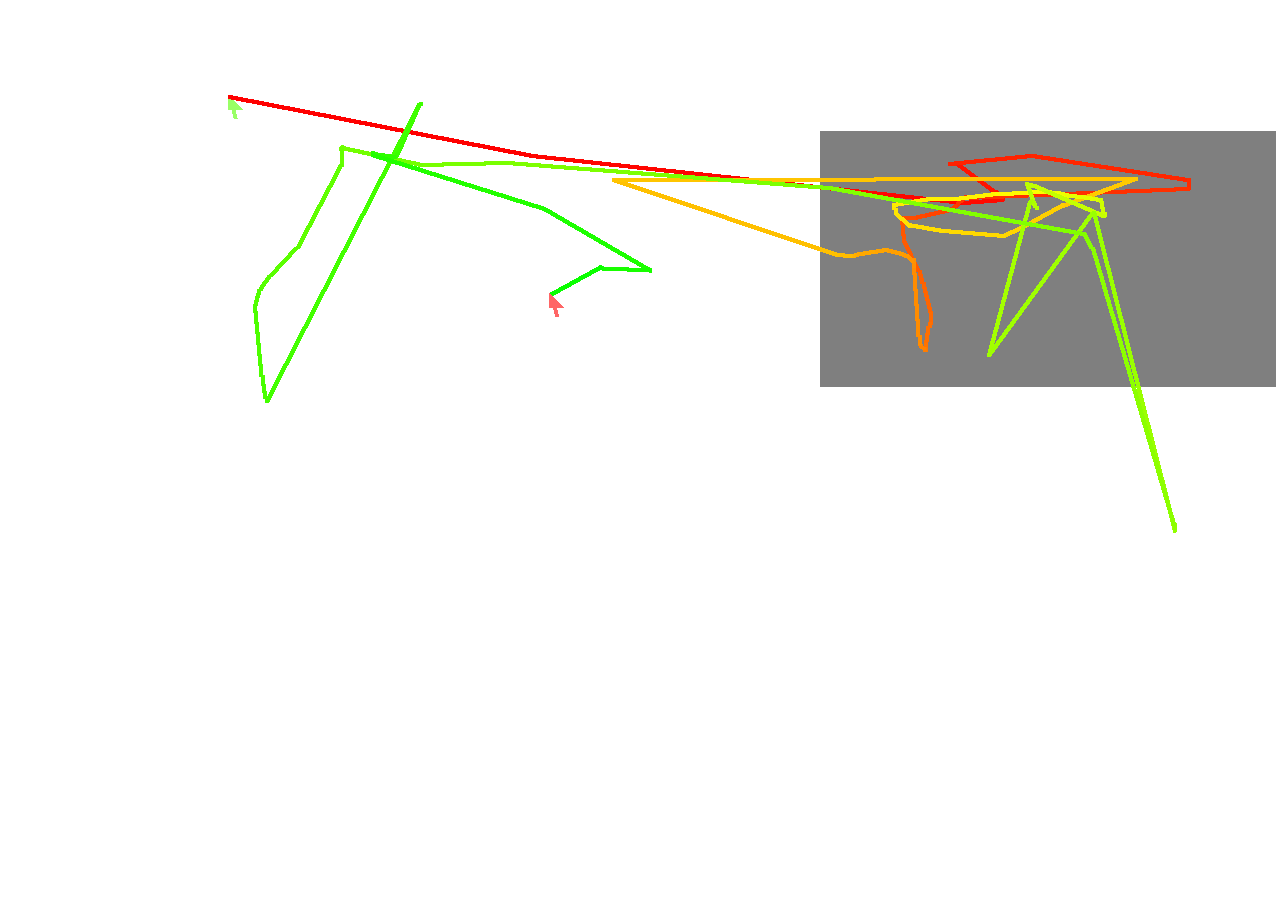}    &
        \fpiclarge{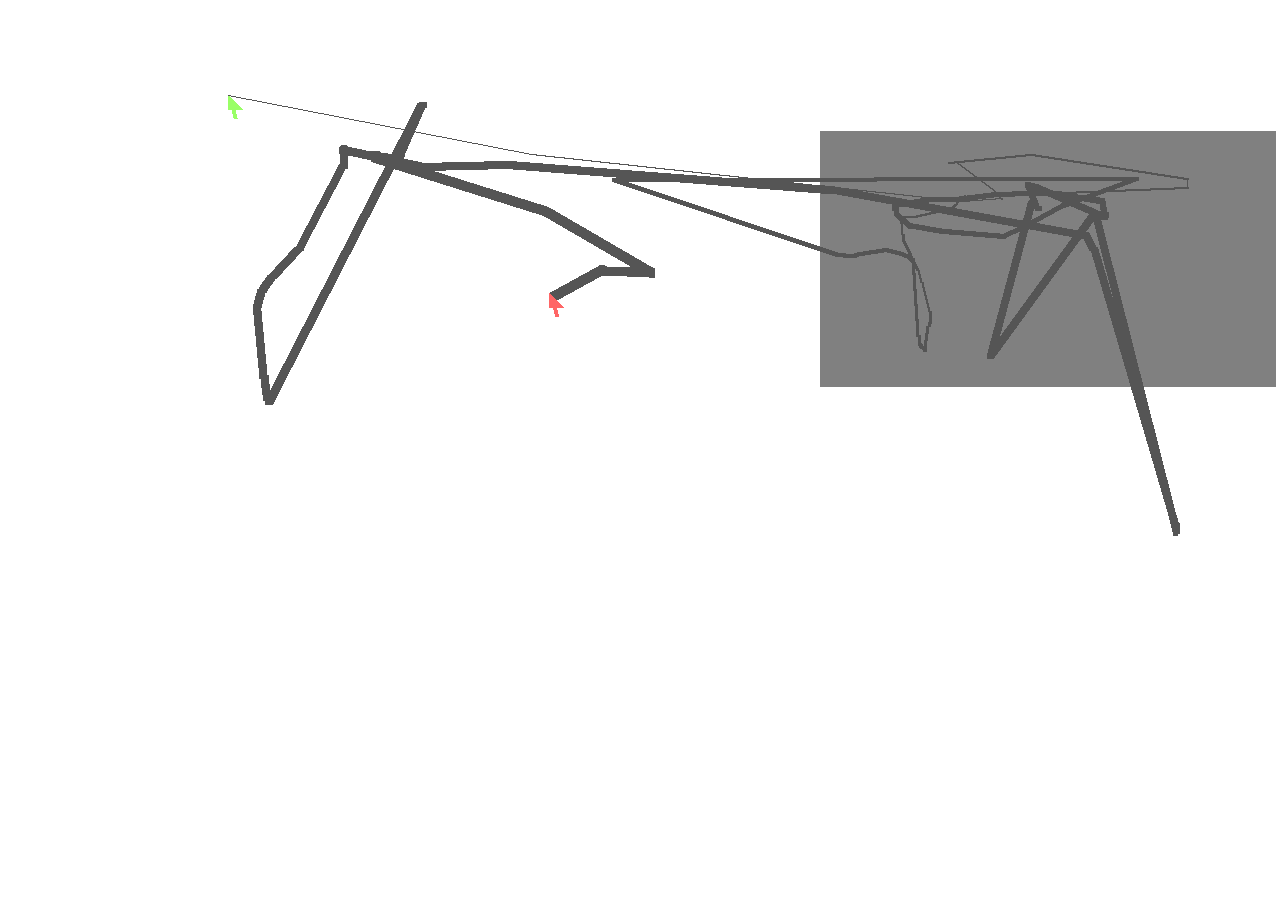}    &
        \fpiclarge{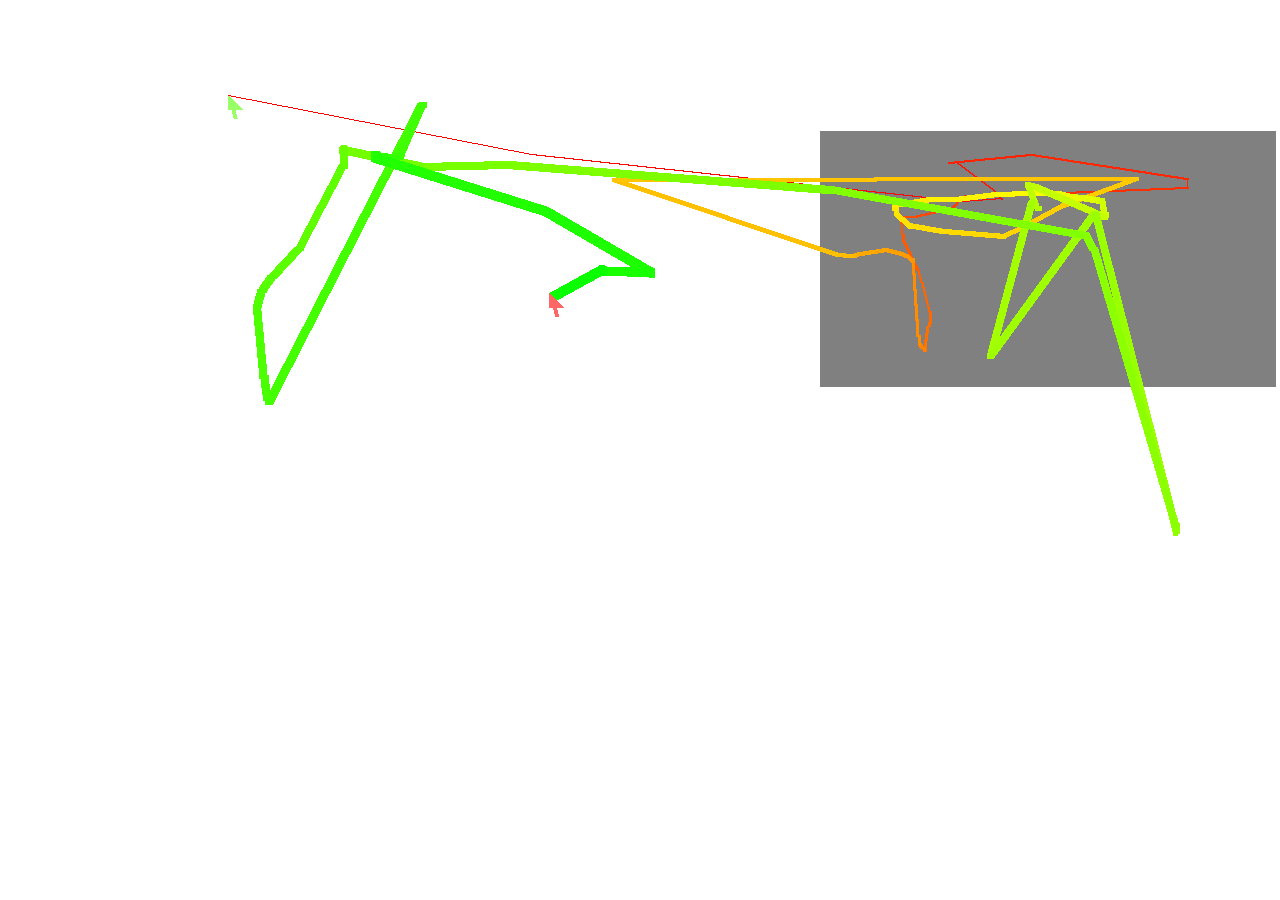}   \\
        \addlinespace
        
        \bottomrule
    \end{tabular}
    }
\end{table*}
\egroup

\subsection{Time Series Representation}
\label{sssec:representation_ts}

In our experiments, a mouse cursor trajectory is modeled as a multivariate time series of 2D cursor coordinates.
The data is ordered by the time they were collected and there are no consecutively duplicated coordinates.
A particular characteristic of this data representation is that events are \emph{asynchronous},
i.e. contrary to regular time series, the sampling rate of \texttt{\small{mousemove}} events is not constant.
This is so because web browser events are first placed in an event queue and then fired as soon as possible~\cite{Resig16}.

An inherent limitation while training RNN models is that the ``memory'' of the network must be fixed, i.e. the maximum number of timesteps that the model can handle must be set to a fixed length.
This will impact model training in two ways.
First, if we choose a small memory footprint (short sequence length) for our models,
we would be truncating longer sequences that otherwise could bear rich behavioral information about the user.
Second, if we choose a large memory footprint,
then we would be wasting computational resources,
as the model would require more weights to optimize
and, in consequence, training time would be unnecessarily longer.
Therefore, to make the training of our RNNs tractable,
we inspected our data and decided to set the maximum sequence length to 50 timestemps,
which roughly corresponds to the mean sequence length observed in our dataset plus one standard deviation.
Since mouse cursor trajectories are variable-length sequences,
shorter sequences were padded to such a fixed length of 50 timesteps
and longer sequences were truncated.
Finally, since each mouse cursor trajectory was performed on different web browsers
with different screen sizes, 
the horizontal coordinates were normalised by each user's viewport width.
The vertical coordinates do not need to be normalised,
since the SERP layout has a fixed width.

\subsection{Visual Representation}
\label{sssec:representation_visual}

According to our data, 90\% of all mouse coordinates happened above the page fold,
i.e. within the browser's visual viewport.
This was somewhat expected given the nature of our task:
in crowdsourcing studies, users often proceed as quickly as possible
in order to maximize their profit~\cite{Eickhoff11, Ipeirotis10}.
However, this suggests that we can expect a visual representation to perform well,
since most of the user interactions would be adequately represented in a fixed-size image.
Therefore, we created five visual encodings (\autoref{tbl:representations}):
\begin{enumerate}[leftmargin=*]

\item \textit{Heatmap.}
The influence of each mouse cursor coordinate is determined with a 2D Gaussian kernel of 25\,px radius.
When various kernels overlap, their values are added together.

\item \textit{Trajectories.}
Every two consecutive coordinates are joined with a straight line.
The first and last coordinates are rendered as cursor-like images, in green and red color, respectively.

\item \textit{Colored trajectories.}
The trajectory line color is mapped to a temperature gradient,
where green areas denote the beginning of the trajectory
and red areas denote the end of the trajectory.

\item \textit{Trajectories with variable line thickness.}
The trajectory line thickness is proportional to the percentage of time,
so that thick areas denote the beginning of the mouse trajectory
and thin ares denote the end of the trajectory.

\item \textit{Colored trajectories with variable line thickness.}
A combination of the two previous representations described.

\end{enumerate}

The mouse cursor data were rendered according to each visual encoding
using the Simple Mouse Tracking system~\cite{Leiva:2013:WBB:2540635.2529996},
which was operated via PhantomJS,\footnote{https://phantomjs.org/} a scriptable headless browser.
Each mouse cursor trajectory was normalized according to the user's viewport.
Finally, no data augmentation or transformation techniques were applied, 
as often performed in computer vision tasks,
and the images were saved as 1280x900\,px PNG files.

\section{Deep Learning Models}
\label{ssec:models}

\subsection{Recurrent Neural Networks}
\label{ssec:rnn_model}

In what follows, we provide an overview of the RNN units, or cells,
that we investigated for our sequence classification task.
These units cover the most popular choices by Deep Learning practitioners.
\begin{enumerate}[leftmargin=*]

\item \textit{SimpleRNN.} Vanilla recurrent cell, i.e.
a fully-connected unit where its output is fed back to its input at every timestep.

\item \textit{LSTM.} This unit introduces an output gate and a forget gate~\cite{Hochreiter97},
to remember long-term dependencies within the data.

\item \textit{GRU.} A simplification of the LSTM cell~\cite{Cho14}, where there is no output gate.
GRU can outperform LSTM units in terms of convergence and generalization~\cite{Chung14}.

\item \textit{BLSTM.}
This unit uses both past and future contexts, by concatenating the outputs of two RNNs~\cite{Graves13}:
one processing the sequence from left to right (forward RNN),
the other one from right to left (backward RNN).
We note that any RNN unit can become bidirectional, however we used the LSTM variant
because of their popularity and to keep our experiments consistent.

\end{enumerate}

All RNN models have an input layer with $50$ neurons (one neuron per timestep),
followed by a hidden layer with $n \in [16, 24, \dots, 128]$ neurons and ReLU activation,
a dropout layer with drop rate $q \in [0.5, 0.4, \dots, 0.1]$ to prevent overfitting,
and a fully-connected layer of 1 output neuron using sigmoid activation.
Each RNN model outputs a probability prediction $p$ of the user's attention to an ad,
where $p>.5$ indicates that the user has noticed the ad.

We trained the models using binary crossentropy as loss function, the popular Adam optimizer (stochastic gradient descent with momentum)
with learning rate $\eta \in [10^{-3}, 10^{-4}, \dots, 10^{-7}]$,
and decay rates $\beta_1=0.9$ and $\beta_2=0.999$.
We set a maximum number of 90 epochs,
using an early stopping of 20 epochs that monitors the validation loss,
and tried different batch sizes $b \in [16, 32, 64]$.

As noted, we explored different combinations of hyperparameters ($n, q, \eta, b$).
Our design space is thus quite large, rendering the classic grid-search approach unfeasible. Therefore, we optimized our RNN models via random search~\cite{Bergstra11},
which has a high probability of finding the most suitable configuration
without having to explore all possible combinations~\cite{Zheng13}.
The best configuration is determined via 3-fold cross-validation,
i.e. a given hyperparameter combination is tested up to three times
over the validation data partition and the final result is averaged.
The best combination is the one with the minimum validation loss.

\subsection{Convolutional Neural Networks}
\label{ssec:cnn_model}

As anticipated in \autoref{ssec:prelim}, we investigated four popular CNN architectures:
\texttt{AlexNet}~\cite{AlexNet}, \texttt{SqueezeNet}~\cite{SqueezeNet}, \texttt{ResNet50}~\cite{ResNet}, and \texttt{VGG19}~\cite{VGGNet},
all of which were trained on the ImageNet database (1M images with 1000 categories).
Our choice of CNNs favoured diversity and was guided by the fact that the above architectures
have been successfully tested in a wide range of computer vision applications,
and have different designs, modules, and number of parameters.
For example, \texttt{ResNet50} and \texttt{VGG19} employ a large number of layers,
hence resulting in \textit{deep} networks, (roughly twice as deep as \texttt{AlexNet}),
use skip connections, and are among the early adopters of batch normalisation.
On the other hand, \texttt{AlexNet} and its simplified variant \texttt{SqueezeNet}
were the first to introduce ReLU activations
and implement a shallow architecture while attaining high accuracy
and requiring less bandwidth to operate.

Using the representations discussed in~\autoref{sssec:representation_visual},
we applied transfer learning to calibrate the CNNs to the particularities our images,
which are quite different from the natural scenery images found in ImageNet.
This way, we can reuse an existing architecture and apply it to our own prediction task
because there are universal, low-level features shared among images.
For each CNN model, we applied the following steps.
First, we initialised the last layer of the CNN model with randomly assigned weights,
while retaining the weights of the initial layers that were pre-trained on \texttt{ImageNet}.
Next, we run a learning rate finder~\cite{smith2015cyclical}
that let $\eta$ values to cyclically vary by linearly increasing it for a few epochs.
Training with cyclical learning rates instead of fixed values improves classification accuracy
without the need to manual fine-tuning and often results in fewer iterations.
We then unfreezed and re-trained all the layers of the CNN model for 300 epochs, using a per-cycle maximal LR.
In addition, we used the early stopping method that terminated training
when the monitored AUC stopped improving after 30 epochs.
Batch size was adjusted accordingly to each architecture, to optimise for the use of GPU memory.
Finally, we trained our CNN models with the Adam optimizer,
using the same decay rates as in the RNN models and binary cross-entropy as loss function.

\bgroup
\def\arraystretch{0.9}
\begin{table*}[!ht]
    \caption{
        Experiment results for organic advertisements.
        Gray cells indicate the top performer in each representation group.
        The overall best performance result (across all groups) is denoted in bold typeface.
        The positive:negative ratio is 447:222.
    }
    \vspace{-5pt}
    \label{tbl:results_top_left_native}
    \centering
    {\footnotesize
    \begin{tabular}[t]{m{\pl} m{1.35cm} l *8c}
      \toprule
        \textbf{Representation}
        & \textbf{Example}
        & \textbf{Architecture}
        & \textbf{Hyperparameters}
        & \textbf{Epoch}
        & \textbf{Adj. Precision}
        & \textbf{Adj. Recall}
        & \textbf{Adj. F-measure}
        & \textbf{AUC} \\
      \midrule
      \multirow{4}{\pl}{Time series}
      &  \multirow{4}{*}{$\begin{matrix}(x_1,y_1),\\ \dots,\\ (x_N,y_N)\end{matrix}$}
      &  \texttt{SimpleRNN}     & $\eta=10^{-3}$, $q=0.4$, $n=64$, $b=64$ & 90 & 0.556      & 0.697     & 0.603     & 0.529 \\
      &&  \texttt{LSTM}         & $\eta=10^{-3}$, $q=0.3$, $n=64$, $b=32$ & 65 & 0.622      & 0.604     & 0.612     & 0.531 \\
      &&  \texttt{BLSTM}        & $\eta=10^{-4}$, $q=0.3$, $n=32$, $b=16$ & 20 & \gb 0.695  & 0.637     & 0.654     & \gb 0.631 \\
      &&  \texttt{GRU}          & $\eta=10^{-3}$, $q=0.2$, $n=32$, $b=32$ & 90 & 0.672      & \gb 0.711 & \gb 0.678 & 0.557 \\
      \arrayrulecolor{gray!50!}\midrule

       \multirow{4}{\pl}{Heatmap}
       &  \multirow{4}{*}{\fpic{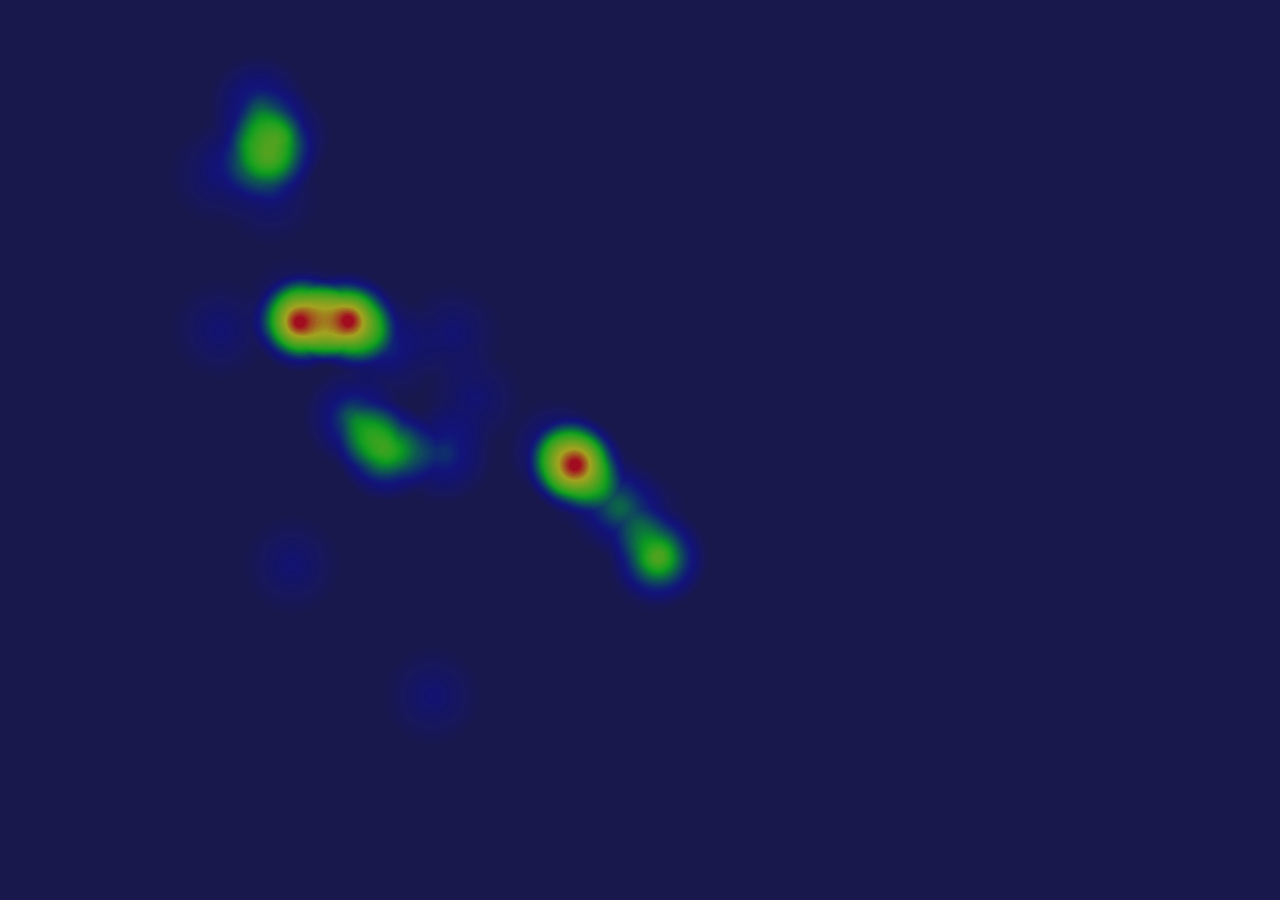}}
       &  \texttt{AlexNet}       & $\eta=2.51\e{-7}$, $b=64$ & 38 & 0.572  &    0.667   &   0.583   & 0.547 \\
       &&  \texttt{SqueezeNet}   & $\eta=6.91\e{-7}$, $b=64$ & 41 & 0.602  &    0.524   &   0.540   & 0.543 \\
       &&  \texttt{ResNet50}     & $\eta=9.12\e{-7}$, $b=32$ & 46 & \gb 0.652  &    \gb 0.679   &   \gb 0.659 & \gb 0.638 \\
       &&  \texttt{VGG19}        & $\eta=1.90\e{-6}$, $b=16$ & 54 & 0.627  &    0.636   &   0.628   & 0.614 \\
       \arrayrulecolor{gray!50!}\midrule

       \multirow{4}{\pl}{Heatmap with ad placeholder}
       &  \multirow{4}{*}{\fpic{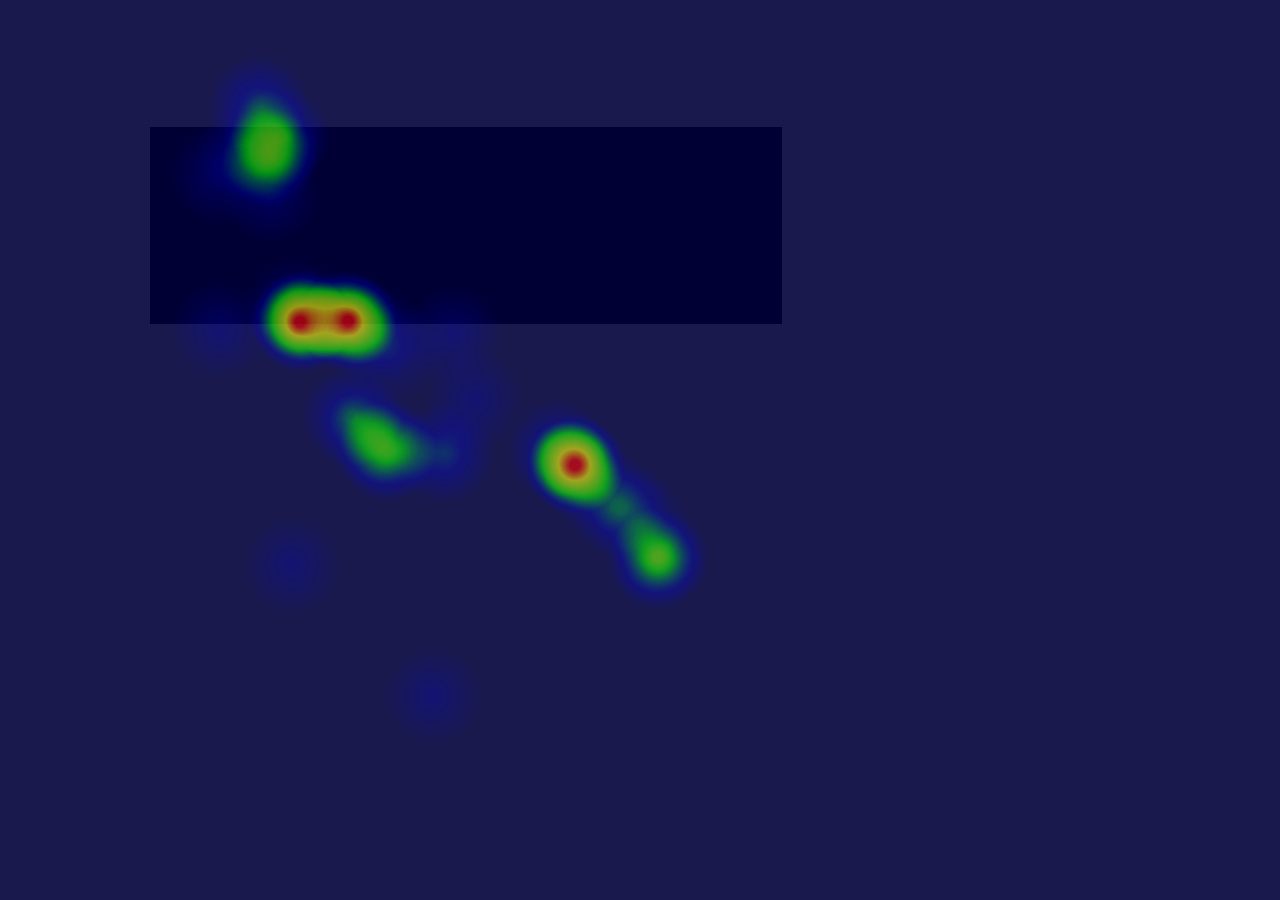}}
       &  \texttt{AlexNet}       & $\eta=2.51\e{-7}$, $b=64$ & 59 & \gb 0.639  &    \gb 0.677   &   \gb 0.650   & \gb 0.656 \\
       &&  \texttt{SqueezeNet}   & $\eta=2.51\e{-7}$, $b=64$ & 34 & 0.617  &    0.657   &   0.621   & 0.587 \\
       &&  \texttt{ResNet50}     & $\eta=8.06\e{-7}$, $b=32$ & 34 & 0.599  &    0.518   &   0.537   & 0.532 \\
       &&  \texttt{VGG19}        & $\eta=3.41\e{-7}$, $b=16$ & 44 & 0.601  &    0.622   &   0.606   & 0.598 \\
       \arrayrulecolor{gray!50!}\midrule

       \multirow{4}{\pl}{Trajectories}
       &  \multirow{4}{*}{\fpic{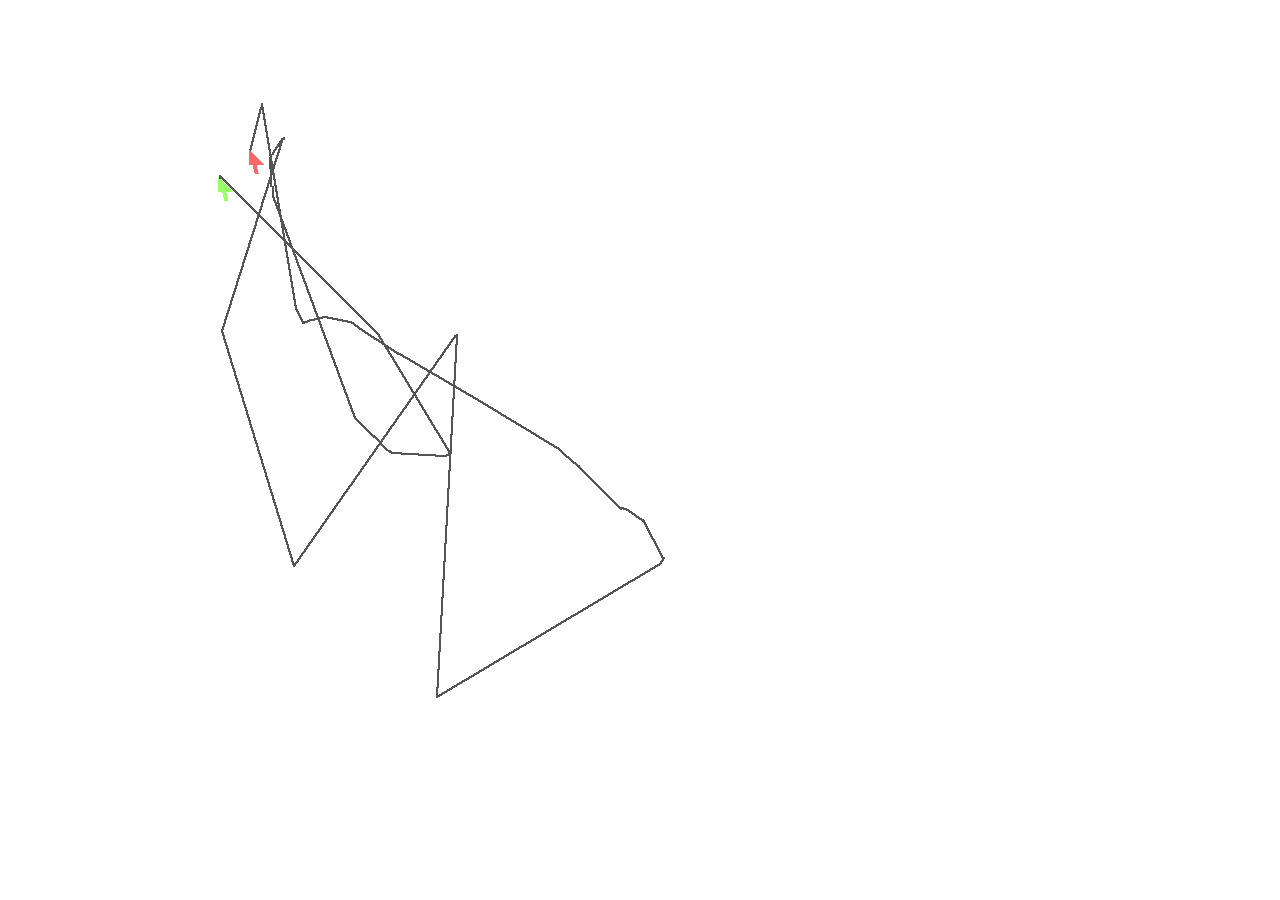}}
       &  \texttt{AlexNet}       & $\eta=4.78\e{-7}$, $b=64$ & 38 & 0.633  &    0.667   &   0.634   & 0.614 \\
       &&  \texttt{SqueezeNet}   & $\eta=3.31\e{-7}$, $b=64$ & 32 & 0.654  &    0.654   &   \gb 0.654   & 0.589 \\
       &&  \texttt{ResNet50}     & $\eta=3.41\e{-7}$, $b=32$ & 81 & 0.637  &    0.627   &   0.630   & \gb 0.626 \\
       &&  \texttt{VGG19}        & $\eta=6.31\e{-7}$, $b=16$ & 33 & \gb 0.658  &    \gb\bf 0.693   &   0.645   & 0.600 \\
       \arrayrulecolor{gray!50!}\midrule

       \multirow{4}{\pl}{Trajectories with ad placeholder}
       &  \multirow{4}{*}{\fpic{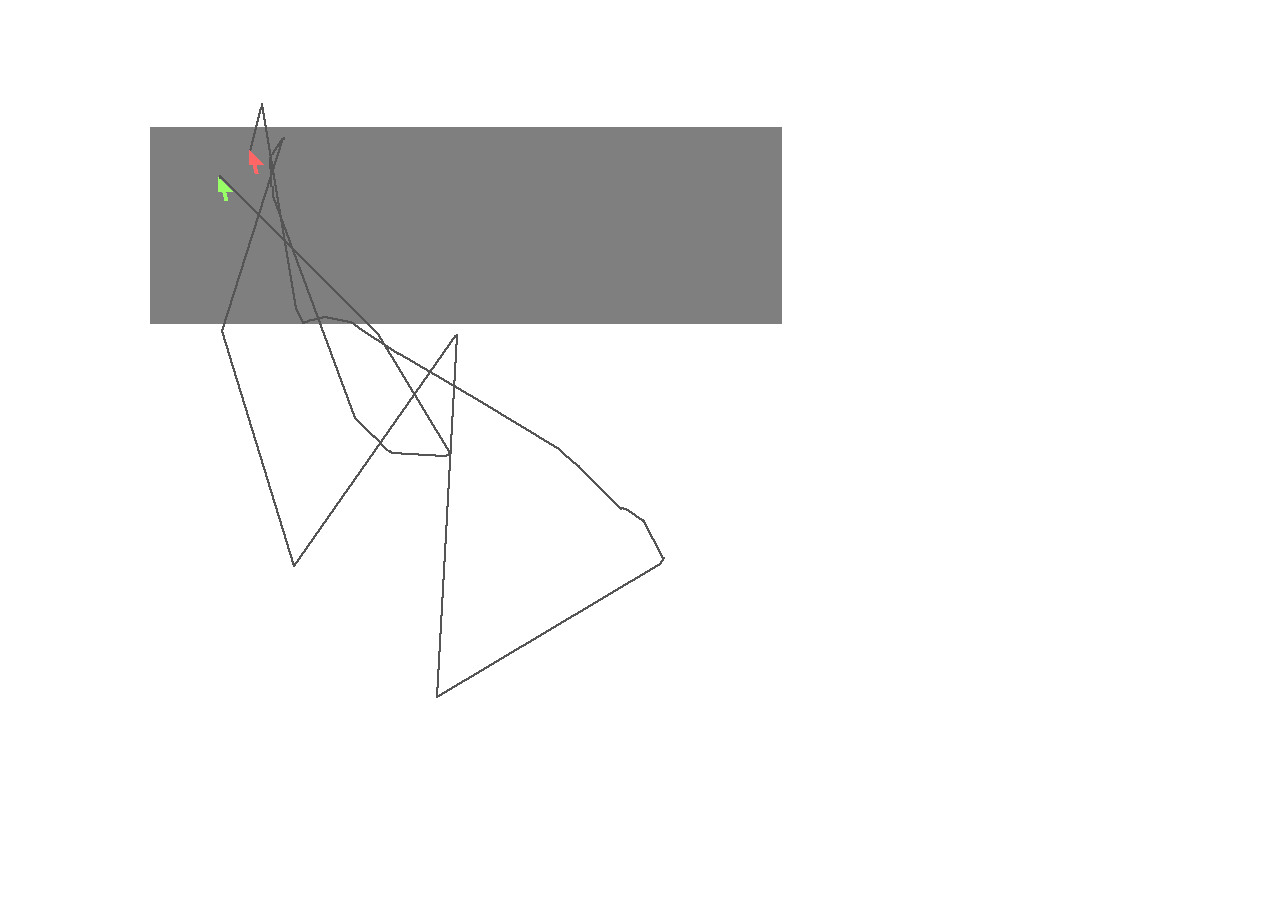}}
       &  \texttt{AlexNet}       & $\eta=8.31\e{-6}$, $b=64$ & 62 & 0.620  &    \gb 0.639   &   \gb 0.626   & 0.610 \\
       &&  \texttt{SqueezeNet}   & $\eta=5.75\e{-7}$, $b=64$ & 59 & 0.614  &    0.619   &   0.615   & 0.578 \\
       &&  \texttt{ResNet50}     & $\eta=1.16\e{-6}$, $b=32$ & 37 & \gb 0.632  &    0.464   &   0.449   & \gb\bf 0.690 \\
       &&  \texttt{VGG19}        & $\eta=1.16\e{-6}$, $b=16$ & 44 & 0.621  &    0.610   &   0.610   & 0.613 \\
       \arrayrulecolor{gray!50!}\midrule

       \multirow{4}{\pl}{Colored trajectories}
       &  \multirow{4}{*}{\fpic{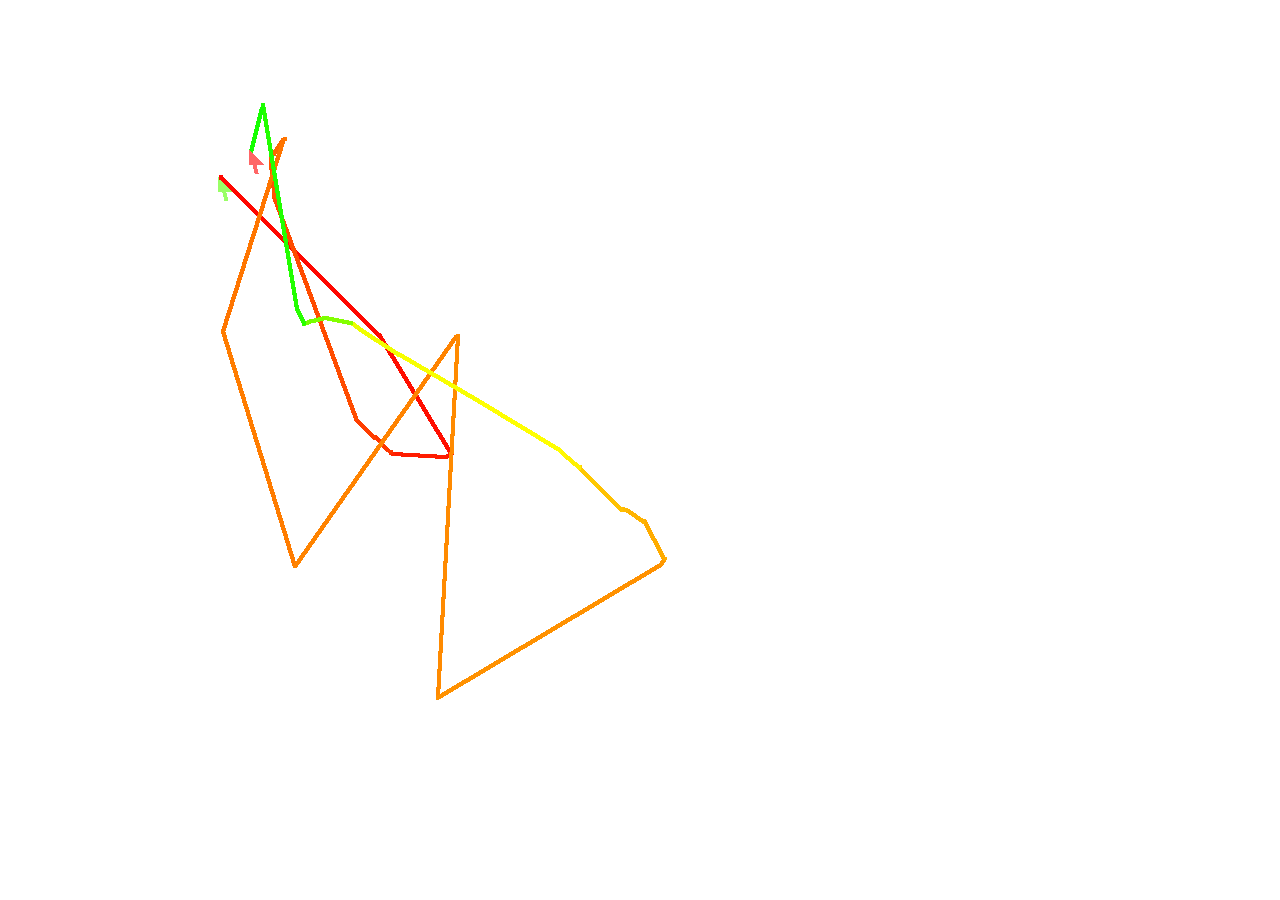}}
       &  \texttt{AlexNet}       & $\eta=3.63\e{-7}$, $b=64$ & 57 & 0.594  &    0.633   &   0.607   & 0.561 \\
       &&  \texttt{SqueezeNet}   & $\eta=2.51\e{-7}$, $b=64$ & 36 & \gb 0.630  &    \gb 0.658   &   \gb 0.639   & 0.605 \\
       &&  \texttt{ResNet50}     & $\eta=5.58\e{-7}$, $b=32$ & 50 & 0.619  &    0.534   &   0.550   & 0.601 \\
       &&  \texttt{VGG19}        & $\eta=3.41\e{-7}$, $b=16$ & 31 & 0.610  &    0.647   &   0.617   & \gb 0.629 \\
       \arrayrulecolor{gray!50!}\midrule

       \multirow{4}{\pl}{Colored trajectories with ad placeholder}
       &  \multirow{4}{*}{\fpic{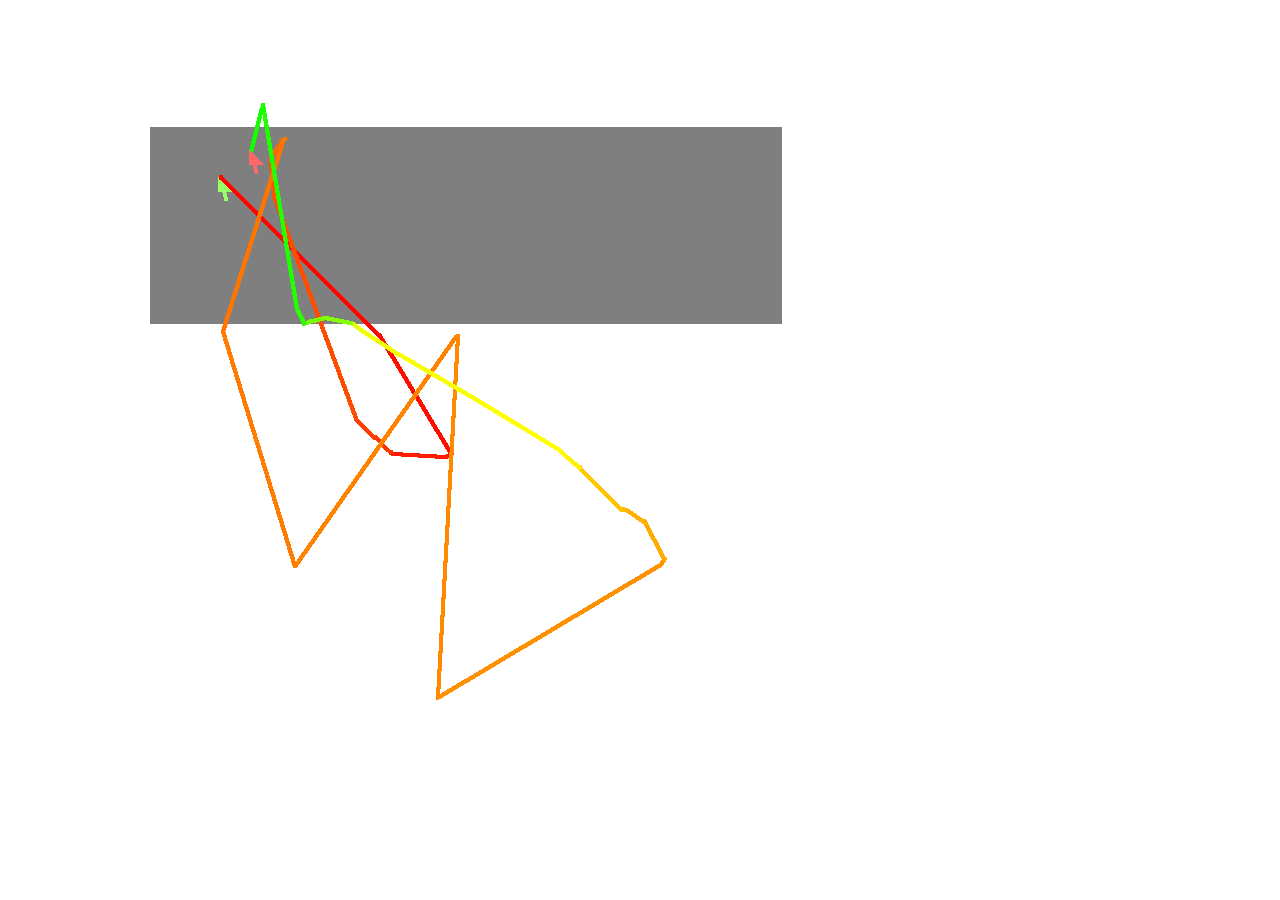}}
       &  \texttt{AlexNet}       & $\eta=3.63\e{-7}$, $=64$ & 57 & 0.640  &    \gb 0.637   &   \gb 0.637   & 0.610 \\
       &&  \texttt{SqueezeNet}   & $\eta=3.98\e{-7}$, $=64$ & 34 & 0.612  &    0.527   &   0.543   & 0.540 \\
       &&  \texttt{ResNet50}     & $\eta=3.41\e{-7}$, $=32$ & 87 & \gb\bf 0.694  &    0.505   &   0.506   & \gb 0.654 \\
       &&  \texttt{VGG19}        & $\eta=3.86\e{-7}$, $=16$ & 50 & 0.563  &    0.623   &   0.580   & 0.582 \\
       \arrayrulecolor{gray!50!}\midrule

       \multirow{4}{\pl}{Trajectories with line thickness}
       &  \multirow{4}{*}{\fpic{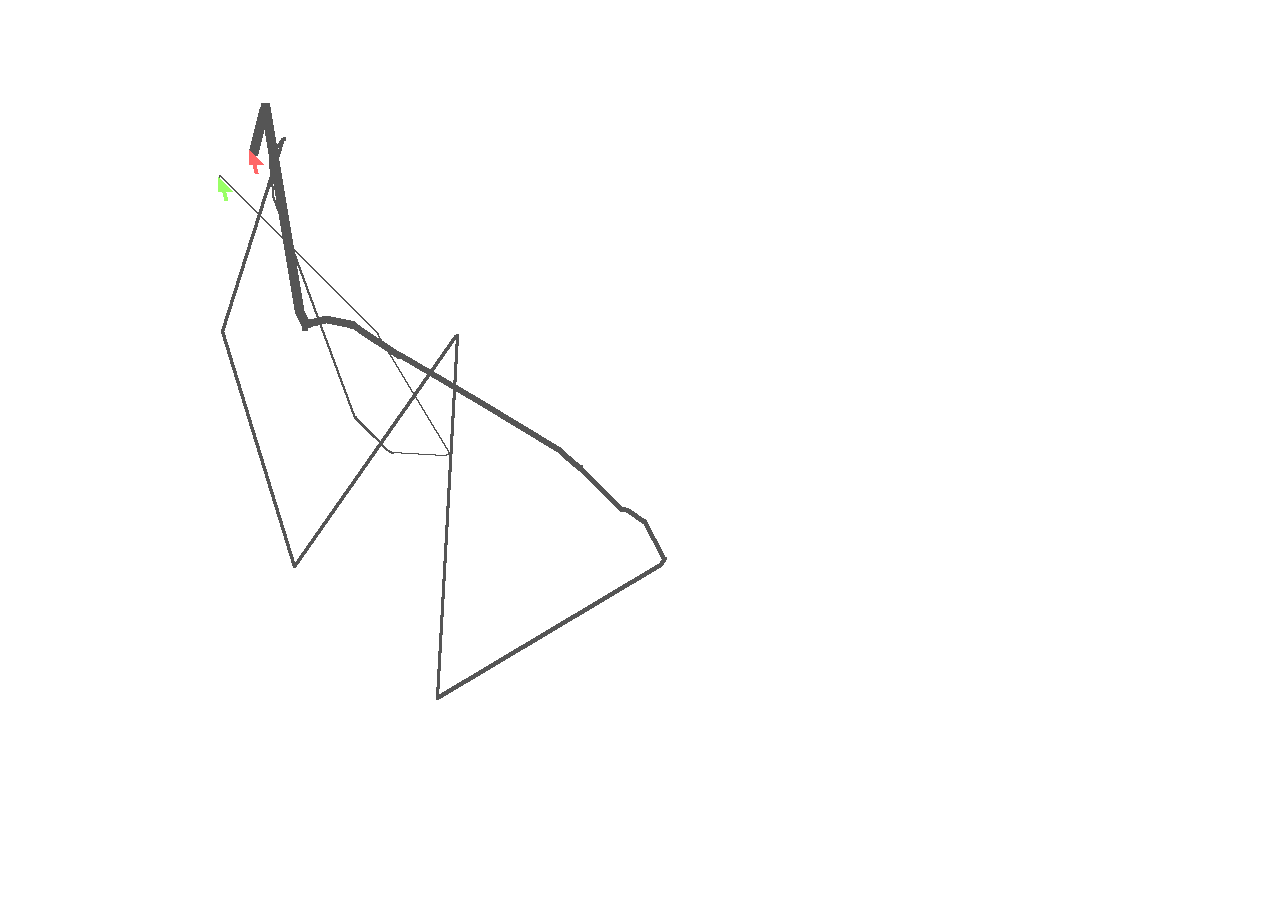}}
       &  \texttt{AlexNet}       & $\eta=5.24\e{-7}$, $b=64$ & 31 & 0.599  &    0.484   &   0.495   & 0.539 \\
       &&  \texttt{SqueezeNet}   & $\eta=5.75\e{-5}$, $b=64$ & 35 & 0.559  &    0.555   &   0.562   & 0.546 \\
       &&  \texttt{ResNet50}     & $\eta=3.86\e{-7}$, $b=32$ & 49 & \gb 0.657  &    \gb 0.660   &   \gb 0.654   & 0.612 \\
       &&  \texttt{VGG19}        & $\eta=4.36\e{-7}$, $b=16$ & 104 & 0.627  &    \gb 0.660   &   0.637   & \gb 0.617 \\
       \arrayrulecolor{gray!50!}\midrule

       \multirow{4}{\pl}{Trajectories with line thickness and ad placeholder}
       & \multirow{4}{*}{\fpic{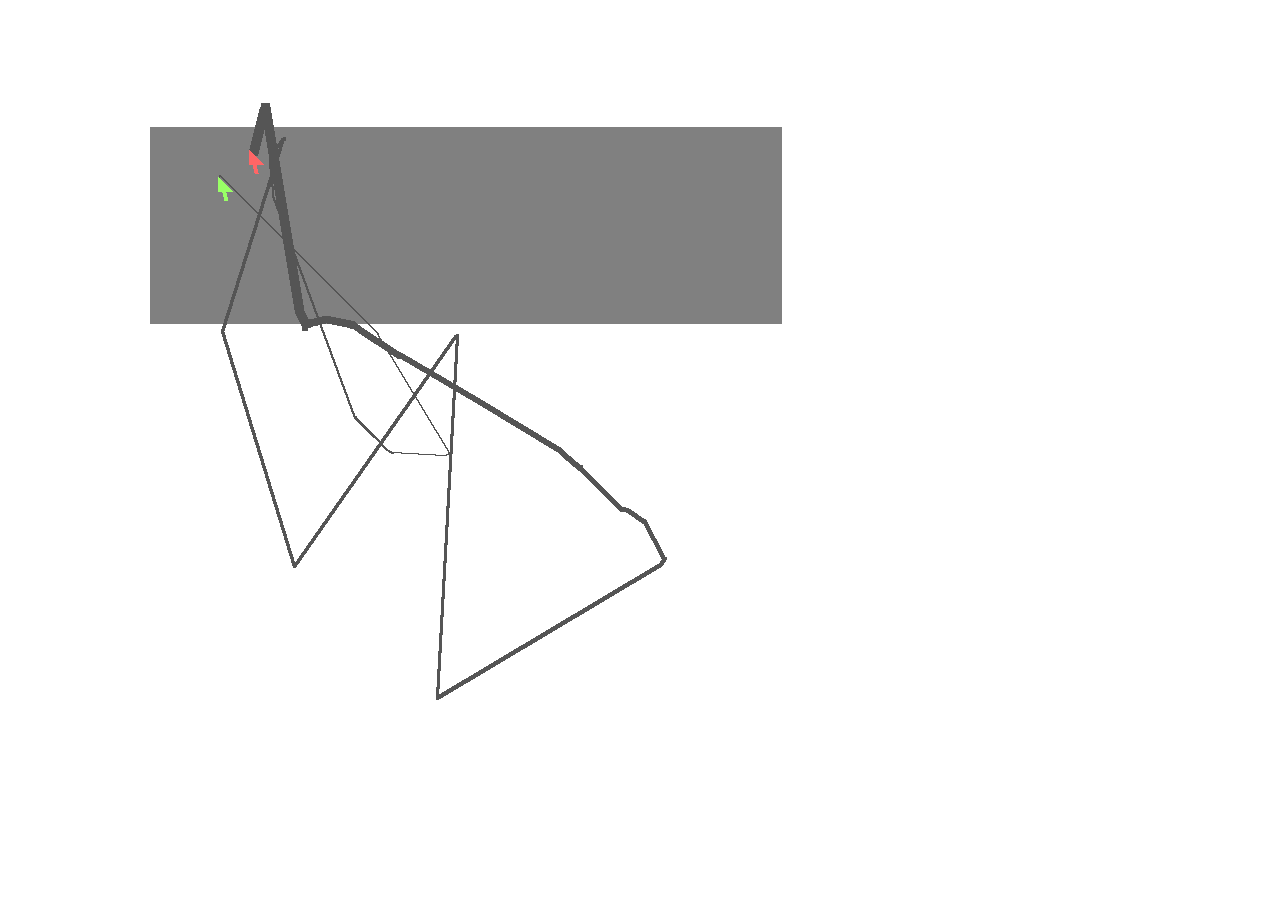}}
       &  \texttt{AlexNet}       & $\eta=2.75\e{-5}$, $b=64$ & 43 & 0.630  &    0.642   &   0.631   & 0.616 \\
       &&  \texttt{SqueezeNet}   & $\eta=3.02\e{-7}$, $b=64$ & 36 & \gb 0.674  &    \gb 0.665   &   \gb\bf 0.700   & \gb 0.657 \\
       &&  \texttt{ResNet50}     & $\eta=8.06\e{-7}$, $b=32$ & 31 & 0.669  &    0.539   &   0.548   & 0.638 \\
       &&  \texttt{VGG19}        & $\eta=4.93\e{-7}$, $b=16$ & 67 & 0.575  &    0.628   &   0.595   & 0.622 \\
       \arrayrulecolor{gray!50!}\midrule

       \multirow{4}{\pl}{Colored trajectories with line thickness}
       & \multirow{4}{*}{\fpic{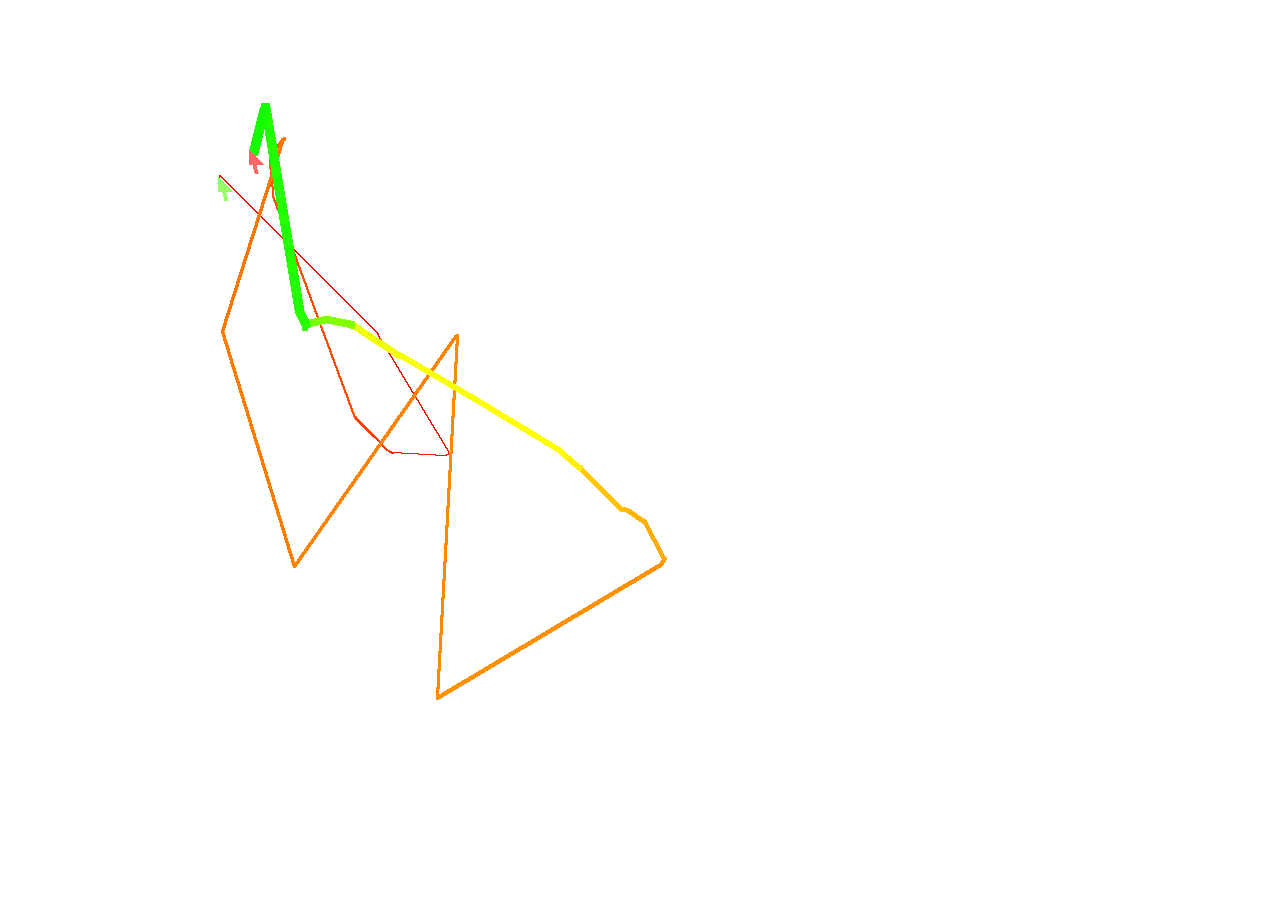}}
       &  \texttt{AlexNet}       & $\eta=2.51\e{-6}$, $b=64$ & 71 & 0.653  &    0.674   &   0.655   & 0.615 \\
       &&  \texttt{SqueezeNet}   & $\eta=5.24\e{-7}$, $b=64$ & 55 & \gb 0.676  &    \gb 0.688   &   \gb 0.680   & \gb\bf 0.690 \\
       &&  \texttt{ResNet50}     & $\eta=4.93\e{-7}$, $b=32$ & 56 & 0.616  &    0.493   &   0.508   & 0.606 \\
       &&  \texttt{VGG19}        & $\eta=8.06\e{-7}$, $b=16$ & 39 & 0.591  &    0.618   &   0.603   & 0.557 \\
       \arrayrulecolor{gray!50!}\midrule

       \multirow{4}{\pl}{Colored trajectories with line thickness and ad placeholder}
       & \multirow{4}{*}{\fpic{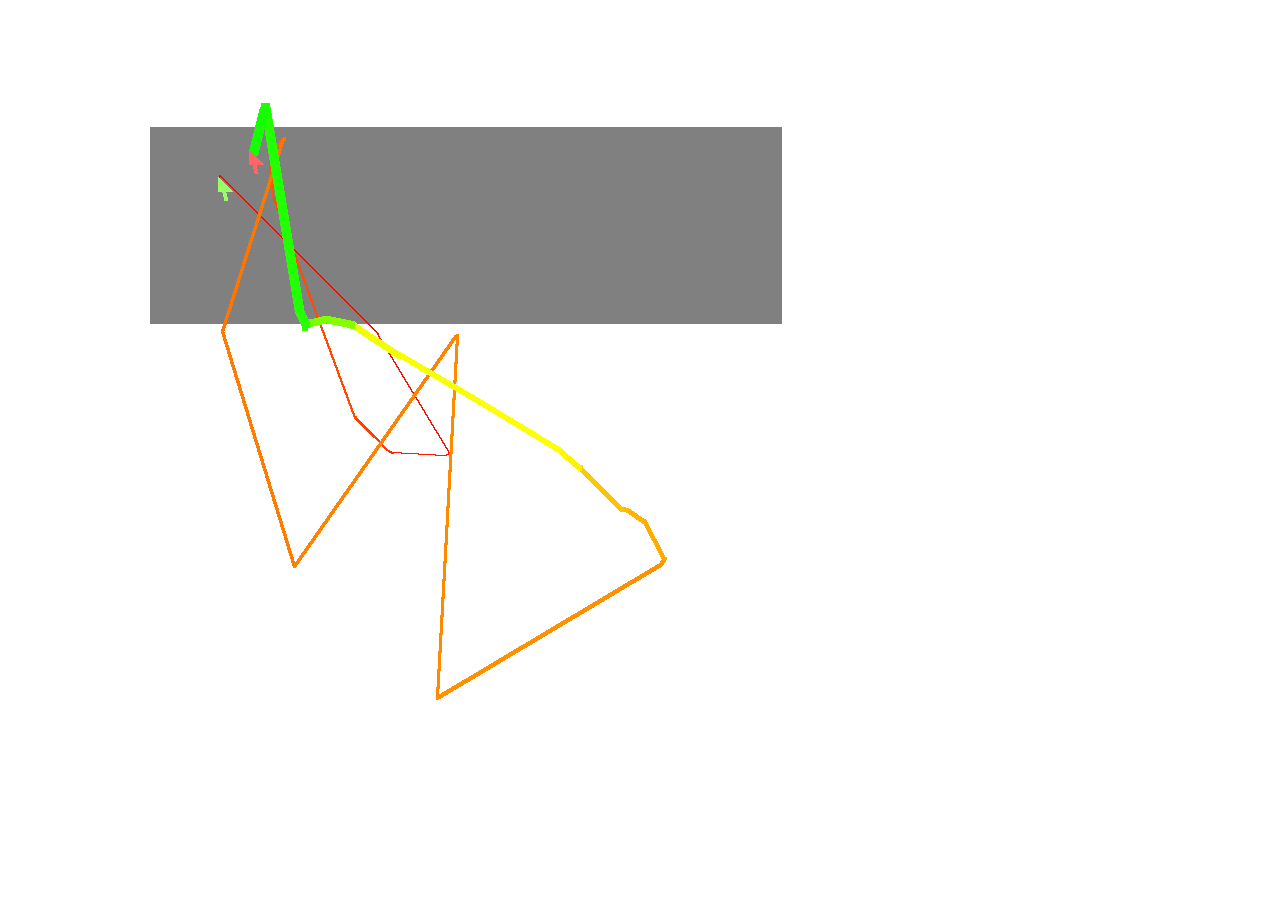}}
       &  \texttt{AlexNet}       & $\eta=2.51\e{-7}$, $b=64$ & 55 & 0.665  &    \gb 0.692   &   \gb 0.666   & 0.657 \\
       &&  \texttt{SqueezeNet}   & $\eta=2.51\e{-7}$, $b=64$ & 61 & 0.614  &    0.626   &   0.618   & 0.576 \\
       &&  \texttt{ResNet50}     & $\eta=4.36\e{-7}$, $b=32$ & 37 & \gb 0.673  &    0.557   &   0.572   & \gb 0.659 \\
       &&  \texttt{VGG19}        & $\eta=1.16\e{-6}$, $b=16$ & 62 & 0.549  &    0.591   &   0.565   & 0.572 \\
       \arrayrulecolor{black}\bottomrule
    \end{tabular}}
\end{table*}
\egroup

\bgroup
\def\arraystretch{0.9}
\begin{table*}[!ht]
    \caption{
        Results for left-aligned direct display ads.
        Gray cells indicate the top performer in each representation group.
        The overall best performance result (across all groups) is denoted in bold typeface.
        The positive:negative ratio is 523:192.
    }
    \vspace{-5pt}
    \label{tbl:results_top_left_dd}
    \centering
    {\footnotesize
    \begin{tabular}[t]{m{\pl} m{1.35cm} l *8c}
      \toprule
        \textbf{Representation}
        & \textbf{Example}
        & \textbf{Architecture}
        & \textbf{Hyperparameters}
        & \textbf{Epoch}
        & \textbf{Adj. Precision}
        & \textbf{Adj. Recall}
        & \textbf{Adj. F-measure}
        & \textbf{AUC} \\
      \midrule
      \multirow{4}{\pl}{Time series}
      &  \multirow{4}{*}{$\begin{matrix}(x_1,y_1),\\ \dots,\\ (x_N,y_N)\end{matrix}$}
      &  \texttt{SimpleRNN} & $\eta=10^{-3}$, $q=0.3$, $n=64$, $b=64$ & 73 & 0.575      & \gb 0.758     & \gb 0.654     & 0.508 \\
      &&  \texttt{LSTM}     & $\eta=10^{-3}$, $q=0.4$, $n=64$, $b=64$ & 74 & 0.607      & 0.656     & 0.628     & 0.542 \\
      &&  \texttt{BLSTM}    & $\eta=10^{-3}$, $q=0.5$, $n=64$, $b=16$ & 58 & \gb 0.658  & 0.647     & 0.652     & 0.548 \\
      &&  \texttt{GRU}      & $\eta=10^{-3}$, $q=0.2$, $n=64$, $b=32$ & 90 & 0.598      & 0.727     & 0.646     & \gb 0.560 \\
      \arrayrulecolor{gray!50!}\midrule

      \multirow{4}{\pl}{Heatmap}
      &  \multirow{4}{*}{\fpic{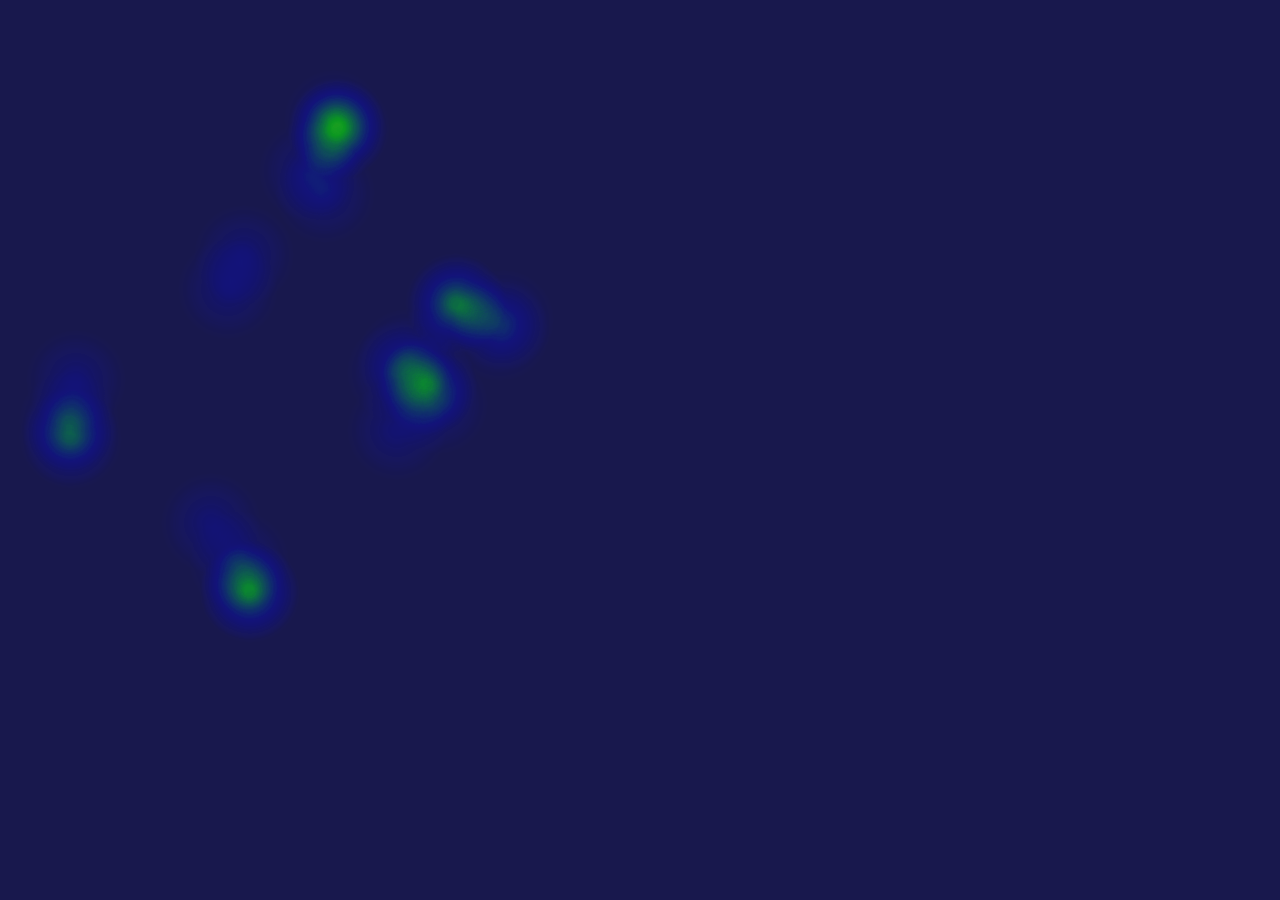}}
      &  \texttt{AlexNet}       & $\eta=9.12\e{-7}$, $b=64$ & 84 & 0.661  &    0.714   &   0.683   & 0.606 \\
      &&  \texttt{SqueezeNet}   & $\eta=3.31\e{-5}$, $b=64$ & 31 & \gb 0.732  &    0.692   &   0.706   & 0.613 \\
      &&  \texttt{ResNet50}     & $\eta=1.44\e{-6}$, $b=32$ & 46 & 0.697  &    \gb 0.743   &   \gb 0.715   & \gb 0.668 \\
      &&  \texttt{VGG19}        & $\eta=7.58\e{-7}$, $b=16$ & 60 & 0.653  &    0.712   &   0.682   & 0.593 \\
      \arrayrulecolor{gray!50!}\midrule

      \multirow{4}{\pl}{Heatmap with ad placeholder}
      &  \multirow{4}{*}{\fpic{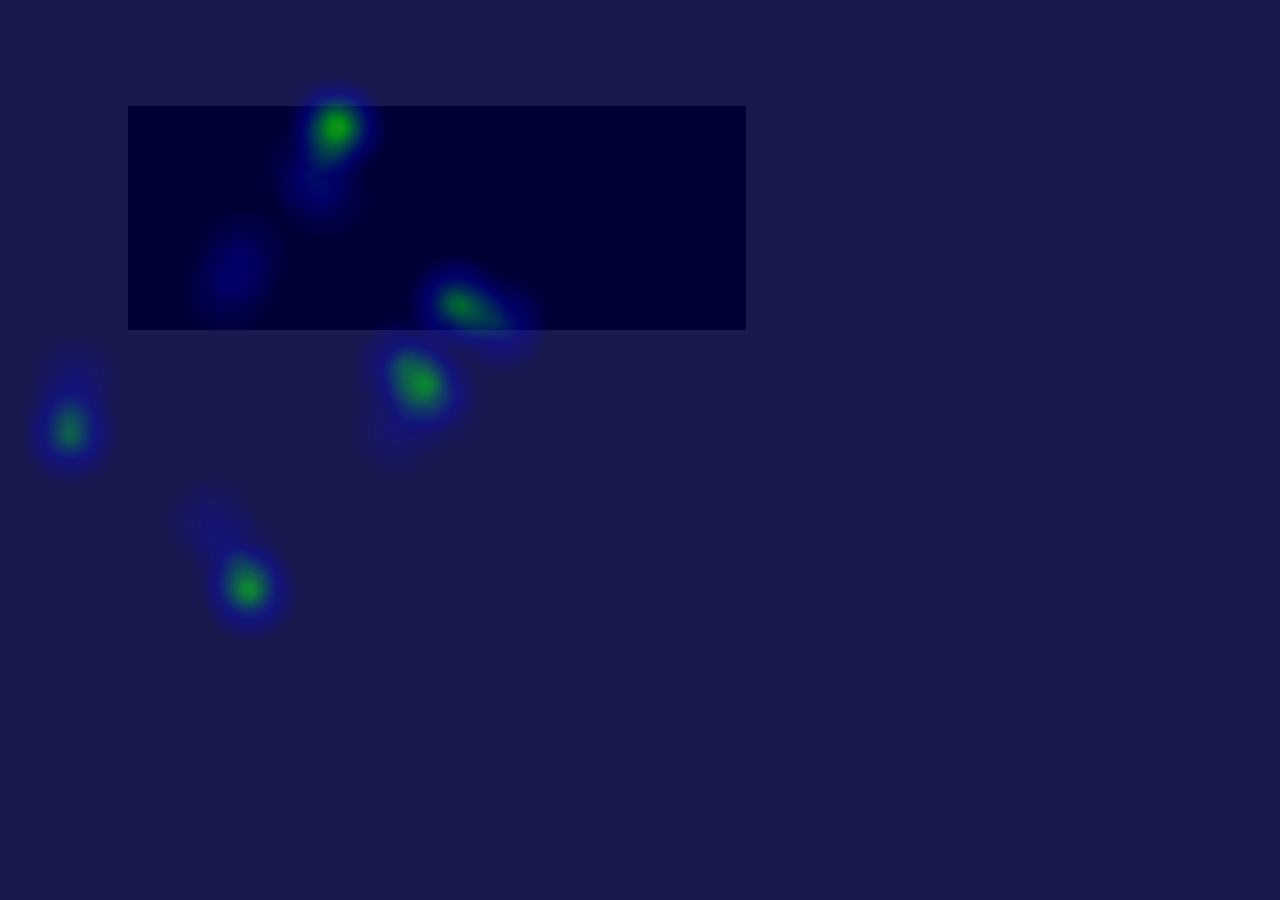}}
      &  \texttt{AlexNet}       & $\eta=2.51\e{-7}$, $b=64$ & 68 & 0.703  &    \gb\bf 0.787   &   0.719   & 0.602 \\
      &&  \texttt{SqueezeNet}   & $\eta=2.51\e{-7}$, $b=64$ & 31 & 0.672  &    0.700   &   0.681   & 0.594 \\
      &&  \texttt{ResNet50}     & $\eta=3.63\e{-7}$, $b=32$ & 31 & 0.732  &    0.607   &   0.643   & 0.628 \\
      &&  \texttt{VGG19}        & $\eta=2.75\e{-7}$, $b=16$ & 66 & \gb 0.749  &    0.712   &   \gb 0.728   & \gb 0.694 \\
      \arrayrulecolor{gray!50!}\midrule

      \multirow{4}{\pl}{Trajectories}
      &  \multirow{4}{*}{\fpic{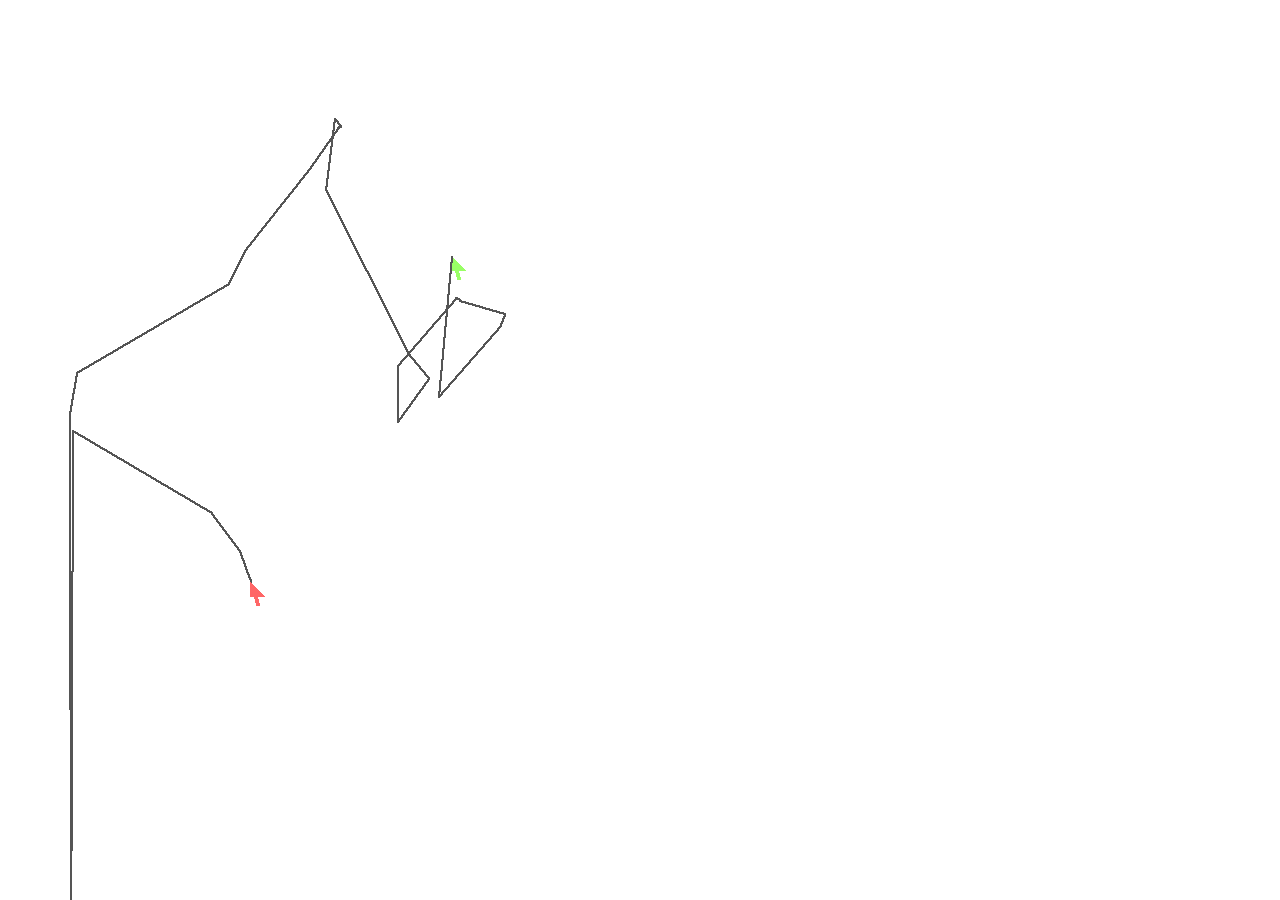}}
      &  \texttt{AlexNet}       & $\eta=3.02\e{-7}$, $b=64$ & 103 & \gb 0.723  &    \gb 0.738   &   \gb 0.727   & 0.596 \\
      &&  \texttt{SqueezeNet}   & $\eta=5.75\e{-7}$, $b=64$ & 31  & 0.684  &    0.690   &   0.684   & 0.628 \\
      &&  \texttt{ResNet50}     & $\eta=7.35\e{-6}$, $b=32$ & 76  & 0.695  &    0.736   &   0.713   & 0.662 \\
      &&  \texttt{VGG19}        & $\eta=5.24\e{-7}$, $b=16$ & 36  & 0.692  &    0.732   &   0.709   & \gb 0.687 \\
      \arrayrulecolor{gray!50!}\midrule

      \multirow{4}{\pl}{Trajectories with ad placeholder}
      &  \multirow{4}{*}{\fpic{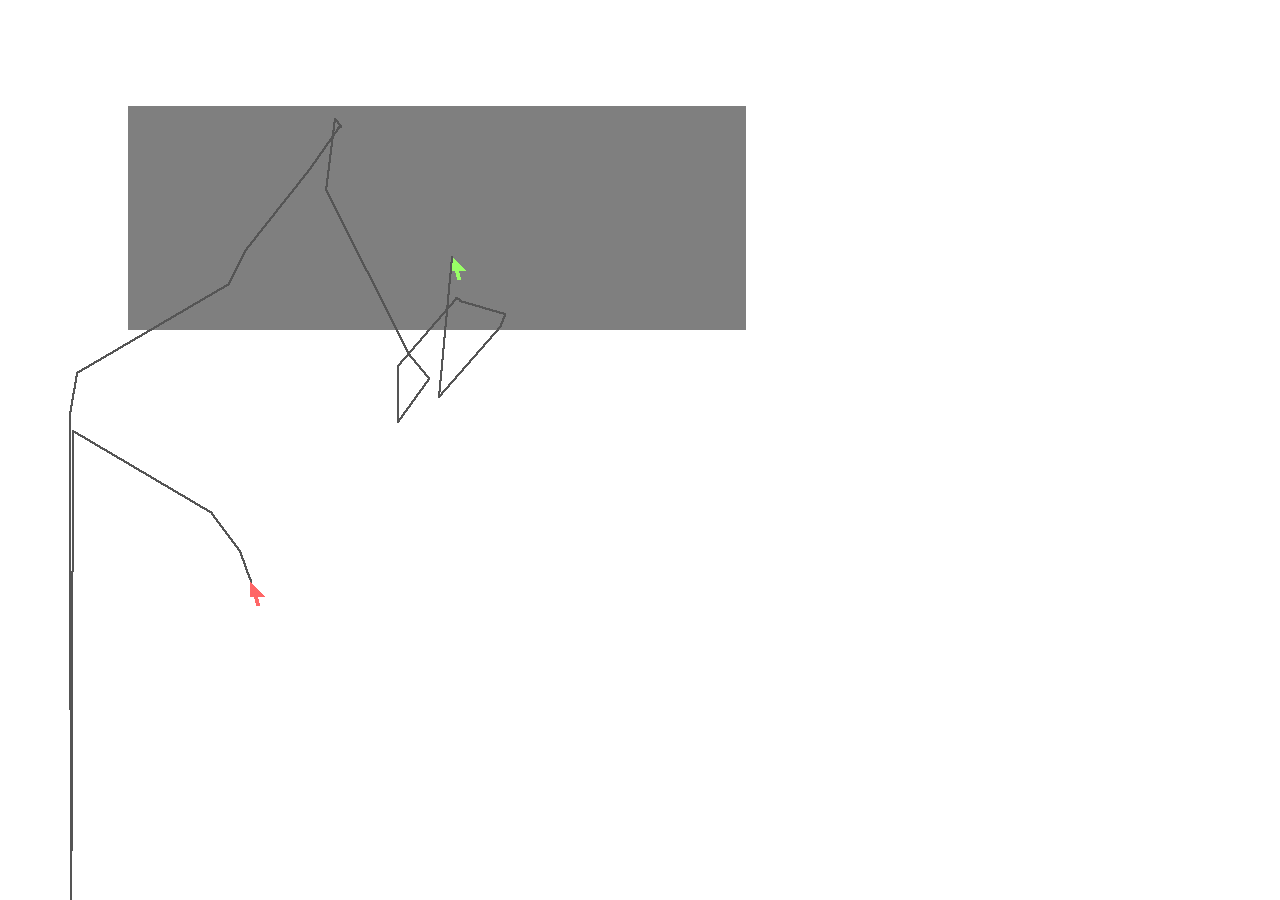}}
      &  \texttt{AlexNet}       & $\eta=3.02\e{-7}$, $b=64$ & 103 & \gb 0.745  &    0.745   &   \gb\bf 0.745   & \gb\bf 0.708 \\
      &&  \texttt{SqueezeNet}   & $\eta=2.75\e{-6}$, $b=64$ & 31  & 0.712  &    0.690   &   0.695   & 0.632 \\
      &&  \texttt{ResNet50}     & $\eta=3.02\e{-7}$, $b=32$ & 39  & 0.717  &    \gb 0.755   &   0.729   & 0.629 \\
      &&  \texttt{VGG19}        & $\eta=7.58\e{-7}$, $b=16$ & 42  & 0.729  &    0.725   &   0.723   & 0.656 \\
      \arrayrulecolor{gray!50!}\midrule

      \multirow{4}{\pl}{Colored trajectories}
      &  \multirow{4}{*}{\fpic{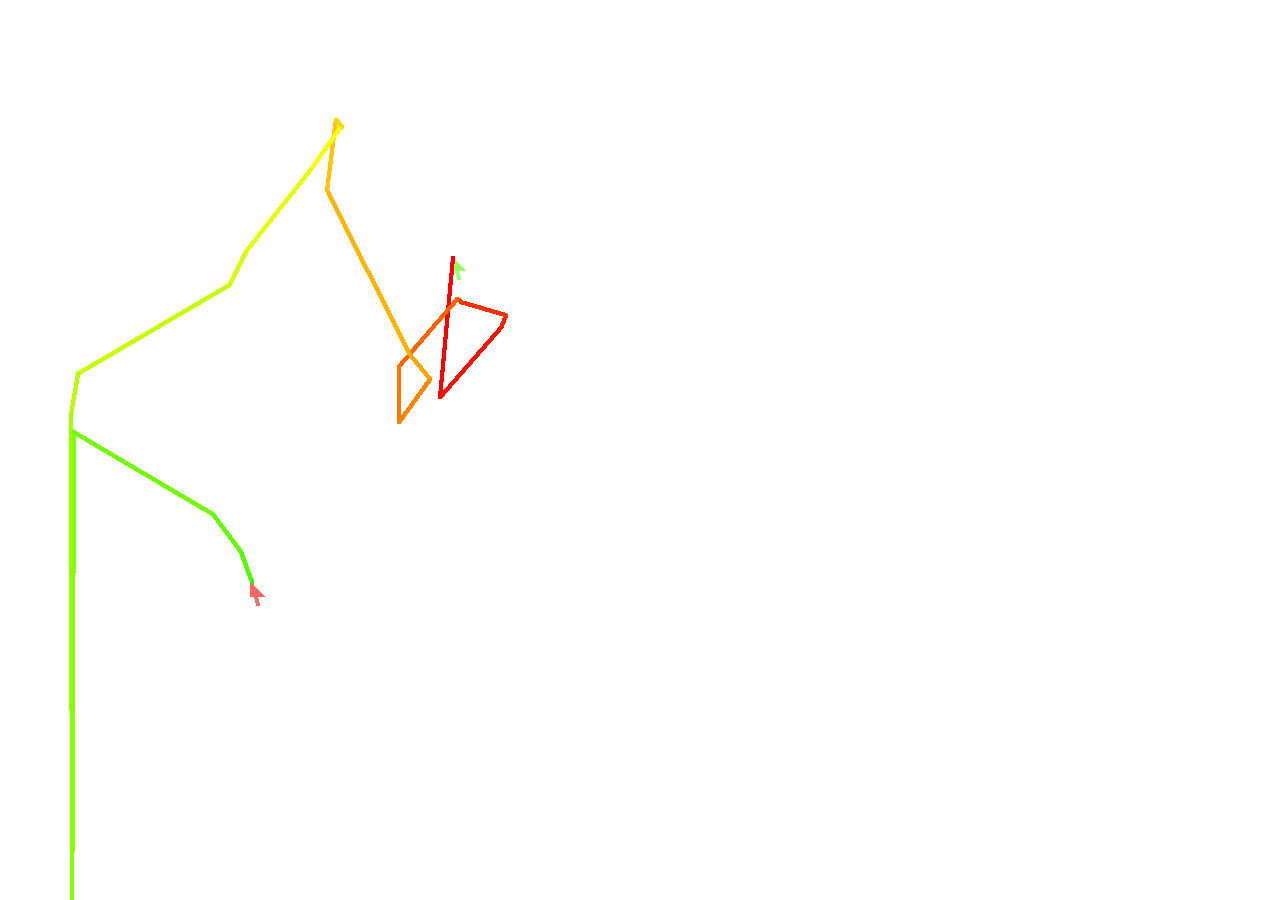}}
      &  \texttt{AlexNet}       & $\eta=5.24\e{-4}$, $b=64$ & 68  & \gb 0.745  &    \gb 0.745   &   \gb\bf 0.745   & \gb 0.677 \\
      &&  \texttt{SqueezeNet}   & $\eta=5.24\e{-7}$, $b=64$ & 31  & 0.715  &    0.528   &   0.568   & 0.597 \\
      &&  \texttt{ResNet50}     & $\eta=8.06\e{-7}$, $b=32$ & 139 & 0.728  &    0.665   &   0.688   & 0.665 \\
      &&  \texttt{VGG19}        & $\eta=3.63\e{-7}$, $b=16$ & 84  & 0.706  &    0.659   &   0.675   & 0.603 \\
      \arrayrulecolor{gray!50!}\midrule

      \multirow{4}{\pl}{Colored trajectories with ad placeholder}
      &  \multirow{4}{*}{\fpic{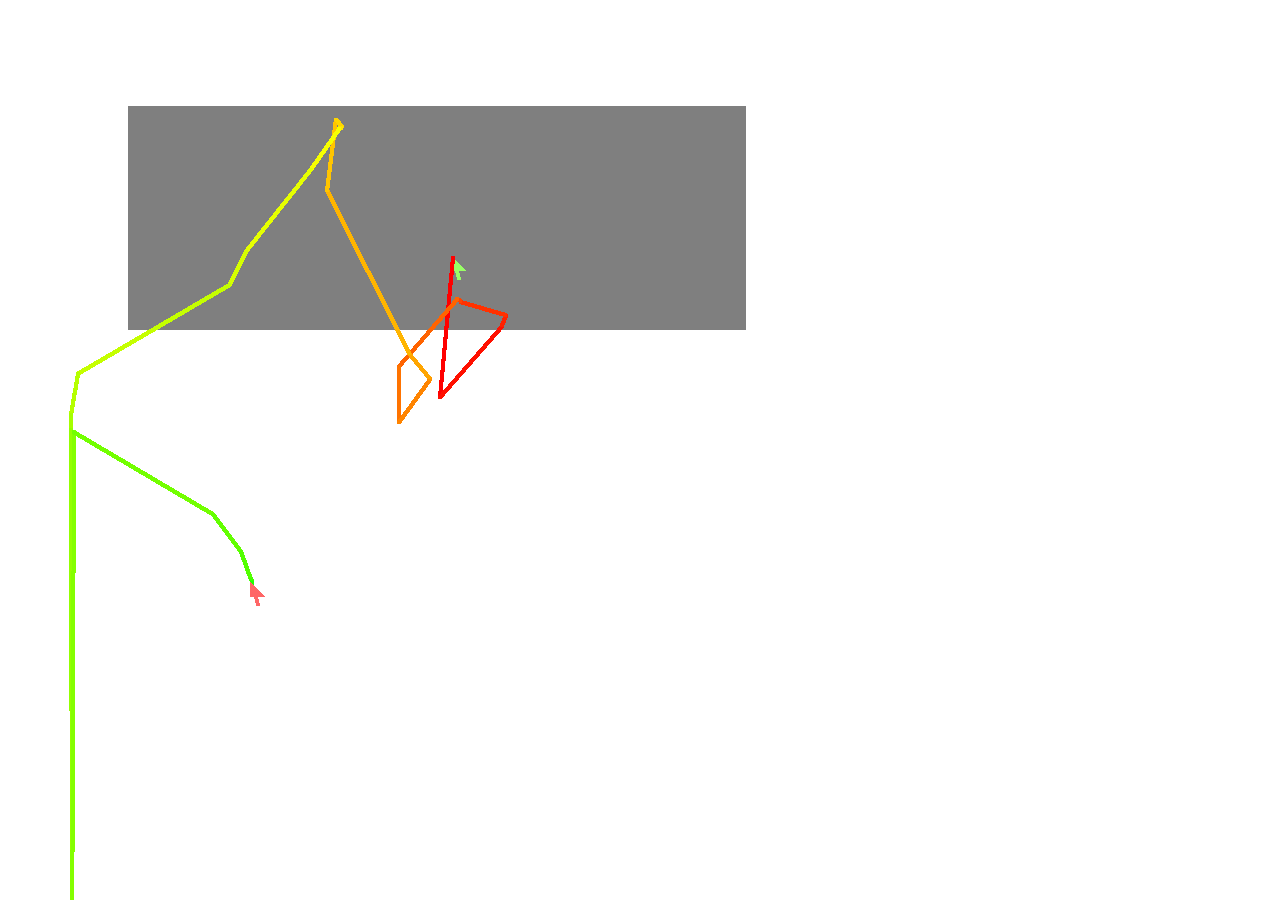}}
      &  \texttt{AlexNet}       & $\eta=2.51\e{-4}$, $b=64$ & 56 & 0.674  &    0.708   &   0.689   & 0.666 \\
      &&  \texttt{SqueezeNet}   & $\eta=2.51\e{-7}$, $b=64$ & 31 & 0.712  &    0.707   &   0.707   & 0.651 \\
      &&  \texttt{ResNet50}     & $\eta=2.51\e{-7}$, $b=32$ & 38 & \gb 0.747  &    0.738   &   \gb 0.742   & \gb 0.675 \\
      &&  \texttt{VGG19}        & $\eta=2.75\e{-7}$, $b=16$ & 72 & 0.703  &    \gb 0.766   &   0.714   & 0.670 \\
      \arrayrulecolor{gray!50!}\midrule

      \multirow{4}{\pl}{Trajectories with line thickness}
      &  \multirow{4}{*}{\fpic{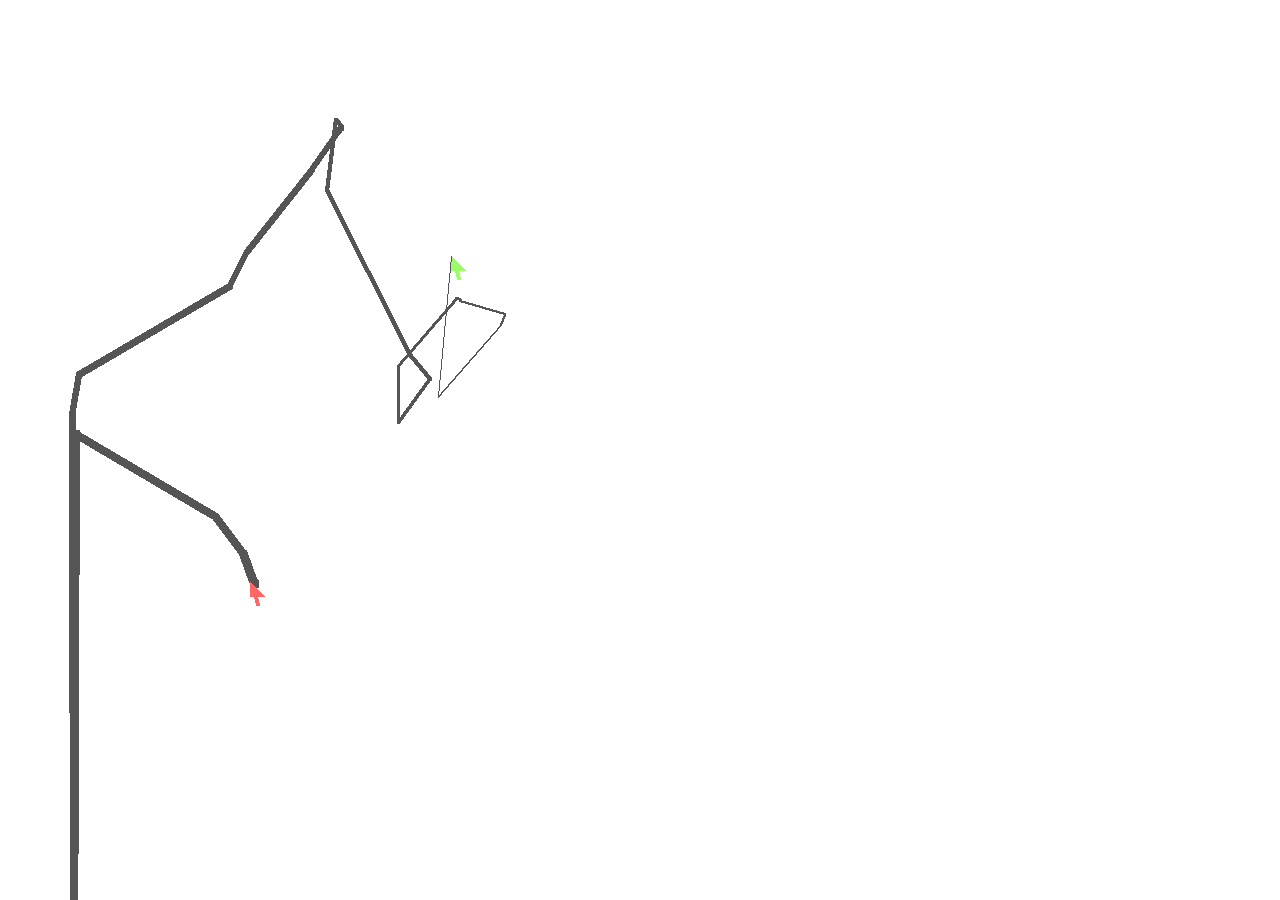}}
      &  \texttt{AlexNet}       & $\eta=1.31\e{-6}$, $b=64$ & 62 & 0.690  &    0.720   &   0.703   & 0.606 \\
      &&  \texttt{SqueezeNet}   & $\eta=3.63\e{-7}$, $b=64$ & 31 & \gb 0.725  &    \gb 0.738   &   \gb 0.737   & \gb 0.637 \\
      &&  \texttt{ResNet50}     & $\eta=8.06\e{-7}$, $b=32$ & 39 & 0.692  &    0.732   &   0.709   & 0.604 \\
      &&  \texttt{VGG19}        & $\eta=2.51\e{-7}$, $b=16$ & 95 & 0.710  &    0.682   &   0.695   & 0.572 \\
      \arrayrulecolor{gray!50!}\midrule

      \multirow{4}{\pl}{Trajectories with line thickness and ad placeholder}
      & \multirow{4}{*}{\fpic{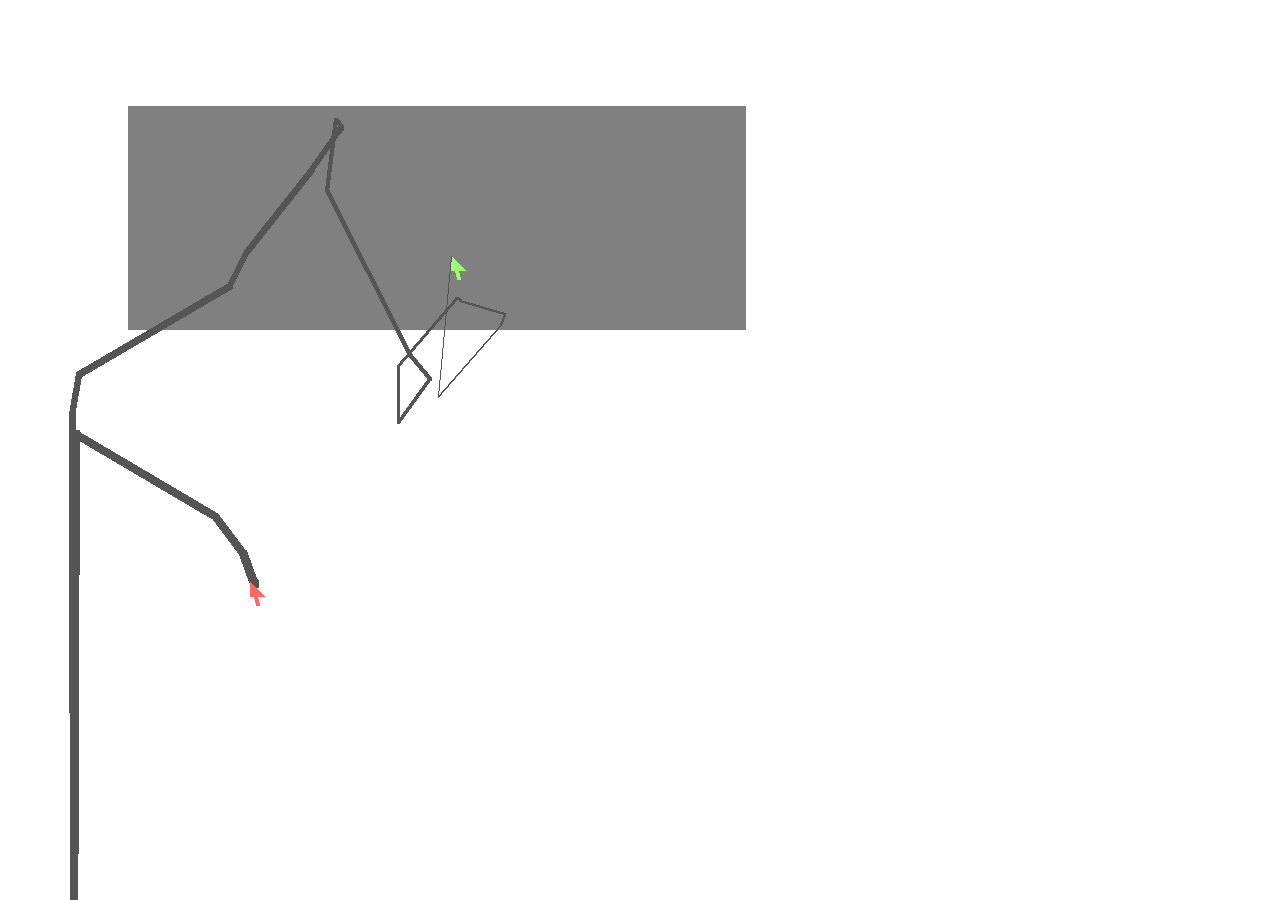}}
      &  \texttt{AlexNet}       & $\eta=3.31\e{-5}$, $b=64$ & 31 & 0.705  &    0.732   &   0.717   & \gb 0.664 \\
      &&  \texttt{SqueezeNet}   & $\eta=2.51\e{-5}$, $b=64$ & 90 & 0.715  &    \gb 0.752   &   0.725   & 0.654 \\
      &&  \texttt{ResNet50}     & $\eta=9.12\e{-7}$, $b=32$ & 72 & 0.709  &    0.740   &   0.720   & 0.653 \\
      &&  \texttt{VGG19}        & $\eta=1.31\e{-6}$, $b=16$ & 38 & \gb 0.733  &    0.730   &   \gb 0.735   & 0.616 \\
      \arrayrulecolor{gray!50!}\midrule

      \multirow{4}{\pl}{Colored trajectories with line thickness}
      & \multirow{4}{*}{\fpic{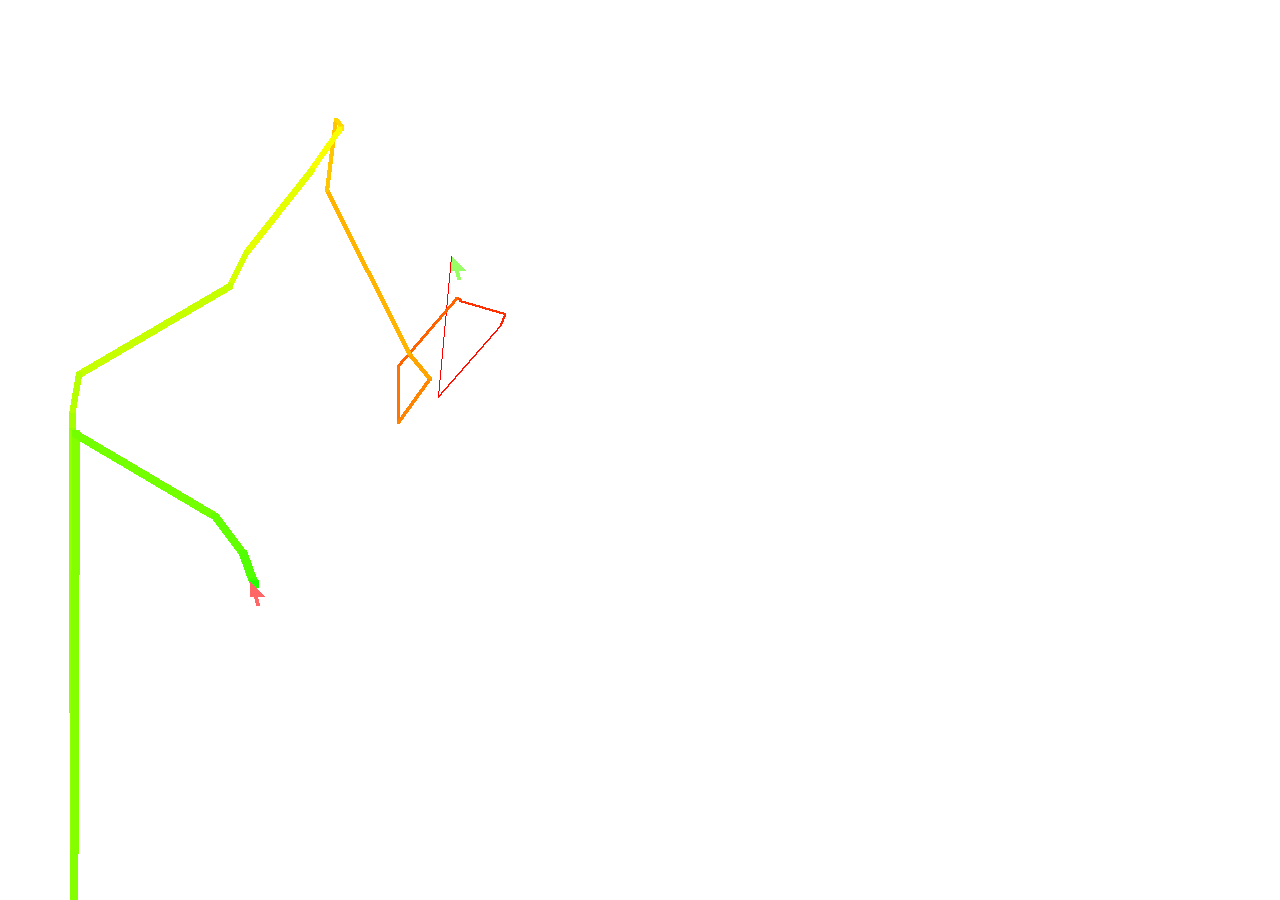}}
      &  \texttt{AlexNet}       & $\eta=1.20\e{-6}$, $b=64$ & 31  & 0.682  &    0.659   &   0.670   & 0.604 \\
      &&  \texttt{SqueezeNet}   & $\eta=9.12\e{-7}$, $b=64$ & 47  & 0.682  &    \gb 0.682   &   \gb 0.682   & 0.593 \\
      &&  \texttt{ResNet50}     & $\eta=3.41\e{-7}$, $b=32$ & 116 & 0.667  &    0.593   &   0.623   & 0.588 \\
      &&  \texttt{VGG19}        & $\eta=2.51\e{-7}$, $b=16$ & 50  & \gb\bf 0.760  &    0.615   &   0.652   & \gb 0.668 \\
      \arrayrulecolor{gray!50!}\midrule

      \multirow{4}{\pl}{Colored trajectories with line thickness and ad placeholder}
      & \multirow{4}{*}{\fpic{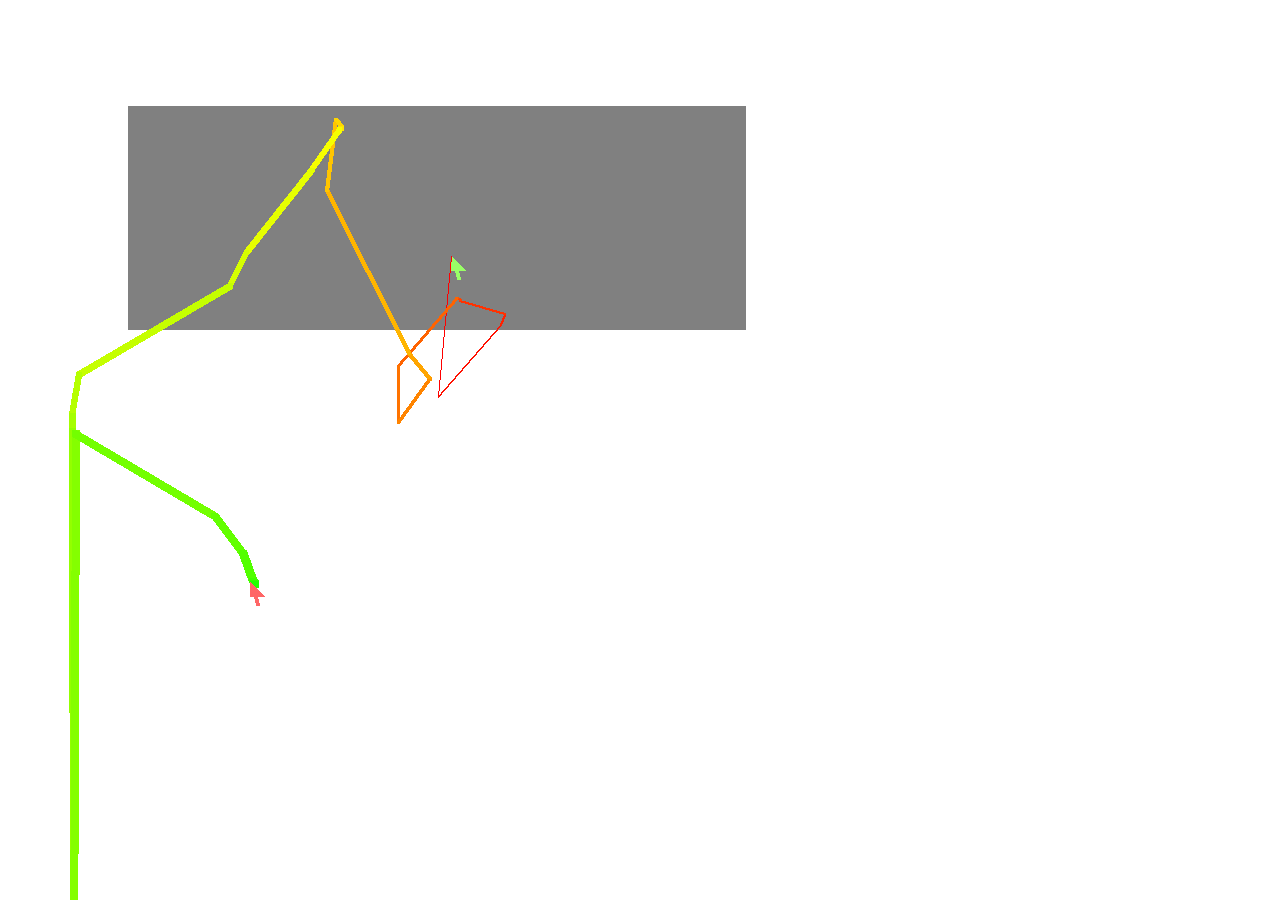}}
      &  \texttt{AlexNet}       & $\eta=1.73\e{-4}$, $b=64$ & 38 & 0.672  &    0.704   &   0.687   & 0.586 \\
      &&  \texttt{SqueezeNet}   & $\eta=1.00\e{-6}$, $b=64$ & 64 & \gb 0.725  &    0.706   &   0.715   & \gb 0.669 \\
      &&  \texttt{ResNet50}     & $\eta=4.36\e{-7}$, $b=32$ & 54 & 0.712  &    0.591   &   0.629   & 0.638 \\
      &&  \texttt{VGG19}        & $\eta=2.51\e{-7}$, $b=16$ & 46 & 0.717  &    \gb 0.726   &   \gb 0.722   & 0.652 \\
      \arrayrulecolor{black}\bottomrule
    \end{tabular}
    }
\end{table*}
\egroup

\bgroup
\def\arraystretch{0.9}
\begin{table*}[!ht]
    \caption{
        Results for right-aligned direct display ads.
        Gray cells indicate the top performer in each representation group.
        The overall best performance result (across all groups) is denoted in bold typeface.
        The positive:negative ratio is 462:178.
    }
    \vspace{-5pt}
    \label{tbl:results_top_right_dd}
    \centering
    {\footnotesize
    \begin{tabular}[t]{m{\pl} m{1.35cm} l *8c}
      \toprule
        \textbf{Representation}
        & \textbf{Example}
        & \textbf{Architecture}
        & \textbf{Hyperparameters}
        & \textbf{Epoch}
        & \textbf{Adj. Precision}
        & \textbf{Adj. Recall}
        & \textbf{Adj. F-measure}
        & \textbf{AUC} \\
      \midrule
      \multirow{4}{\pl}{Time series}
      &  \multirow{4}{*}{$\begin{matrix}(x_1,y_1),\\ \dots,\\ (x_N,y_N)\end{matrix}$}
      &  \texttt{SimpleRNN} & $\eta=10^{-3}$, $q=0.3$, $n=32$, $b=64$ & 56 & 0.577      & 0.677     & 0.572     & 0.530 \\
      && \texttt{LSTM}      & $\eta=10^{-3}$, $q=0.4$, $n=32$, $b=32$ & 27 & 0.560      & 0.643     & 0.566     & 0.511 \\
      &&  \texttt{BLSTM}    & $\eta=10^{-3}$, $q=0.5$, $n=48$, $b=16$ & 82 & \gb 0.608  & 0.658     & \gb 0.615 & \gb 0.614 \\
      &&  \texttt{GRU}      & $\eta=10^{-3}$, $q=0.4$, $n=32$, $b=64$ & 48 & 0.550      & \gb 0.678 & 0.564     & 0.561 \\
      \arrayrulecolor{gray!50!}\midrule

      \multirow{4}{\pl}{Heatmap}
      &  \multirow{4}{*}{\fpic{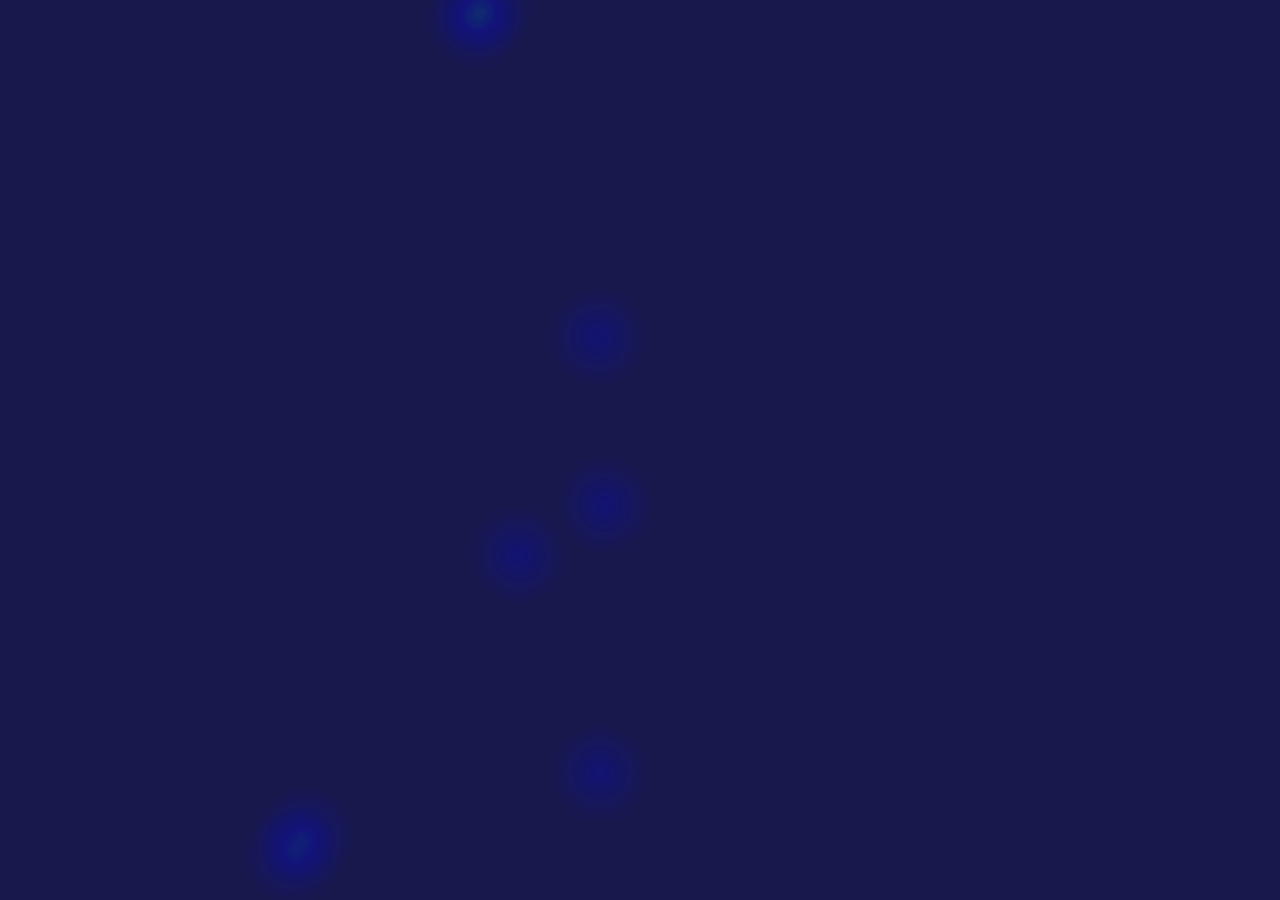}}
      &  \texttt{AlexNet}       & $\eta=4.78\e{-7}$, $b=64$ & 60 & \gb\bf 0.743  &    \gb 0.721   &   0.630   & 0.566 \\
      &&  \texttt{SqueezeNet}   & $\eta=3.63\e{-7}$, $b=64$ & 59 & 0.647  &    0.708   &   0.636   & \gb 0.599 \\
      &&  \texttt{ResNet50}     & $\eta=8.07\e{-7}$, $b=32$ & 39 & 0.668  &    0.698   &   \gb 0.668   & \gb 0.599 \\
      &&  \texttt{VGG19}        & $\eta=3.02\e{-7}$, $b=16$ & 31 & 0.611  &    0.394   &   0.369   & 0.525 \\
      \arrayrulecolor{gray!50!}\midrule

      \multirow{4}{\pl}{Heatmap with ad placeholder}
      &  \multirow{4}{*}{\fpic{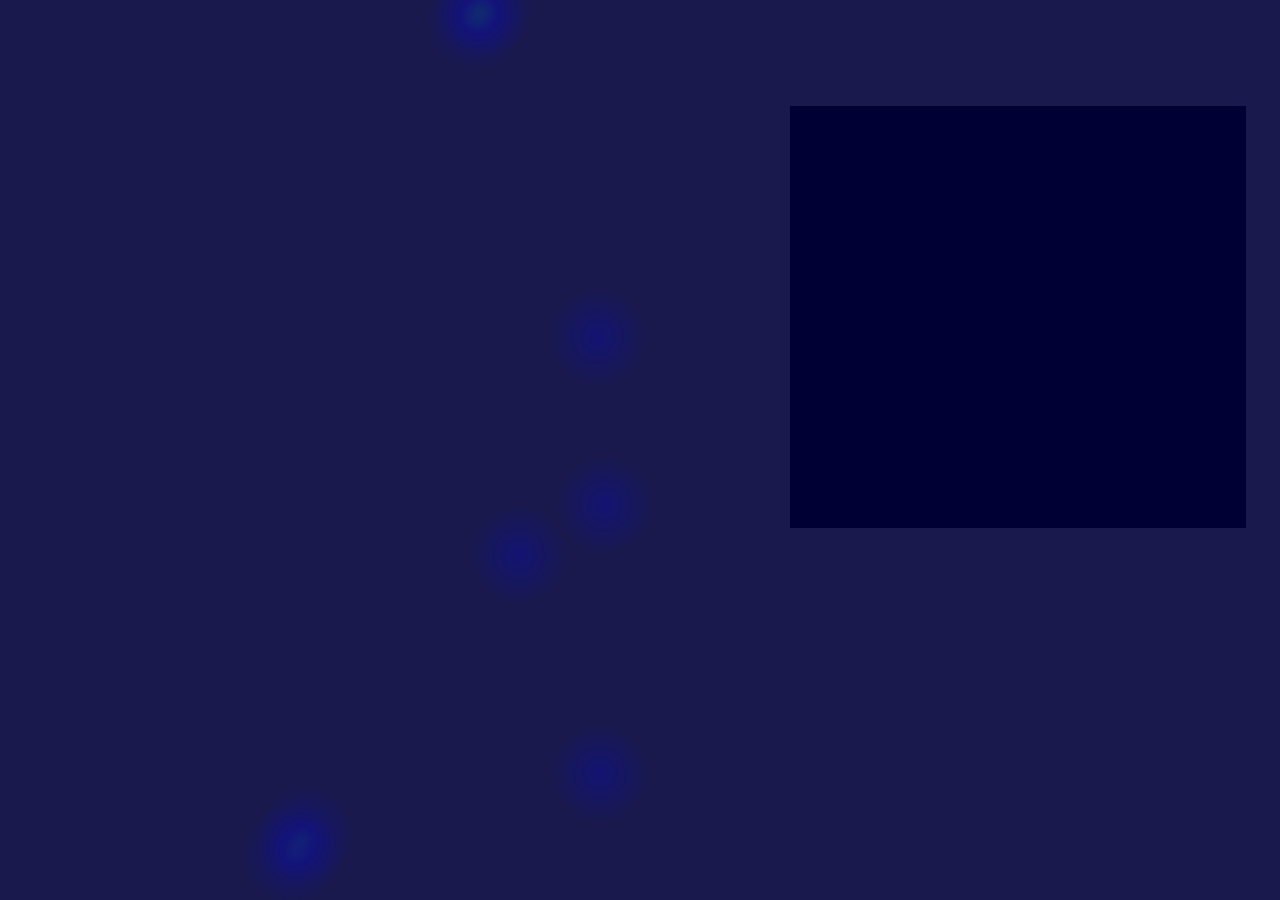}}
      &  \texttt{AlexNet}       & $\eta=4.78\e{-7}$, $b=64$ & 60 & 0.587  &    \gb 0.697   &   0.598   & 0.566 \\
      &&  \texttt{SqueezeNet}   & $\eta=6.91\e{-7}$, $b=64$ & 70 & 0.568  &    0.652   &   0.593   & 0.609 \\
      &&  \texttt{ResNet50}     & $\eta=1.44\e{-6}$, $b=32$ & 57 & 0.642  &    0.633   &   0.638   & 0.607 \\
      &&  \texttt{VGG19}        & $\eta=2.75\e{-7}$, $b=16$ & 32 & \gb 0.679  &    0.685   &   \gb 0.681   & \gb 0.680 \\
      \arrayrulecolor{gray!50!}\midrule

      \multirow{4}{\pl}{Trajectories}
      &  \multirow{4}{*}{\fpic{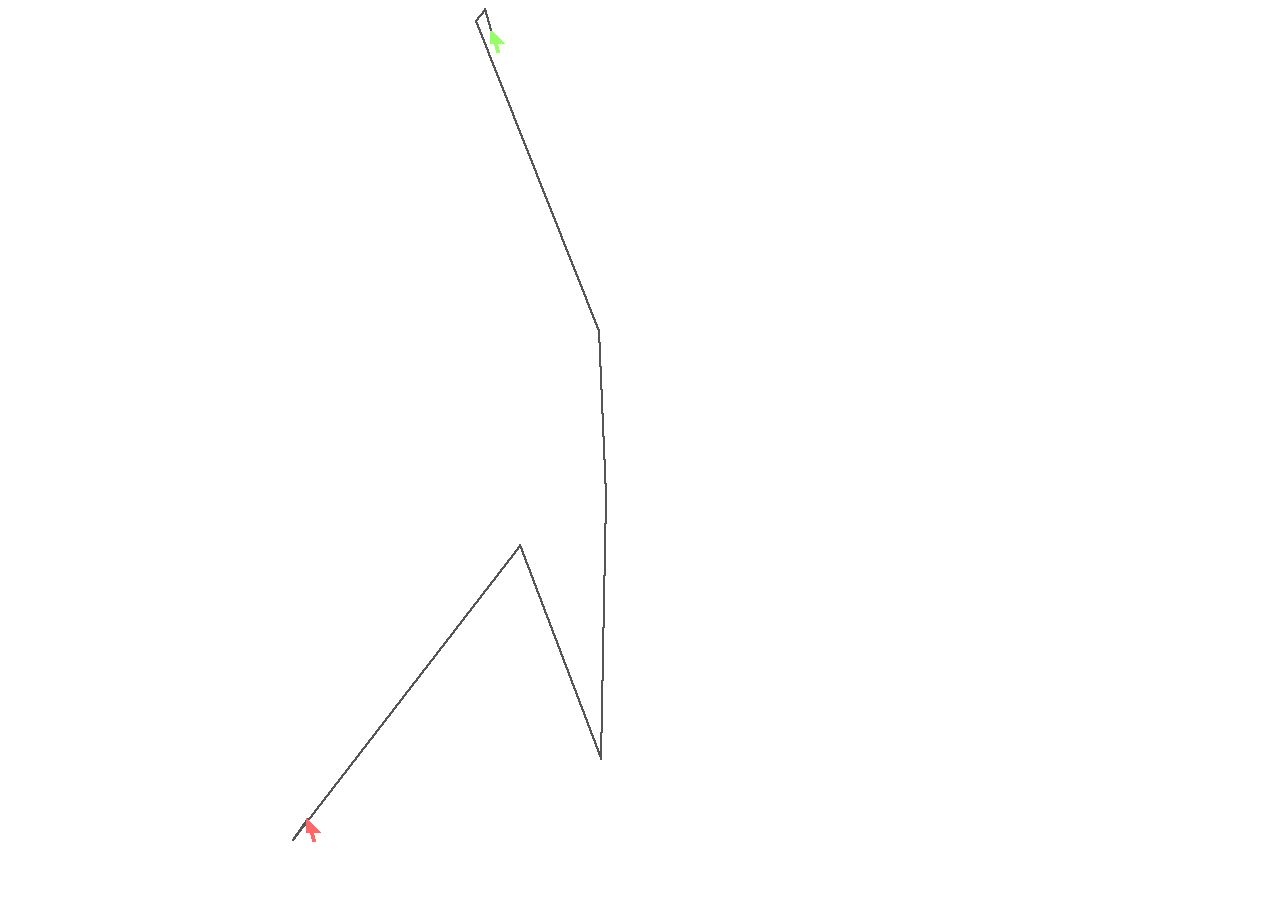}}
      &  \texttt{AlexNet}       & $\eta=5.75\e{-7}$, $b=64$ & 79 & 0.623  &    0.687   &   0.626   & 0.578 \\
      &&  \texttt{SqueezeNet}   & $\eta=2.75\e{-7}$, $b=64$ & 56 & 0.606  &    0.632   &   0.613   & 0.584 \\
      &&  \texttt{ResNet50}     & $\eta=1.49\e{-6}$, $b=32$ & 57 & \gb 0.726  &    \gb\bf 0.732   &   \gb\bf 0.731   & \gb\bf 0.739 \\
      &&  \texttt{VGG19}        & $\eta=3.98\e{-7}$, $b=16$ & 73 & 0.618  &    0.673   &   0.626   & 0.581 \\
      \arrayrulecolor{gray!50!}\midrule

      \multirow{4}{\pl}{Trajectories with ad placeholder}
      &  \multirow{4}{*}{\fpic{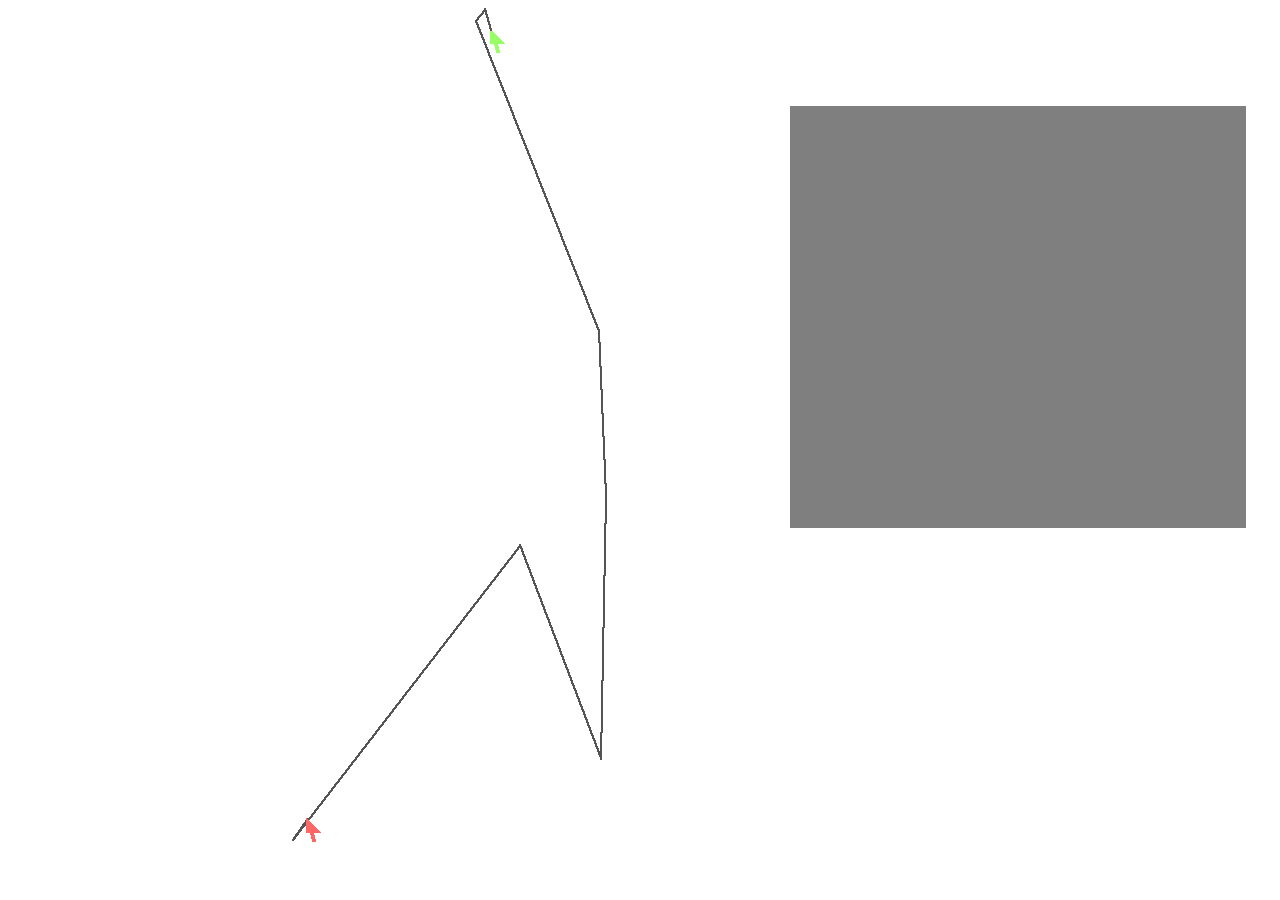}}
      &  \texttt{AlexNet}       & $\eta=5.75\e{-7}$, $b=64$ & 32 & 0.612  &    0.662   &   0.629   & 0.561 \\
      &&  \texttt{SqueezeNet}   & $\eta=2.51\e{-7}$, $b=64$ & 39 & 0.616  &    0.612   &   0.609   & \gb 0.607 \\
      &&  \texttt{ResNet50}     & $\eta=7.35\e{-7}$, $b=32$ & 31 & \gb 0.646  &    0.676   &   \gb 0.658   & 0.602 \\
      &&  \texttt{VGG19}        & $\eta=4.93\e{-7}$, $b=16$ & 36 & 0.608  &    \gb 0.690   &   0.614   & 0.596 \\
      \arrayrulecolor{gray!50!}\midrule

      \multirow{4}{\pl}{Colored trajectories}
      &  \multirow{4}{*}{\fpic{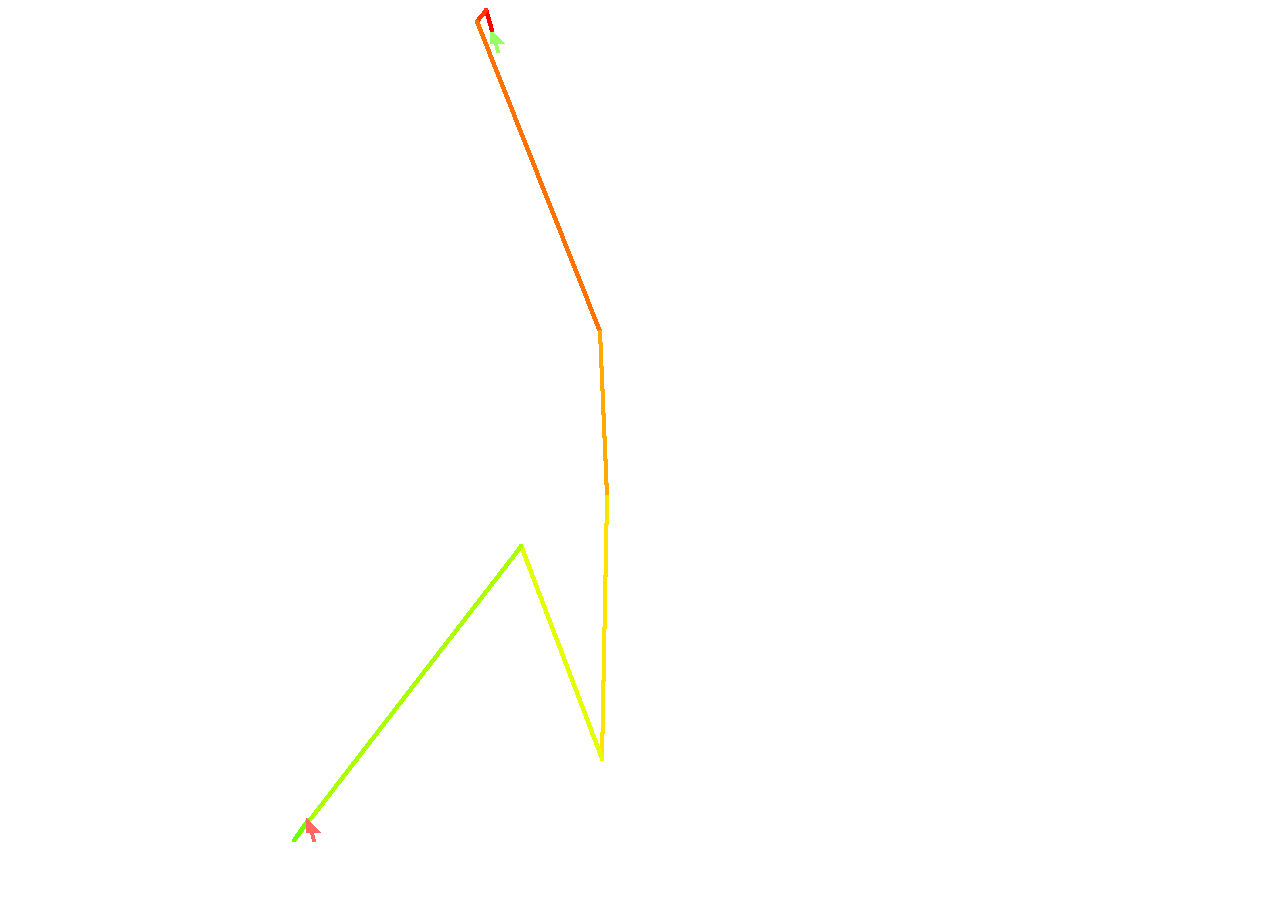}}
      &  \texttt{AlexNet}       & $\eta=2.51\e{-7}$, $b=64$ & 118 & 0.620  &    0.676   &   0.632   & 0.607 \\
      &&  \texttt{SqueezeNet}   & $\eta=6.91\e{-7}$, $b=64$ & 39 & 0.634  &    0.527   &   0.550   & 0.564 \\
      &&  \texttt{ResNet50}     & $\eta=2.15\e{-6}$, $b=32$ & 47 & \gb 0.639  &    0.687   &   \gb 0.636   & \gb 0.658 \\
      &&  \texttt{VGG19}        & $\eta=5.58\e{-7}$, $b=16$ & 53 & 0.599  &    \gb 0.695   &   0.606   & 0.644 \\
      \arrayrulecolor{gray!50!}\midrule

      \multirow{4}{\pl}{Colored trajectories with ad placeholder}
      &  \multirow{4}{*}{\fpic{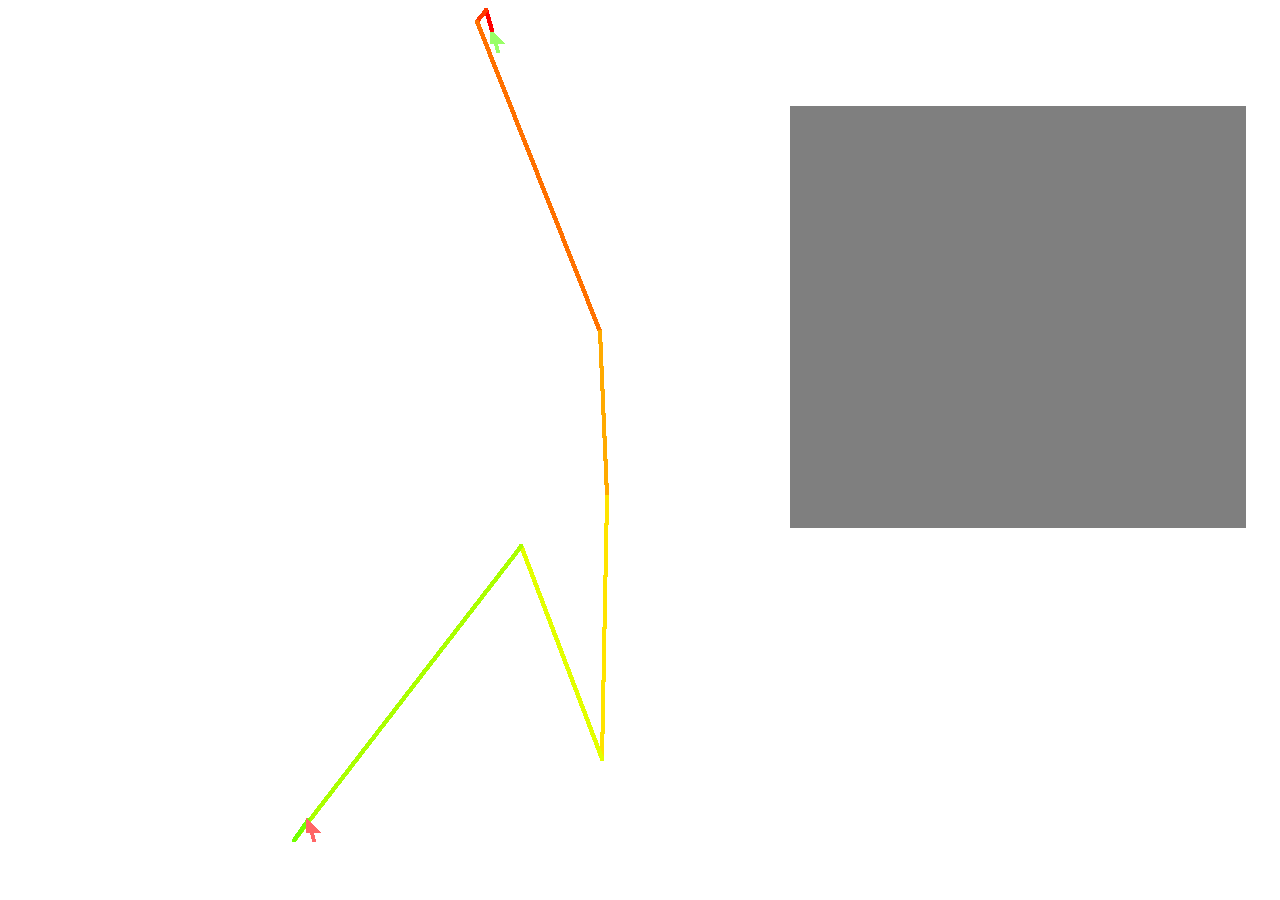}}
      &  \texttt{AlexNet}       & $\eta=2.51\e{-7}$, $b=64$ & 69 & 0.606  &    0.645   &   0.621   & 0.570 \\
      &&  \texttt{SqueezeNet}   & $\eta=2.75\e{-7}$, $b=64$ & 54 & 0.628  &    0.659   &   0.636   & 0.570 \\
      &&  \texttt{ResNet50}     & $\eta=1.49\e{-6}$, $b=32$ & 59 & \gb 0.697  &    \gb 0.725   &   0.652   & 0.640 \\
      &&  \texttt{VGG19}        & $\eta=4.36\e{-7}$, $b=16$ & 59 & 0.683  &    0.712   &   \gb 0.688   & \gb 0.679 \\
      \arrayrulecolor{gray!50!}\midrule

      \multirow{4}{\pl}{Trajectories with line thickness}
      &  \multirow{4}{*}{\fpic{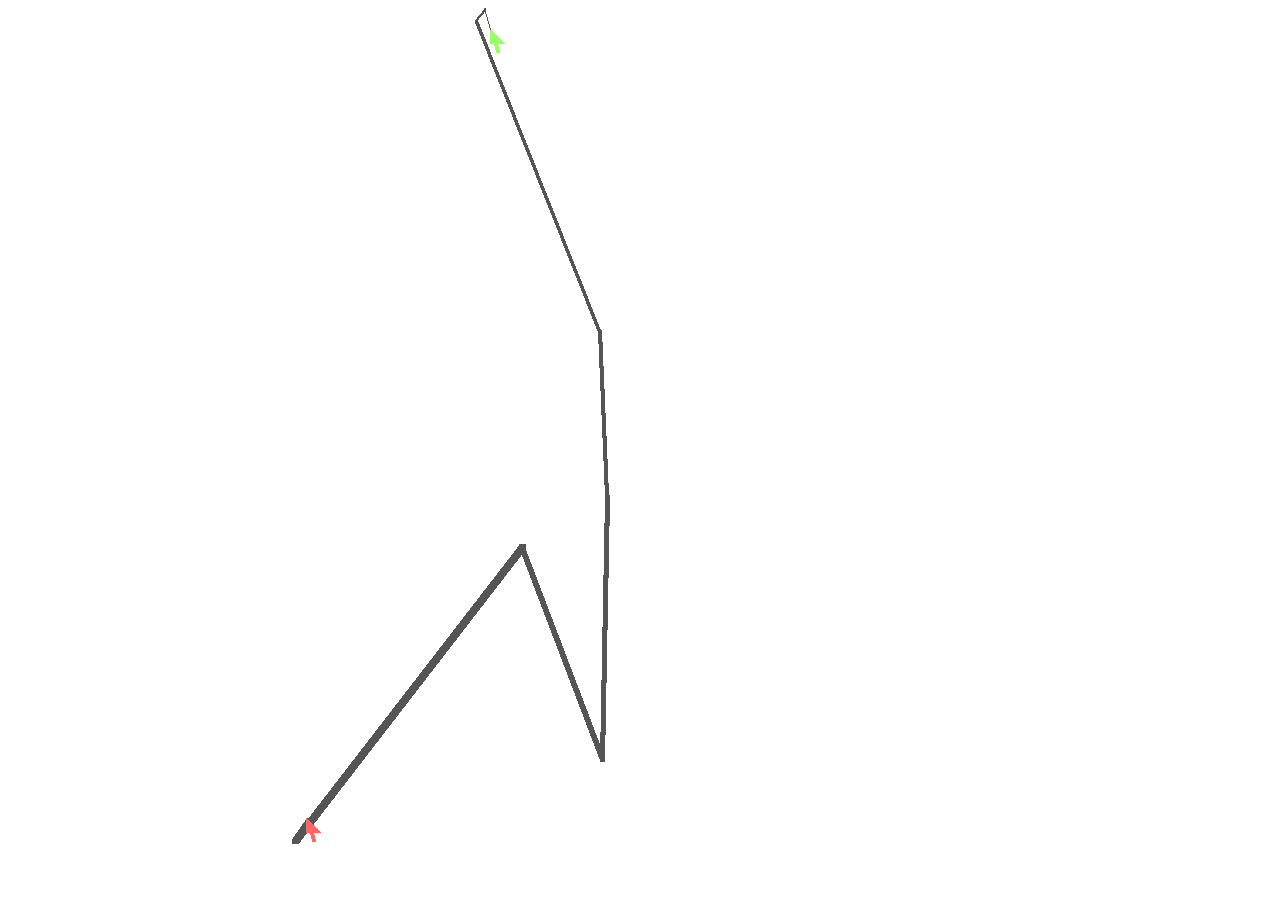}}
      &  \texttt{AlexNet}       & $\eta=4.78\e{-7}$, $b=64$ & 38 & 0.584  &    0.639   &   0.604   & 0.598 \\
      &&  \texttt{SqueezeNet}   & $\eta=2.51\e{-7}$, $b=64$ & 62 & \gb 0.708  &    0.678   &   \gb 0.683   & \gb 0.601 \\
      &&  \texttt{ResNet50}     & $\eta=4.93\e{-7}$, $b=32$ & 57 & 0.626  &    0.690   &   0.632   & 0.582 \\
      &&  \texttt{VGG19}        & $\eta=5.58\e{-7}$, $b=16$ & 67 & 0.655  &    \gb 0.691   &   0.669   & 0.567 \\
      \arrayrulecolor{gray!50!}\midrule

      \multirow{4}{\pl}{Trajectories with line thickness and ad placeholder}
      &  \multirow{4}{*}{\fpic{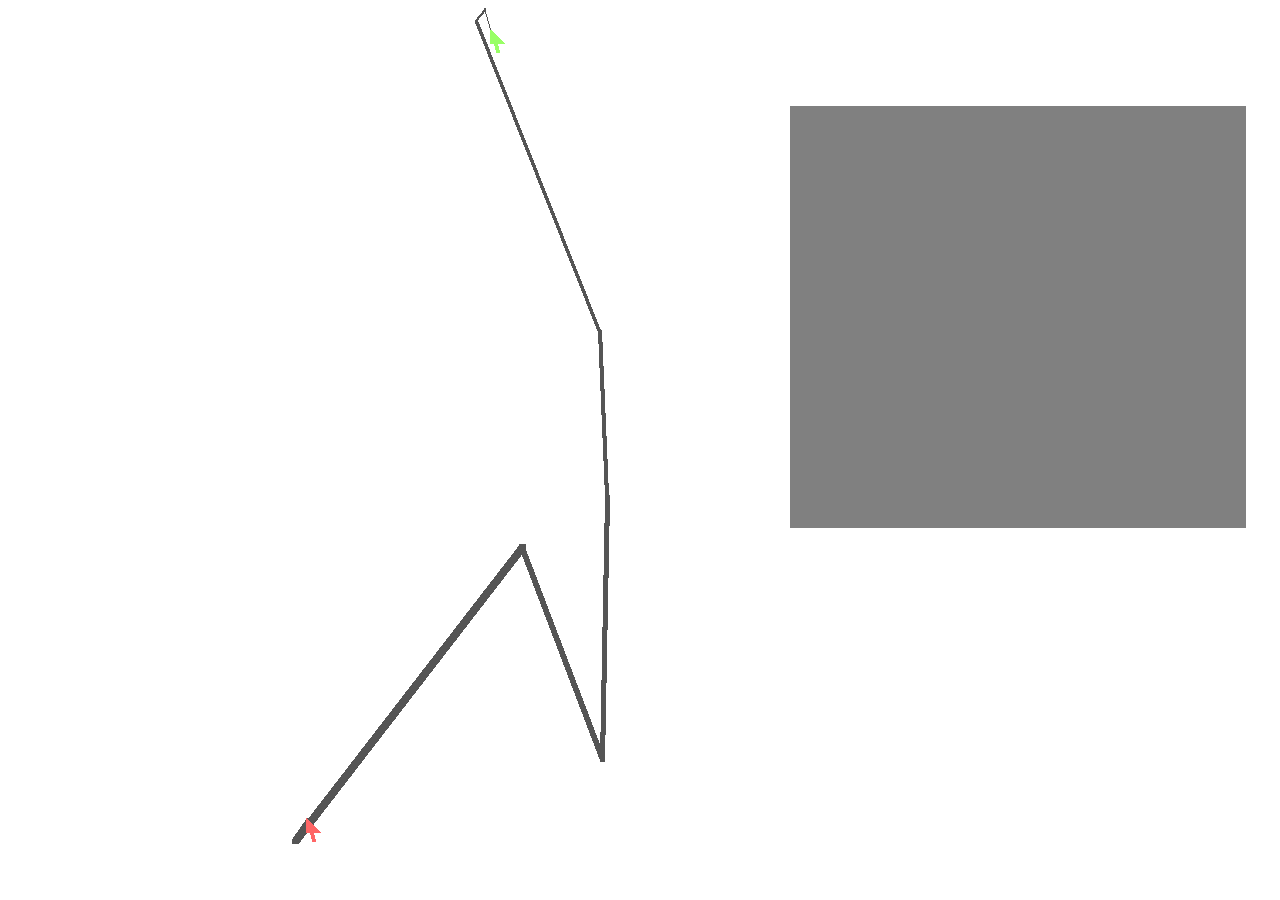}}
      &  \texttt{AlexNet}       & $\eta=6.31\e{-5}$, $b=64$ & 31 & 0.646  &    0.454   &   0.460   & 0.572 \\
      &&  \texttt{SqueezeNet}   & $\eta=3.02\e{-7}$, $b=64$ & 47 & 0.669  &    0.688   &   \gb 0.676   & 0.576 \\
      &&  \texttt{ResNet50}     & $\eta=6.31\e{-7}$, $b=32$ & 33 & 0.593  &    0.661   &   0.615   & \gb 0.596 \\
      &&  \texttt{VGG19}        & $\eta=2.43\e{-6}$, $b=16$ & 34 & \gb 0.670  &    \gb 0.716   &   0.637   & 0.573 \\
      \arrayrulecolor{gray!50!}\midrule

      \multirow{4}{\pl}{Colored trajectories with line thickness}
      & \multirow{4}{*}{\fpic{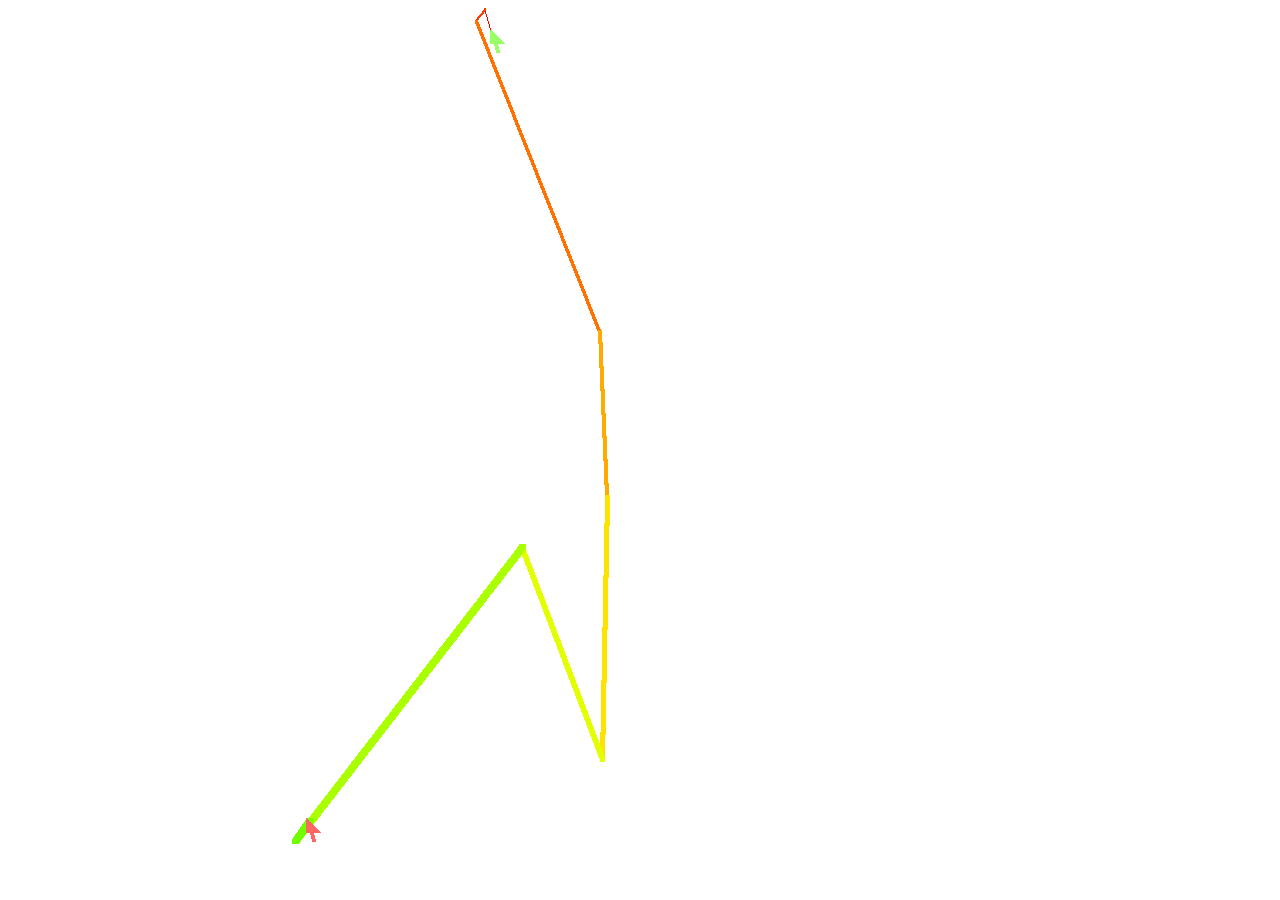}}
      &  \texttt{AlexNet}       & $\eta=5.75\e{-7}$, $b=64$ & 72 & 0.553  &    0.612   &   0.581   & 0.528 \\
      &&  \texttt{SqueezeNet}   & $\eta=2.51\e{-7}$, $b=64$ & 34 & 0.618  &    0.673   &   0.626   & 0.548 \\
      &&  \texttt{ResNet50}     & $\eta=1.03\e{-6}$, $b=32$ & 61 & \gb 0.694  &    \gb 0.711   &   \gb 0.705   & \gb 0.685 \\
      &&  \texttt{VGG19}        & $\eta=2.51\e{-7}$, $b=16$ & 73 & 0.626  &    0.590   &   0.605   & 0.588 \\
      \arrayrulecolor{gray!50!}\midrule

      \multirow{4}{\pl}{Colored trajectories with line thickness and ad placeholder}
      & \multirow{4}{*}{\fpic{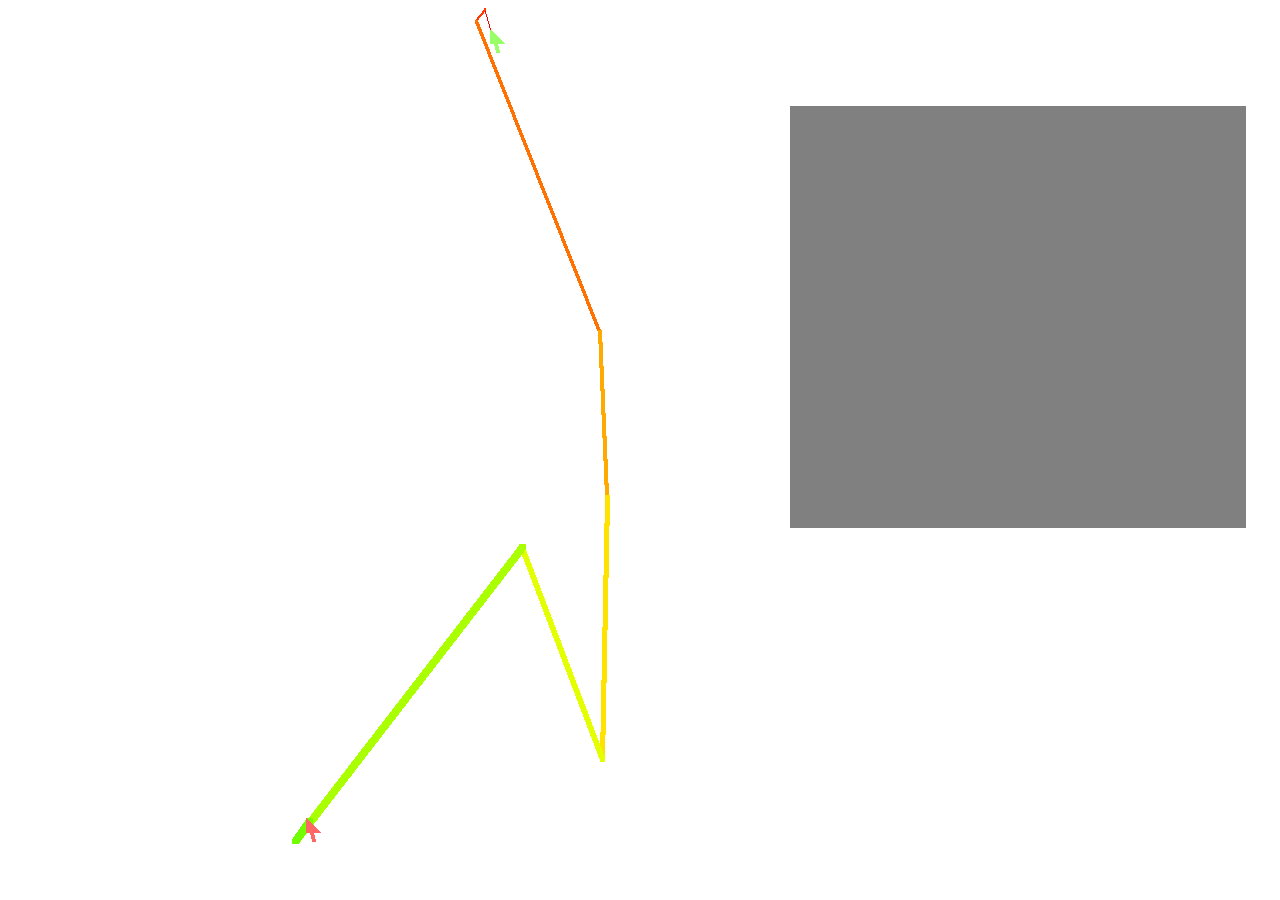}}
      &  \texttt{AlexNet}       & $\eta=4.36\e{-7}$, $b=64$ & 83 & 0.625  &    0.662   &   0.637   & 0.592 \\
      &&  \texttt{SqueezeNet}   & $\eta=1.90\e{-6}$, $b=64$ & 33 & 0.636  &    0.604   &   0.622   & 0.566 \\
      &&  \texttt{ResNet50}     & $\eta=3.86\e{-7}$, $b=32$ & 49 & 0.590  &    0.665   &   0.609   & 0.606 \\
      &&  \texttt{VGG19}        & $\eta=4.36\e{-7}$, $b=16$ & 63 & \gb 0.646  &    \gb 0.675   &   \gb 0.654   & \gb 0.633 \\
      \arrayrulecolor{black}\bottomrule
    \end{tabular}
    }
\end{table*}
\egroup

\section{Results}
\label{ssec:resutls}

We report the performance of our ANNs trained on the different data representations,
and for the different ad formats.
We use the standard IR metrics of Precision, Recall, and F-Measure (F1 score),
weighted according to the target class distributions in each case.
The F-Measure provides an aggregated insight about the functionality of a classifier,
however remaining sensitive to data distributions.
Therefore, we also report the Area Under the ROC curve (AUC),
which is insensitive to class distribution and error costs~\cite{Hand2001},
and use it as our key metric to determine the top performing classifier for each setup.
To investigate further the performance differences across models and conditions,
we run Friedman's ANOVA as an omnibus test and, if the result is statistically significant,
we use the Wilcoxon signed-rank test for pairwise comparisons, with correction for multiple testing.
Tables~\ref{tbl:results_top_left_native} to~\ref{tbl:results_top_right_dd} show the results of our experiments,
including the hyperparameter configuration used for each model.
The \texttt{Epoch} column indicates the maximum number of epochs used for training each model,
as we used early stopping to prevent overfitting.
Gray table cells indicate the top performer in each data representation group,
whereas the overall best performance result (across all representations) is denoted in bold typeface.

\subsection{Effect of Model Type}

We note that, under our experimental settings,
CNN models outperform RNN models across all ad format conditions, sometimes by a large margin.
When considering the best overall performing models for each type of ANN architecture,
we can observe noticeable improvements in terms of the F1 and AUC metrics.
More specifically, the best CNN model represents an increment over the best RNN model
by $3.24\%$ in terms of F1 and by $9.35\%$ in AUC for the organic ads (\autoref{tbl:results_top_left_native}).
Similarly, we observe an increment of $13.91\%$ (F1) and $26.42\%$ (AUC)
for the left-aligned direct display advertisements (\autoref{tbl:results_top_left_dd}).
Lastly, we note an increment of $18.65\%$ (F1) and $20.35\%$ (AUC)
for the right-aligned direct display advertisements (\autoref{tbl:results_top_right_dd}).

\subsection{Effect of Ad Placeholder}

We run statistical analysis to determine whether the presence of the ad placeholder
had any effect on the models' performance. 
For the organic ads and the right-aligned direct display advertisements,
the Wilcoxon signed-rank test showed that the presence of the ad placeholder in the representations
did not elicit a significant change in the AUC or F1 scores.
However, in the left-aligned direct display condition, F1 scores were significantly higher
for the models trained on the representations without the ad placeholder (Mdn=0.718),
as opposed to those trained on the representations with the ad placeholder (Mdn=0.691): $W=50, p = 0.041, r = -0.27$.

\subsection{Effect of Ad Format}

In organic ads, 
we note that the top performers are \texttt{SqueezeNet} (AUC=0.690), trained on the colored trajectories with varied line thickness,
and \texttt{ResNet50} (AUC=0.690), trained on the trajectories with ad placeholder.
Considering the remaining performance metrics, \texttt{SqueezeNet},
despite its shallower architecture (3 hidden layers),
seems to generalise better. 
In left-aligned direct display ads, 
the top performer is \texttt{AlexNet} (AUC=0.708), trained on the trajectories with ad placeholder,
followed closely by \texttt{VGG19} (AUC=0.694), trained on the heatmap with ad placeholder representation.
The Wilcoxon signed-rank test showed that the presence of the ad placeholder in the representations
did not elicit a significant change in the AUC or F-Measure scores, for either group.
Also, in right-aligned direct display ads, 
where the advertisement is clearly separated from the SERP results,
the top performing model is the \texttt{ResNet50}, trained on trajectories without ad placeholder,
This model holds the best AUC (0.739) and F-Measure (0.731) scores overall.

The Wilcoxon signed-rank test on all pairs of ad formats revealed a significant difference in terms of AUC
between the organic ad (Mdn=0.610) and the left-aligned direct display (Mdn=0.634):
$W=185.5, p < .01, r = -0.62$.
Similarly, we found a significant difference between the left-aligned (Mdn=0.634) and right-aligned direct displays (Mdn=0.594):
$W=716, p < .0001, r = -0.88$.
When examining the F-Measure, the Wilcoxon signed-rank test showed a significant difference between organic ads ($Mdn = 0.616$)
and the left-aligned direct display (Mdn=0.708): $W=17, p < .0001, r = -1.15$.
Furthermore, we observed a significant difference between the left-aligned (Mdn=0.708) and right-aligned direct displays (Mdn=0.629):
$W=788, p < .0001, r = -1.10$.
The observed effect sizes are rather large,
thus suggesting a practical importance of the results.

\section{Discussion and Future Work}
\label{sec:discussion}

This work has served as a first exploration on the feasibility of ANNs to predict user attention to ads on SERPs.
We have shown that, using relatively few training data, it is possible to train RNN models from scratch
and fine-tune existing CNNs via transfer learning.
Our findings indicate that the mouse cursor representations and the tested model architectures achieve competitive performance in detecting user attention for all ad formats. Note that none of our models use handcrafted features, which require domain expertise,
nor page-level information, since they are trained on raw sequences of mouse cursor movements.
Taken together, our experiments raise the bar in the IR community and
can inform researchers and practitioners when it comes to choosing one model or network configuration over another.

Having explored multiple representations of the same mouse cursor data, we have obtained several new insights and perspectives. For example, a times series representation of mouse movements is the obvious choice, if we already know that user interactions consist of a small number of mouse movements. On the contrary, if we foresee that users are going to dwell for a relatively long time on a page, e.g. due to query difficulty or the nature of the search task,
then an image-based representation would be a more apt choice.

Interestingly, our CNN models outperformed RRN models in most cases.
However, we note that this might be due to the fact that our RNN models had a limited sequence length,
in order to make training tractable on a single GPU,\footnote{
Even if a computing cluster were used, a single GPU is still required to do a single forward/backward pass during training.}
thereby limiting the learning capacity of these models.
On the contrary, the CNN models had almost full coverage of the mouse cursor movements,
since most user interactions happened above the fold,
and they were rendered as a static image, which can be easily trained on commodity hardware.

Regarding the CNN models, our experimental results indicate that, in most cases, the presence of the ad placeholder in the visual representation seems to benefit the models' performance, although that finding was not always statistically significant. In addition, the visual representations based on \emph{trajectories} and \emph{colored trajectories with variable line thickness} are consistently found amongst the top-ranked performers. We presume that embedding the temporal dimension into the representations plays a role in accurate prediction of visual attention. Furthermore, we observe that the CNNs that implement shallow architectures (e.g. \texttt{AlexNet} and \texttt{SqueezeNet}) appear to perform equally well, if not better, than their \emph{deeper} counterparts. This suggests that such CNN implementations can attain high accuracy while requiring less bandwidth to operate. Also, the application of transfer learning proved to be useful; hence, reusing existing architectures allows for a quick and inexpensive solution to visual attention prediction, with relatively few training data.

Finally, we should mention that our diagnostic technology was tested for the desktop setting,
and currently half of the web traffic is mobile.
However, user engagement is still higher on desktop~\cite{Arapakis20_ppaa}
and amounts for a profitable and sizeable percentage of web traffic.
A potential extension is to account for touch-based interactions like,
for example, zoom/pinch gestures and scroll activity~\cite{Guo:2013:MTI:2484028.2484100}.
Further ideas that could be explored in future work include:
benchmark custom CNN architectures (or even combine RNNs and CNNs),
analyze other color schema (e.g. for the ad placeholder color),
improve the prediction capabilities of our RNNs models
(e.g. stacking recurrent layers, using other activation functions,
or implementing self-attention mechanisms),
and train a general model to predict attention to any direct display.
Ultimately, modeling user attention on SERPs has wide-ranging applications in web search ranking and UI design, and this work paves the way to many exciting future directions for research in this topic.

\begin{acks}
I. Arapakis acknowledges the support of NVIDIA Coorporation with the donation of a Titan Xp GPU used for this research.
\end{acks}

\end{document}